\documentclass{article}

\PassOptionsToPackage{numbers}{natbib}
\usepackage[final,dandb]{neurips_2025}

\usepackage[utf8]{inputenc} 
\usepackage[T1]{fontenc}    
\usepackage{hyperref}       
\usepackage{url}            
\usepackage{booktabs}       
\usepackage{amsfonts}       
\usepackage{nicefrac}       
\usepackage{microtype}      
\usepackage{xcolor}         
\usepackage{listings}
\usepackage{makecell}
\usepackage{amsmath}
\usepackage{tabularx}
\usepackage{booktabs}
\usepackage{ragged2e}
\usepackage{array} 
\usepackage{multirow}
\usepackage{graphicx}
\usepackage{tcolorbox}

\definecolor{delim}{RGB}{20,105,176}
\definecolor{numb}{RGB}{106, 109, 32}
\definecolor{string}{rgb}{0.64,0.08,0.08}

\lstdefinelanguage{json}{
    numbers=left,
    numberstyle=\small,
    frame=single,
    rulecolor=\color{black},
    showspaces=false,
    showtabs=false,
    breaklines=true,
    postbreak=\raisebox{0ex}[0ex][0ex]{\ensuremath{\color{gray}\hookrightarrow\space}},
    breakatwhitespace=true,
    basicstyle=\ttfamily\small,
    upquote=true,
    morestring=[b]",
    stringstyle=\color{string},
    literate=
     *{0}{{{\color{numb}0}}}{1}
      {1}{{{\color{numb}1}}}{1}
      {2}{{{\color{numb}2}}}{1}
      {3}{{{\color{numb}3}}}{1}
      {4}{{{\color{numb}4}}}{1}
      {5}{{{\color{numb}5}}}{1}
      {6}{{{\color{numb}6}}}{1}
      {7}{{{\color{numb}7}}}{1}
      {8}{{{\color{numb}8}}}{1}
      {9}{{{\color{numb}9}}}{1}
      {\{}{{{\color{delim}{\{}}}}{1}
      {\}}{{{\color{delim}{\}}}}}{1}
      {[}{{{\color{delim}{[}}}}{1}
      {]}{{{\color{delim}{]}}}}{1},
}

\title{PANORAMA: A Dataset and Benchmarks Capturing Decision Trails and Rationales in Patent Examination}

\author{%
  \textnormal{
  \textbf{Hyunseung Lim}\textsuperscript{1}\quad
  \textbf{Sooyohn Nam}\textsuperscript{1}\quad
  \textbf{Sungmin Na}\textsuperscript{1}\quad
  \textbf{Ji Yong Cho}\textsuperscript{2}\quad
  \textbf{June Yong Yang}\textsuperscript{2}}\\
  \textnormal{
  \textbf{Hyungyu Shin}\textsuperscript{1}\quad
  \textbf{Yoonjoo Lee}\textsuperscript{1}\quad
  \textbf{Juho Kim}\textsuperscript{1}\quad
  \textbf{Moontae Lee}\textsuperscript{2,3}\quad
  \textbf{Hwajung Hong}\textsuperscript{1}}\\
  {\normalfont\centering\textsuperscript{1}KAIST\quad\quad\textsuperscript{2}LG AI Research\quad\quad\textsuperscript{3}University of Illinois Chicago}\\
}

\begin{document}

\maketitle

\begin{abstract}
 Patent examination remains an ongoing challenge in the NLP literature even after the advent of large language models (LLMs), as it requires an extensive yet nuanced human judgment on whether a submitted \textit{claim} meets the statutory standards of \textit{novelty} and \textit{non-obviousness} against previously granted claims—\textit{prior art}—in expert domains. Previous NLP studies have approached this challenge as a prediction task (e.g., forecasting grant outcomes) with high-level proxies such as similarity metrics or classifiers trained on historical labels. However, this approach often overlooks the step-by-step evaluations that examiners must make with profound information, including rationales for the decisions provided in \textit{office actions} documents, which also makes it harder to measure the current state of techniques in patent review processes. To fill this gap, we construct PANORAMA, a dataset of 8,143 U.S. patent examination records that preserves the full decision trails, including original applications, all cited references, \textit{Non-Final Rejections}, and \textit{Notices of Allowance}. Also, PANORAMA decomposes the trails into sequential benchmarks that emulate patent professionals' patent review processes and allow researchers to examine large language models' capabilities at each step of them. Our findings indicate that, although LLMs are relatively effective at retrieving relevant prior art and pinpointing the pertinent paragraphs, they struggle to assess the novelty and non-obviousness of patent claims. We discuss these results and argue that advancing NLP, including LLMs, in the patent domain requires a deeper understanding of real-world patent examination. Our dataset is openly available at \url{https://huggingface.co/datasets/LG-AI-Research/PANORAMA}.

\end{abstract}

\section{Introduction}\label{sec:introduction}

\begin{figure}
    \centering
    \includegraphics[width=1\linewidth]{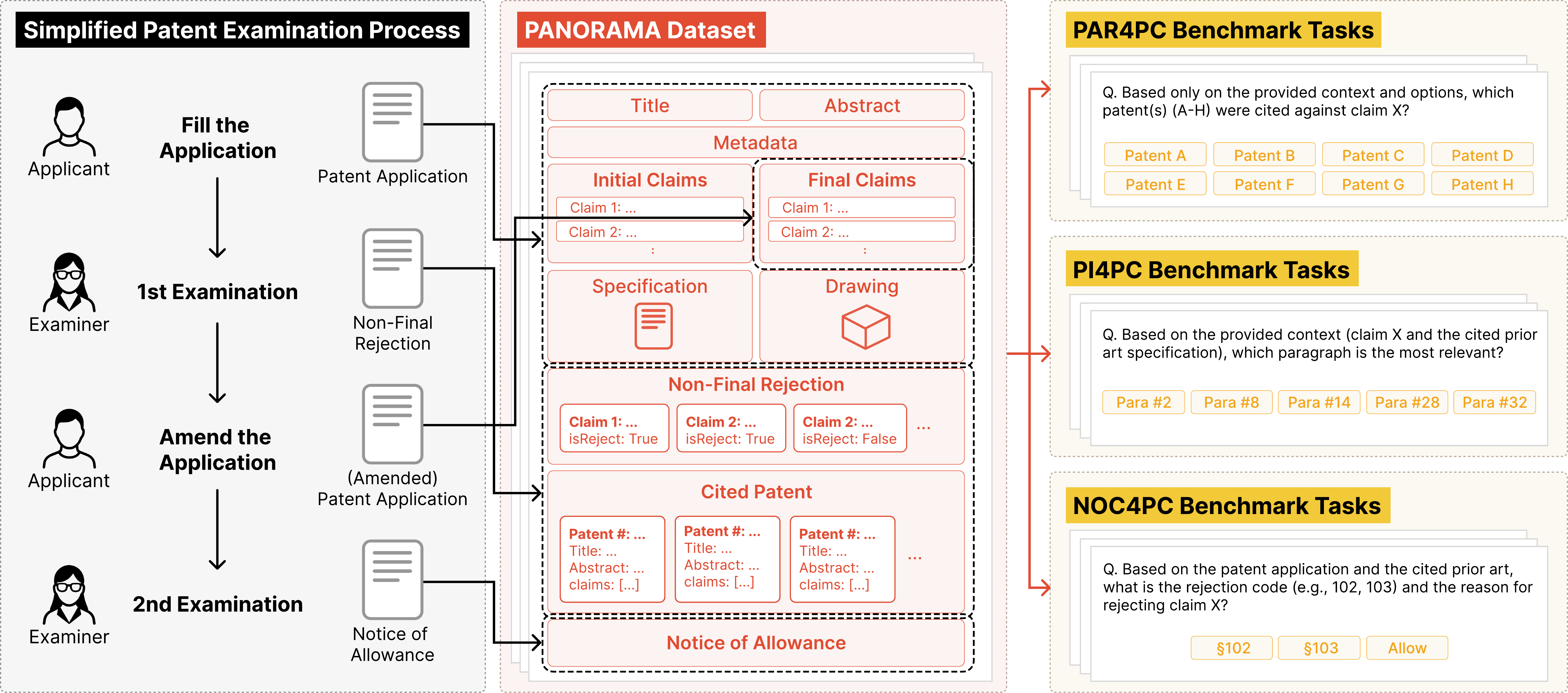}
    \caption{The overall framework of the PANORAMA dataset and benchmark construction. The PANORAMA dataset is constructed from documents appearing in patent examination. It comprises patent documents and office actions, such as non-final rejections. The benchmark tasks are designed to emulate the sequential nature of the patent examination process.}
    \label{fig:thumbnail}
    \vspace{-0.2cm}
\end{figure}

Patents play a fundamental role in driving innovation by granting inventors exclusive rights to their creations for a fixed period. Before a patent is issued, each application undergoes a rigorous examination process to assess its novelty and validity against existing prior art by patent examiners~\cite{ralf2021survey}. Amid the rapid global increase in patent applications and the introduction of large language models (LLMs), a growing interest lies in developing techniques to evaluate patentability efficiently and accurately, potentially with LLMs, including predicting grant outcomes \cite{Suzgun2022, Jiang2023}, patent valuation~\cite{Hu2023, Du2021, Lin2018, Chung2020, Liu2023}, litigation prediction~\cite{Campbell2016, Liu2018, Wu2023}, and determining novelty~\cite{Verhoeven2016, Plantec2021, Schmitt2023, Jeon2022, Wang2019, Zanella2023, Arts2021, beaty2021, Shibayama2021, Siddharth2020, Chikkamath2020, Jang2023}.

Despite many efforts to develop techniques to evaluate patentability, primarily but not limited to judging if an application is \textit{novel} and \textit{non-obvious}, most often fall short of performing human patent professionals' nuanced patentability evaluation ~\cite{Jiang2024, Schmitt2023}. In practice, determining whether a patent is novel or non-obvious requires deep domain expertise and comprehensive prior art searches involving detailed comparisons between existing inventions and new applications to support robust assessments~\cite{Schmitt2023}. It is essential to break down the human experts' workflow and decision trails into meaningful, feasible data and tasks to advance patent evaluation techniques.  

Our work aims to provide resources that unpack the complex evaluative processes of patent examiners. This perspective shifts attention from prediction accuracy to enabling models capable of performing patentability evaluation aligned with human expert judgment. We present \textbf{PANORAMA} (Patent Applications' Novelty and non-Obviousness Reasoning Analysis Model Assessment), a dataset that is curated with 8,143 authentic examples of Office Action (OA) documents written by patent examiners from the United States Patent and Trademark Office (USPTO) database spanning the past ten years (2015-2024) across domains. Each application contains not only a trajectory of a patent application---(1) initial patent applications, (2) cited patents, (3) Non-Final Rejections, (4) (revised) final patent claims, and (5) Notices of Allowance issued during the examination process Specifically---but also all the elements from the patent application and cited inventions, including detailed descriptions and drawings. This extensive and profound data offers patent examiners' decision trails and rationales behind their evaluations of patent applications. 

We also offer a suite of challenging sequential benchmark tasks designed to reflect the real-world patent examination workflow to showcase our dataset. They can provide a standardized evaluation framework for LLMs in patentability evaluation by (1) retrieving prior art pertinent to a specific claim (\textbf{PAR4PC}); (2) identifying the key paragraph(s) within cited prior art (\textbf{PI4PC}); and (3) making judgments of novelty and non-obviousness (\textbf{NOC4PC}).

We conduct baseline evaluations of proprietary and open-source LLMs on three tasks under zero-shot and chain-of-thought prompting and demonstrate supervised fine-tuning of open-source LLMs using our dataset.

In summary, the contributions of our study are threefold:

\begin{itemize}
    \item We curate PANORAMA, a dataset with rich information from diverse, authentic sources for patentability evaluation, emulating the human patent professionals' evaluation trails with rationales behind judgments.
    \item We propose three sequential benchmark tasks that break down the complex patentability evaluation workflow into measurable steps, offering a standard for evaluating LLMs' capabilities in patentability evaluation.
    \item We provide baseline evaluations of leading LLMs on these benchmark tasks, establishing initial performance levels and highlighting key areas for future model development in complex examination in patent domains.
\end{itemize}

\section{PANORAMA Dataset}\label{sec:dataset}

We present the PANORAMA dataset, which contains the decision trails and rationales by patent examiners to assess the patentability of applications, including novelty and non-obviousness. In this section, we describe an overview of the dataset and its curation process.

\subsection{Patent Examination Process and Dataset Overview}
The USPTO patent examination process involves evaluating a patent application and issuing an office action. An office action is an official document from an examiner during the patent or trademark examination process, outlining any objections, rejections, or required clarifications based on legal grounds (e.g., 35 U.S.C. §101, §102, §103, §112). Instead of an outright rejection, the inventor may respond by either amending the claims for reconsideration or by appealing the examiners' decision.

We identify the rationale behind rejected claims using two documents from the office action: \textit{Non-Final Rejection} and \textit{Notice of Allowance}. A Non-Final Rejection document outlines the reasons for rejecting an application and specifies the legal basis for each claim's rejection, while a Notice of Allowance indicates that the application has been approved and explains the examiner's rationale for allowing the revised claims. Our dataset maps initially rejected claims, reasons for rejection, revised claims, and reasons for final decision.

Reflecting the real-world patent examination process, we construct procedural data from 8,143 patent applications reviewed over the past ten years (2015 to 2024), with each record including the details of the patent application, cited patents, and corresponding office actions (Table~\ref{tab:data_structure}).

\begin{table}
    \renewcommand{\arraystretch}{1.4}
    {\footnotesize
    \begin{tabular}{c|p{7.8cm}|c}
        \Xhline{1.1pt}
        \textbf{Column} & \textbf{Description} & \textbf{Type} \\ \Xhline{1.1pt}
        Metadata & Metadata of patent application and office actions & JSON \\ \hline
        Title & Title of the invention & STRING \\ \hline
        Abstract & A brief summary of the invention and its purpose & STRING \\ \hline
        Initial Claims & Initial claims in the patent application (claims prior to receiving a non-final rejection) & STRING[] \\ \hline
        Final Claims & Final claims in the patent application (claims prior to receiving NOA(Notice of Allowance)) & STRING[] \\ \hline
        Specification & Specification document of the invention, which includes background and detailed description of the invention & STRING \\ \hline
        Drawing & Drawings of the invention & PDF \\ \hline
        Non-Final Rejection & Non-final rejection document of the application & STRING \\ \hline
        Notice of Allowance & Notice of allowance document of the application & STRING \\ \hline
        Cited Patents & Cited patents mentioned in Non-final rejection documents (each cited patent JSON includes title, abstract, claims, specification, and drawing) & JSON \\ \Xhline{1.1pt}
        Parsed Non-Final Rejection & Data parsed from the non-final rejection document into items such as whether the claim was rejected (\textbf{isRejected}), the legal basis code (\textbf{sectionCode}), the cited prior arts (\textbf{citedPatents}), and the rejection reasons (\textbf{reason}). & JSON \\ \Xhline{1.1pt}
    \end{tabular}
    }
    \caption{Brief description of PANORAMA dataset.}
    \label{tab:data_structure}
    \vspace{-0.5cm}
\end{table}

To extract evaluation rationales from Non-Final Rejection documents, we use LLMs to parse each document and identify the relevant reasoning. Since patent applications are examined at the individual claim level, we segment the rejection reasons by claim and extract their legal basis (e.g., §101, §102, §103, §112). As a single claim can be rejected for multiple reasons, it can be associated with multiple rejection bases. As both §102 and §103 require comparisons with prior art, we also collect prior art cited by the examiner and descriptions of comparisons, as well as justifications of the rejection. We generate a JSON-structured dataset, Parsed Non-Final Rejection, which contains the extracted evaluation rationales from the non-final rejection documents, including if the claim is rejected (\textbf{isRejected}), its legal basis code (\textbf{sectionCode}), cited patents (i.e., prior art) (\textbf{citedPatents}), and rejection rationales (\textbf{reason}).

\subsection{Data Curation}

\begin{figure}
    \centering
    \includegraphics[width=1\linewidth]{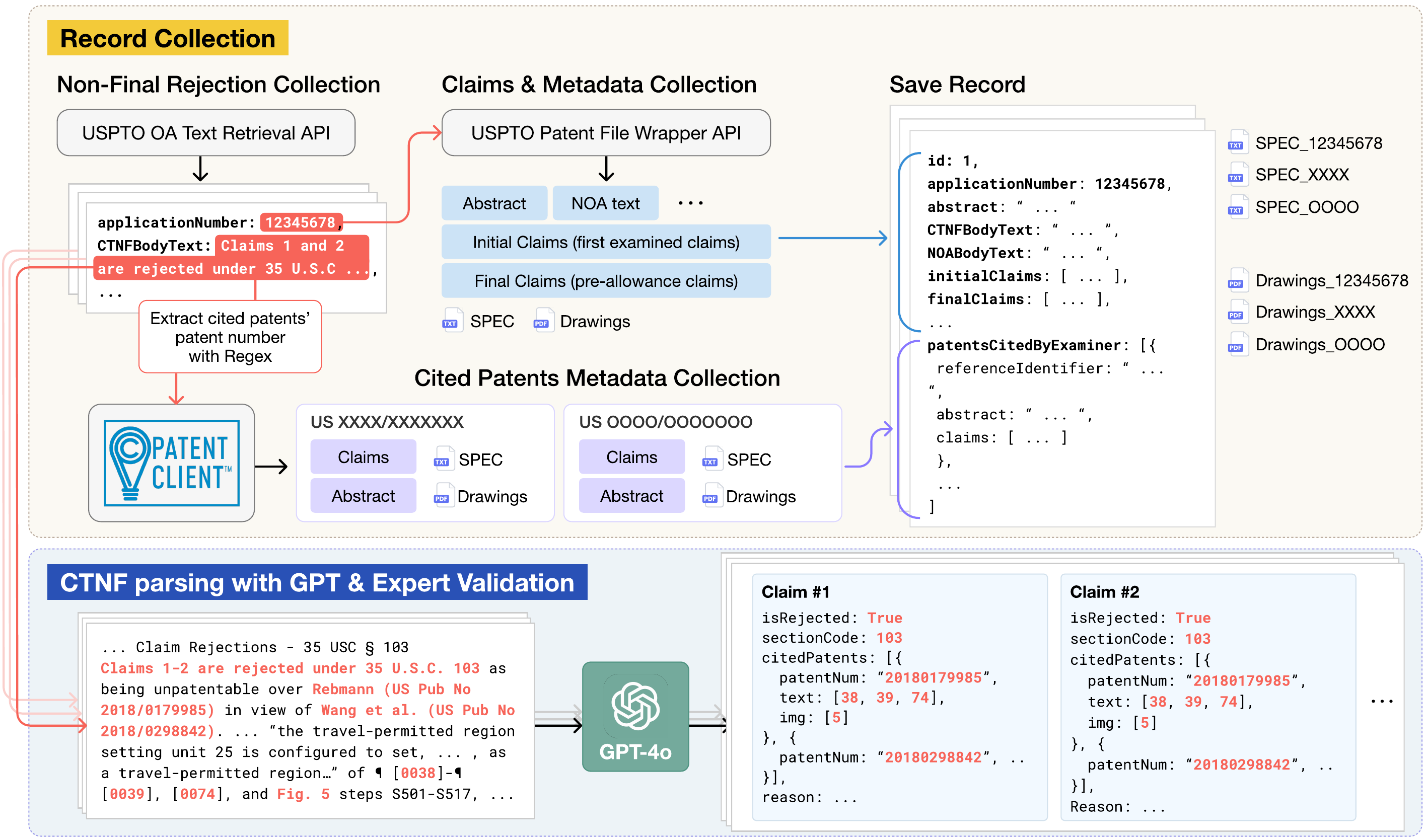}
    \caption{An overview of the PANORAMA dataset curation pipeline.}
    \label{fig:enter-label}
    \vspace{-0.2cm}
\end{figure}

Curating our data involves three steps: (1) We initially collect raw data from verified sources, and (2) then curate Non-Final Rejection documents–text documents–into JSON with the help of LLMs. (3) Lastly, we conduct a study with patent professionals to validate the data.

\subsubsection{Data Collection}
All patent documents are publicly accessible by law and available via the USPTO API. However, since the data we aimed to collect is distributed across different APIs within the USPTO API library, and mapping each data point accurately to its corresponding application was essential, we develop our own customized process to accomplish this task.

We use the USPTO OA Text Retrieval API to collect Non-final Rejection documents. Since this API returns various office actions--such as Non-final Rejection, Restriction/Election Requirement, and Final Rejection--we filter the results to include only Non-Final Rejection documents. Additionally, because an application may have multiple rejection notices if submitted multiple times, we ensure consistency by collecting only the first Non-Final Rejection, avoiding redundancy.

Next, we extract the patents cited by the examiner from the Non-Final Rejection document and collected the corresponding cited patents. Although these patents are also listed in the application, we extract them separately to gather the minimum number of cited patents necessary for the rejection. We use regular expressions to identify patent numbers in the format (e.g., Patent Application No. US 2025/1234567; Patent No. US 12345678) and the patent\_client Library to extract the specification, abstract, and claims of the cited patents. For drawings, which are not provided by the  patent\_client library, we extracted them separately using the USPTO ODP beta API.

Following this, we collect the metadata of the application using the USPTO ODP Beta API, including the title, abstract, specification, drawing, and Notice of Allowance document, based on the patent application number. We then gather the claims--those written just before the Non-Final Rejection and Notice of Allowance documents--as initial claims and final claims, respectively. All data, except for the specification and drawing, are saved in a single JSON file, while the specification and drawing are stored in separate folders. In total, 12,839 data entries are collected, with details provided in the App.\ref{data_collection}.

\subsubsection{Parsing Non-Final Rejection Documents}
\label{sec:parsing}
We parse the Non-Final Rejection document to extract and structure the rationale for evaluating the patent applications. Because patent applications are evaluated claim by claim, patent examiners record, for each claim, whether it is rejected and the grounds for that decision. Reflecting this structure, we convert the examiner's rationale into claim-level data drawn from the Non-Final Rejection documents.

We used LLMs (GPT-4o~\cite{openai2024gpt4ocard}) to parse the Non-Final Rejection documents consistently, as they contain standard terms and tacit knowledge, though their organization may vary depending on the patent application and examiner. First, we categorize the documents by patent claims and classify each claim as rejected or accepted (\textbf{isRejected}). Then, we categorize the claims into specific legal basis codes (\textbf{sectionCode})--§101, §102, §103, and §112--which are the most common codes for rejecting claims. For §102 and §103 rejections, we additionally extracted every cited prior art (\textbf{citedPatents}), the exact paragraphs the examiner used, and, when indicated, the specific figure elements used in the prior art comparison. In addition, we capture the full written rationale for each rejection (\textbf{reason}), which, for §102 and §103, explicitly incorporates the identified prior art. Details of the procedure and the prompt driving this parsing are provided in the App.\ref{parsed_ctnf}.

After parsing with LLMs, we conduct a systematic refinement process to identify and correct issues, such as missing evidence for specific claims, incorrectly cited inventions, and other inconsistencies. This process results in a final collection of 8,143 data entries, with details provided in the App.\ref{validation_procedure}.

\subsection{Expert Validation}
We validate the parsed data with seven domain experts to ensure that our data schema accurately reflects the decision trails and rationales in Non‑Final Rejection documents and that the LLM parses are accurate. We randomly sample 100 applications by selecting 20 patent applications (which contain 10 to 26 claims, a range within the mean $\pm 1\sigma$) across the five most frequent technology domains: Circuit‑Signal, Device‑Hardware, IT‑Data Processing, Manufacturing‑Mechanics, and Chemistry‑Bio. The validation tasks are divided into 60 bundles (100 claims × 3 reviews / 5 documents), ensuring that every document was examined by at least three experts. Each bundle contains five Non‑Final Rejection documents from the same domain and requires approximately 1.5 hours to review.

To facilitate a comparative review between the original Non‑Final Rejection documents and the parsed output, we develop a web‑based evaluation interface (App.~\ref{expert_evaluation_system}) that displays both side‑by‑side. For every claim, patent professionals are asked to flag errors in four fields: the rejection status (\textbf{isReject}), cited prior art (\textbf{citedPatents}), legal basis code (\textbf{sectionCode}), and accompanying rationale (\textbf{reason}).

On average, each expert reviews 8.57 bundles (SD = 6.76). Expert review indicated that the parser accurately extracted data for 92.5\% of claims. Across 100 applications (1,874 claims), experts flag 141 claims as erroneous, of which 132 involved an incorrect legal basis code. Inter‑rater reliability is substantial (Fleiss's kappa = 0.751), yet reviewers still disagree on 32.55\% of claims. These discrepancies highlight both occasional LLM‑induced hallucinations and frequent ambiguities in Non‑Final Rejection documents themselves, particularly in how examiners denote citedPatent and sectionCode. Details of results and a discussion of the challenges inherent in parsing Non‑Final Rejections are provided in the App.\ref{expert_validation_result}.
\section{Benchmark Tasks}
The PANORAMA dataset captures the end‑to‑end examination workflow and the underlying reasons for patent applications. Based on the framework of \citet{Schmitt2023}, we divide this workflow into three benchmark tasks that replicate the main steps taken by real‑world examiners using the PANORAMA dataset, especially the parsed Non-Final Rejection. Our benchmarks primarily focus on patent examination under §102 (novelty) and §103 (non‑obviousness), where patentability is decided by comparing the claim with the prior art. Unlike previous novelty prediction tasks~\cite{Suzgun2022, Arts2021, Jiang2023, ikoma2025aiexamine} conducted at the application level, we frame each task at the granularity of individual claims, closely mirroring the key procedural steps of patent examination. We formalize the examination sequence as follows, treating each step as an independent benchmark:

\begin{itemize}
    \item \textbf{Prior‑Art Retrieval for Patent Claims (PAR4PC)}: Select the document(s) from a pool of candidate prior-art documents that must be consulted to determine whether the target claim should be rejected.
    \item \textbf{Paragraph Identification for Patent Claims (PI4PC)}: Given a claim and a prior‑art document, identify the paragraph number within the document that should be compared with the claim when assessing patentability.
    \item \textbf{Novelty and Non‑Obviousness Classification for Patent Claims (NOC4PC)}: Given a claim and the cited prior‑art documents with the relevant paragraphs, determine whether the claim is novel and non‑obvious in relation to that prior art.
\end{itemize}

\subsection{Evaluation Settings}
To establish baseline performance in these benchmark tasks, we experiment with 12 different LLMs using two prompting strategies. We evaluate three proprietary models (GPT‑4o~\cite{openai2024gpt4ocard}, Claude‑3.7-sonnet~\cite{anthropic_claude}, and Gemini‑2.0-flash~\cite{gemmateam2024gemmaopenmodelsbased}), seven open‑source models (Llama-3.1-8B-Instruct~\cite{touvron2023llama2openfoundation}, Qwen-2.5-8B-Instruct~\cite{qwen2025qwen25technicalreport}, and EXAONE-3.5-7.8B-Instruct~\cite{research2024exaone35serieslarge}, Gemma-3-12B-Instruct~\cite{gemmateam2024gemmaopenmodelsbased}, Qwen-2.5-32B-Instruct~\cite{qwen2025qwen25technicalreport}, EXAONE-3.5-32B-Instruct~\cite{research2024exaone35serieslarge},  Gemma-3-27B-Instruct~\cite{gemmateam2024gemmaopenmodelsbased}), and two reasoning models (QWQ-32B~\cite{qwen2025qwen25technicalreport}, EXAONE-Deep-32B~\cite{research2025exaonedeepreasoningenhanced}). We also study the efficacy of PANORAMA as a training dataset by conducting supervised fine-tuning on each task for three models (Llama-3.1-8B-Instruct~\cite{touvron2023llama2openfoundation}, EXAONE-3.5-7.8B-Instruct~\cite{research2024exaone35serieslarge}, Qwen2.5-7B-Instruct~\cite{qwen2025qwen25technicalreport}) and report their respective performance (Section \ref{sec:sft}).

We consider two prompting strategies: zero-shot and chain-of-thought (CoT) prompting. (For the reasoning models, we only conducted the CoT prompting experiment.) In the zero-shot setting, the model receives minimal task-specific instructions and is expected to produce the final answer directly. In the CoT setting, the model first generates an explicit reasoning process, followed by the final answer. Preliminary testing shows that a free-reasoning CoT prompt slightly outperforms a CoT prompt with step-by-step instruction based on USPTO patent examination guidelines (App.~\ref{app:chain-of-thought}). We therefore report all results using this prompt. Prompts for each task are detailed in the appendix: PAR4PC in App.\ref{PAR4PC_prompt}, PI4PC in App.\ref{PI4PC_prompt}, and NOC4PC in App.\ref{NOC4PC_prompt}. Experimental details are in App.\ref{resource}.

\begin{table}[htbp]
\centering
\renewcommand{\arraystretch}{1.2}
\begin{tabularx}{\textwidth}{lcccccc}
\toprule
\multirow{2}{*}{\textbf{Model}} & \multicolumn{2}{c}{\textbf{PAR4PC}} & \multicolumn{2}{c}{\textbf{PI4PC}} & \multicolumn{2}{c}{\textbf{NOC4PC}} \\
\cmidrule(lr){2-3} \cmidrule(lr){4-5} \cmidrule(lr){6-7}
 & Zero-shot & CoT & Zero-shot & CoT & Zero-shot & CoT \\
\midrule
Baseline (random) & \multicolumn{2}{c}{5.63} & \multicolumn{2}{c}{27.10} & \multicolumn{2}{c}{32.33}\\
\midrule
GPT-4o & 47.34 & 56.95 & \textbf{63.33} & \textbf{62.62} & 34.69 & 32.19 \\
Claude 3.7 Sonnet & 40.12 & 40.29 & 57.33 & 60.55 & \textbf{35.84} & \textbf{45.40}  \\
Gemini 2.0 flash & 37.56 & 43.61 & 61.96 & 61.72 & 21.06 & 31.79 \\
\midrule
Llama-3.1-8B-Instruct & 13.45 & 37.99 & 9.61 & 0.00 & 15.71 & 19.56 \\
Qwen2.5-7B-Instruct & 66.11 & 67.42 & 29.25 & 48.41 & 28.92 & 20.31 \\
EXAONE-3.5-7.8B-Instruct & 0.00 & 22.52 & 44.55 & 41.34 & 15.00 & 24.99 \\
Gemma-3-12B-Instruct & 56.47 & \textbf{77.30} & 44.34 & 31.11 & 32.54 & 17.67 \\
Qwen2.5-32B-Instruct & \textbf{68.94} & 55.05 & 60.55 & 59.94 & 26.88 & 33.85 \\
EXAONE-3.5-32B-Instruct & 51.46 & 44.93 & 49.40 & 51.06 & 23.05 & 28.47 \\
Gemma-3-27B-Instruct & 50.19 & 55.36 & 54.66 & 56.22 & 24.00 & 22.45 \\
\midrule
QWQ-32B & - & 59.03 & - & 58.98 & - & 34.73 \\
EXAONE-Deep-32B & - & 42.59 & - & 35.80 & - & 21.23 \\
\bottomrule
\end{tabularx}
\vspace{0.2cm}
\caption{Performance comparison of 12 LLMs across three tasks (PAR4PC, PI4PC, NOC4PC). The baseline score is the average of 20 trials of random responses. The model with best performance are \textbf{bolded}.} \label{tab:task_results}
\renewcommand{\arraystretch}{1.0}
\end{table}

\subsection{TASK 1: Prior‑Art Retrieval for Patent Claims}
As the first fine‑grained task in patent examination, we introduce a task for retrieving prior art to evaluate a specific patent claim. Framed as a multiple‑choice question, the task presents a target claim alongside eight candidate prior‑art documents. The model must select the document(s) cited by the examiner in the office action: one document for §102 rejections or multiple for §103 rejections.
Unlike previous prior‑art retrieval benchmarks~\cite{Helmers2019, Hofstatter2019, Althammer2021, Trappey2021, Vowinckel2023}, our benchmark uniquely requires identifying the precise references that can be directly used to assess—and, when appropriate, reject—a claim.

For each instance, the prompt provides the title, abstract, and claims of the target application, along with the same three fields for each candidate prior art. Because §103 (non‑obviousness) rejections may rely on combinations of references, we allow multiple correct answers. In addition to gold answers (cited against the target claim), we include silver answers (cited against other claims in the same office action but not the target claim) to help models distinguish between fully correct and merely relevant options. Each question contains eight choices labeled gold, silver, or negative; at least one gold reference is always present, and the combined number of gold + silver references never exceeds five.

\paragraph{Results.}
Table \ref{tab:task_results} reports summary results for the 2,896 instances in the PAR4PC benchmark (full scores appear in App.~\ref{task1_result_detail}). Among the evaluated models, Qwen2.5-32B-Instruct achieves the highest score in the zero-shot setting (47.34), whereas Gemma-3-12B-Instruct obtains the best result with chain-of-thought (CoT) prompting (77.30). Most systems outperform the random-guess baseline, and GPT-4o's best accuracy reaches 51.04\%.

Since each claim rejected under §102 is linked to exactly one gold reference, while §103 rejections can cite two or more, scores differ between these subsets. Most models achieve higher accuracy on §102 claims and lower accuracy on §103 claims. Qwen 2.5-7B-Instruct and Gemma 3-12B-Instruct, however, showed comparatively better results on the §103 items, suggesting that performance varies with the rejection type.

In general, CoT prompting yields higher scores than zero-shot prompting, with the largest gains observed on the §103 subset. An exception was Qwen 2.5-32B-Instruct, whose accuracy decreased, likely because it tended to select a single answer through its own reasoning—an approach suited to §102 but less effective for §103, where multiple prior-art references may be needed.

\subsection{TASK 2: Paragraph Identification for Patent Claims}

Next, we introduce a benchmark task that asks LLMs to locate the specific portions of each cited reference that matter for a target claim. Given the set of prior-art documents retrieved in the previous step, the benchmark evaluates whether a model can identify the paragraph(s) most pertinent to assessing the claim's patentability. The task mirrors a critical phase of real-world examination, where examiners must pinpoint the exact disclosures that could anticipate a claim or render it obvious, thereby forming the basis for rejection under §102 or §103.

This task supplies the LLMs with a single claim from the target application and five candidate paragraphs from one cited reference. The model must select the paragraph most relevant for comparison with the claim during examination. Each prompt includes the title, abstract, and claim text, followed by the same three fields for the cited patent and the list of five paragraphs. The paragraph cited by the USPTO examiner against the target claim in the Office Action is the gold answer; a paragraph cited in the same Office Action but for a different claim is marked silver; any paragraph never cited is negative. Every question contains exactly one gold option, may include one silver option, and fills the remaining slots with negatives. Detailed answer definitions appear in App.\ref{PI4PC_construction}. 

\paragraph{Results.}
Table \ref{tab:task_results} presents the brief evaluation results of various LLMs on the 3,402 instances in the PI4PC benchmark (Detailed results are illustrated in App.\ref {task2_result_detail}). GPT-4o achieves the highest performance across all metrics, scores a 63.33 in the zero-shot setting and 62.62 with CoT prompting. Overall, most models outperform the random-guess baseline, and GPT-4o's score represents a 56.06\% accuracy gain over chance. 

According to our analysis, models also perform slightly better on claims rejected under §102 than on those rejected under §103, although the difference is modest. Surprisingly, unlike other benchmarks, zero-shot prompts outperform CoT prompts for most LLMs. This suggests that locating the pertinent paragraphs of a cited invention for patent examination involves complex considerations and may demand high-quality reasoning.

\subsection{TASK 3: Novelty and non‑Obviousness Classification for Patent Claims}
The last task is to evaluate the capability of LLMs to determine the patentability of given patent applications, with the comparison of prior art. This task also focuses on §102 and §103, which involve structured comparative analyses between patent applications and cited inventions, making them clear and consistent to evaluate. In contrast, §101 and §112 pose challenges due to their complex and frequently evolving legal interpretations, making assessments complex and less consistent.

The task supplies LLMs with a patent‑application claim and the relevant prior art, then asks it to assess whether the claim is rejected. The prompt contains the title, abstract, and claims of the target application and the prior arts' title, abstract, claims, and cited paragraph in the actual Non‑final Rejection. (If a particular claim was not actually rejected, we randomly insert the paragraph that was cited to reject another claim from the same application.) The LLMs must answer with one of three labels: rejected by §102, rejected by §103, or allowed. A detailed description of the task construction in App.\ref{NOC4PC_construction}. 

\paragraph{Results.}
Table \ref{tab:task_results} presents the brief evaluation results of various LLMs on the 2,884 instances in the NOC4PC benchmark (detailed results are illustrated in App.\ref{task3_result_detail}).  Claude-3.7 achieves the best performance, attaining 35.84 in the zero-shot and 45.40 with CoT. 

Overall, our results indicate that assessing a claim's novelty or non-obviousness remains challenging for current LLMs. The random-guess baseline for this three-way classification task is 32.33, and most models score under or only slightly above that level. Our results reveal that several LLMs collapsed their predictions into a single decision category: EXAONE-3.5-7.8B-Instruct with the zero-shot returned almost exclusively §102 rejections, GPT-4o with the zero-shot prompt favors §103 rejections, and GPT-4o with the CoT prompt largely issues allowance decisions. In general, CoT prompts consistently outperform their zero-shot counterparts. Our analysis finds that LLMs are generally less biased toward specific answers in CoT prompting.

For CoT Prompting only, we further evaluate how similar the LLM-generated reasoning was to the actual patent examiner's reasoning (App.\ref{task3_result_detail}). While the analysis does not show a high degree of similarity between the LLM-generated reasoning and the actual examiner's reasoning, we do observe that higher similarity tends to result in higher scores.

\subsection{Supervised Fine-Tuning}\label{sec:sft}

\begin{table}[tbp]
\footnotesize
\centering
\setlength{\tabcolsep}{10pt}
\renewcommand{\arraystretch}{1.2}
\begin{tabular}{lcccccc}
\toprule
\multirow{2}{*}{\textbf{Model}} & \multicolumn{2}{c}{\textbf{PAR4PC}} & \multicolumn{2}{c}{\textbf{PI4PC}} & \multicolumn{2}{c}{\textbf{NOC4PC}} \\
\cmidrule(lr){2-3} \cmidrule(lr){4-5} \cmidrule(lr){6-7}
 & Baseline & SFT & Baseline & SFT & Baseline & SFT \\
\midrule
Llama-3.1-8B-Instruct & 13.45 & 77.98 & 9.61 & \textbf{69.95} & 15.71 & 49.39 \\
Qwen2.5-7B-Instruct & 66.11 & 81.06 & 29.25 & 68.87 & 28.92 & 48.34 \\
EXAONE-3.5-7.8B-Instruct & 0.00 & \textbf{82.92} & 44.55 & 65.84 & 15.00 & \textbf{51.71} \\
\bottomrule
\end{tabular}
\vspace{0.2cm}
\caption{Supervised fine-tuning results of three LLMs across three tasks (PAR4PC, PI4PC, NOC4PC), where models prior to SFT are compared as baselines. For each task, the model with the best performance is \textbf{bolded}.} \label{tab:task_sft_results}
\end{table}
We also investigate the utility of the PANORAMA dataset as a resource for LLM training. Specifically, we perform supervised fine-tuning (SFT) on three instruction-tuned LLMs (Llama-3.1-8B-Instruct, EXAONE-3.5-7.8B-Instruct, and Qwen-2.5-7B-Instruct) using the task-specific training split of PANORAMA. Due to the absence of ground-truth CoTs, we train and evaluate each task and model under the zero-shot setting only. Further experimental details are provided in App.\ref{resource}. Table~\ref{tab:task_sft_results} reports the model performance before and after SFT. Although the three baseline models exhibit notably low scores in their off-the-shelf form, SFT on the PANORAMA training set leads to consistent and substantial improvements across all tasks and evaluation metrics.
\section{Limitations and Future Work}\label{sec:limitation}
While PANORAMA and its benchmarks advance in evaluating complex patent reasoning, limitations exist. Although our dataset is derived from authentic evaluation documents written by domain experts, converting these materials into a benchmark with quantifiable metrics for LLMs necessarily abstracts certain aspects of the real-world examination workflow. As a result, the benchmark may still diverge from the multifaceted tasks performed by human examiners. In particular, for the NOC4PC, while it demonstrates training signals, further investigation is needed to understand why LLMs still struggle to achieve high scores. We must therefore verify (1) how difficult the proposed tasks are for expert practitioners and (2) how well benchmark performance predicts success on the actual patent examination. Since comprehensive expert involvement demands significant resources, future research should explore efficient strategies for integrating professional expertise into scalable benchmark creation and validation processes. In future work, we will investigate this gap and iteratively refine the benchmark to more faithfully mirror professional practice, ultimately enabling LLMs to support better—or even partially automate—examiner responsibilities.

Our dataset has several additional limitations. First, while we have processed examination data from 8K high-quality patent applications, this scale could be significantly expanded. If there were multiple authorized API keys and a substantial budget for LLM parsing and patent agent recruiting fees, it is possible to extend the dataset, given that the USPTO database contains 7M patent applications. Second, our data is limited to the USPTO. Since patent examination procedures and standards vary considerably across different jurisdictions, future research should comprehensively address these international variations. Finally, because our dataset is based on office actions and deliberately excludes synthetic data, it lacks rationales explaining why allowed claims were accepted. This limitation prevented us from performing SFT in the chain-of-thought condition, and we suggest that future work develop more comprehensive datasets that include not only reasons for claim rejections but also explanations for allowances.

\paragraph{Potential NLP Tasks Using PANORAMA Dataset.}
Beyond these benchmarks, the PANORAMA dataset supports a broad spectrum of downstream tasks. We propose potential benchmark tasks that can be derived from our dataset, and we encourage the research community to pursue further challenges associated with contemporary patent evaluations. Real-world patent examination involves various evaluation criteria--such as §101 and §112--which can evolve over time. Examination is also an ongoing process, closely tied to later stages like revision. Beyond the three benchmark tasks we propose, it is essential to explicitly define additional tasks in the evaluation process and develop models and frameworks that simulate the full patent-examination workflow. In this section, we outline additional tasks within the patent-examination process that can be benchmarked using the PANORAMA dataset. \textbf{Claim Revision Task}: PANORAMA's aligned prosecution records--original claim, examiner's rejection, applicant's amendment, and notice of allowance--support a task in which a model must propose revised claim language that overcomes the cited prior art. \textbf{§101/§112 Classification Task}: With 16,831 claims rejected under §101 and 23,193 under §112, PANORAMA enables a task that predicts whether a claim should be rejected on subject-matter eligibility or specification grounds. \textbf{Drawing Component Extraction Task}: Because the dataset links cited figures in each Non-Final Rejection to the corresponding drawings, it supports a multimodal task that requires a model to locate and extract the specific drawing elements relevant to the rejection.
\section{Related Works}

\paragraph{Patent Examination in NLP.}
NLP community has long been fascinated by learning problems in the patent domain and has produced a diverse slate of task‑specific datasets, ranging from subject‑classification that predict IPC/CPC technology codes~\cite{molla2018overview, nguyen2024cinpatent, lee2019patent, fall2003automated}, through grant‑outcome prediction that forecast allowance decisions~\cite{hongxun2023deep}, litigation‑risk benchmarks that estimate post‑grant legal exposure~\cite{wu2024multi}, text‑based prior‑art retrieval that ranks relevant documents against claim language~\cite{risch2020patent, jiang2025enrich, shomee2024advance}, image‑based retrieval that matches technical drawings~\cite{gong2020image, kucer2022deeppatent, wang2024learning, lo2024large}, long‑document summarization for condensing full specifications~\cite{sharma2019bigpatent}, and generation‑oriented resources supporting claim drafting~\cite{jiang2025large} and claim revision to overcome rejections~\cite{jiang2025patent}, as well as novelty‑related tasks~\cite{ikoma2025ai}. 

Researchers have explored methods to analyze patent content to predict its novelty and non-obviousness, fundamental aspects of patentability assessment~\cite{Verhoeven2016, Plantec2021, Schmitt2023, Jeon2022, Wang2019, Zanella2023, Arts2021, beaty2021, Shibayama2021, Siddharth2020, Chikkamath2020, Jang2023}. Various NLP techniques, including indicator-based methods~\cite{Verhoeven2016, Plantec2021, Schmitt2023}, outlier detection~\cite{Jeon2022, Wang2019, Zanella2023}, similarity measurement~\cite{Arts2021, beaty2021, Shibayama2021, Siddharth2020}, and supervised learning~\cite{Chikkamath2020, Jang2023}, have been applied to assess patentability. These prior works approach novelty and non-obviousness prediction as a binary classification task, leveraging trained models to enhance prediction accuracy~\cite{Arts2021, beaty2021, Chikkamath2020, Jang2023}. However, \citet{Jiang2024} highlights the limitations in these approaches: They primarily focus on predicting novelty, rely on generic patent content (e.g., titles and abstracts) rather than informative patent claims, and have yet to explore the potential of LLMs for automating patentability assessment.

\paragraph{Dataset for Patent Examination Related-Tasks.}

HUPD~\cite{Suzgun2022}, a large‑scale, multi‑purpose dataset, allows language models to be evaluated on patent acceptance prediction, while \citet{Arts2021} proposed a dataset for assessing novelty. \citet{Jiang2023} introduced a deep‑learning dataset for patentability prediction, combining content (abstract, claim) and network features (citation, inventor, assignee). \citet{ikoma2025aiexamine} are the only researchers who utilized Non-final Rejection documents to evaluate LLMs' novelty‑examination abilities, focusing on the application level. In contrast, we break down office actions by individual claims, creating a benchmark for claim‑level evaluation. Unlike prior work, which mainly focuses on novelty, our dataset covers multiple tasks, including both novelty and non‑obviousness assessments.
\section{Conclusion}
In conclusion, we present PANORAMA, a multi-purpose dataset that captures patent examiners' evaluation trails and rationales behind their evaluations of patent applications. By structuring Non-Final Rejection documents within the dataset, we derive three sequential benchmark tasks that let LLMs conduct patent examination on a claim-by-claim basis, closely mirroring the workflow of real examiners. Our results show that evaluating the patentability of each claim in a patent application posed significant challenges for LLMs compared to the existing application-level examination. Additionally, our findings confirm that assessing non-obviousness is a more complex task for LLMs and requires intricate reasoning and specialized knowledge. However, we want to emphasize that novelty rejection and non-obviousness assessments are not conducted independently in actual patent examination, and both must be addressed to support real-world patent examination. We hope PANORAMA serves as a step toward LLMs that better capture the complexity and nuance of real-world patent evaluation.

\begin{ack}
This work was supported by LG AI Research.
\end{ack}

\bibliographystyle{plainnat}
\small
\bibliography{contents/10bibliography}

\begin{thebibliography}{58}
\providecommand{\natexlab}[1]{#1}
\providecommand{\url}[1]{\texttt{#1}}
\expandafter\ifx\csname urlstyle\endcsname\relax
  \providecommand{\doi}[1]{doi: #1}\else
  \providecommand{\doi}{doi: \begingroup \urlstyle{rm}\Url}\fi

\bibitem[Althammer et~al.(2021)Althammer, Hofstätter, and Hanbury]{Althammer2021}
Sophia Althammer, Sebastian Hofstätter, and Allan Hanbury.
\newblock \emph{Cross-Domain Retrieval in the Legal and Patent Domains: A Reproducibility Study}, pages 3--17.
\newblock 03 2021.
\newblock ISBN 978-3-030-72239-5.
\newblock \doi{10.1007/978-3-030-72240-1_1}.

\bibitem[{Anthropic}(2024)]{anthropic_claude}
{Anthropic}.
\newblock {Meet Claude}, 2024.
\newblock URL \url{https://www.anthropic.com/claude}.

\bibitem[Arts et~al.(2021)Arts, Hou, and Gomez]{Arts2021}
Sam Arts, Jianan Hou, and Juan~Carlos Gomez.
\newblock Natural language processing to identify the creation and impact of new technologies in patent text: Code, data, and new measures.
\newblock \emph{Research Policy}, 50\penalty0 (2):\penalty0 104144, 2021.
\newblock ISSN 0048-7333.
\newblock \doi{https://doi.org/10.1016/j.respol.2020.104144}.
\newblock URL \url{https://www.sciencedirect.com/science/article/pii/S0048733320302195}.

\bibitem[Beaty and Johnson(2021)]{beaty2021}
Roger~E Beaty and Dan~R Johnson.
\newblock Automating creativity assessment with semdis: An open platform for computing semantic distance.
\newblock \emph{Behavior research methods}, 53\penalty0 (2):\penalty0 757--780, 2021.

\bibitem[Campbell et~al.(2016)Campbell, Li, Dagli, Greenfield, Wolf, and Campbell]{Campbell2016}
William Campbell, Lin Li, C~Dagli, K~Greenfield, E~Wolf, and J~Campbell.
\newblock Predicting and analyzing factors in patent litigation.
\newblock 01 2016.

\bibitem[Chikkamath et~al.(2020)Chikkamath, Endres, Bayyapu, and Hewel]{Chikkamath2020}
Renukswamy Chikkamath, Markus Endres, Lavanya Bayyapu, and Christoph Hewel.
\newblock An empirical study on patent novelty detection: A novel approach using machine learning and natural language processing.
\newblock In \emph{2020 Seventh International Conference on Social Networks Analysis, Management and Security (SNAMS)}, pages 1--7, 2020.
\newblock \doi{10.1109/SNAMS52053.2020.9336557}.

\bibitem[Chung and Sohn(2020)]{Chung2020}
Park Chung and So~Young Sohn.
\newblock Early detection of valuable patents using a deep learning model: Case of semiconductor industry.
\newblock \emph{Technological Forecasting and Social Change}, 158:\penalty0 120146, 2020.
\newblock ISSN 0040-1625.
\newblock \doi{https://doi.org/10.1016/j.techfore.2020.120146}.
\newblock URL \url{https://www.sciencedirect.com/science/article/pii/S0040162520309720}.

\bibitem[Du et~al.(2021)Du, Wang, Xu, and Ma]{Du2021}
Wei Du, Yibo Wang, Wei Xu, and Jian Ma.
\newblock A personalized recommendation system for high-quality patent trading by leveraging hybrid patent analysis.
\newblock \emph{Scientometrics}, 126\penalty0 (12):\penalty0 9369–9391, December 2021.
\newblock ISSN 0138-9130.
\newblock \doi{10.1007/s11192-021-04180-x}.
\newblock URL \url{https://doi.org/10.1007/s11192-021-04180-x}.

\bibitem[Fall et~al.(2003)Fall, T\"{o}rcsv\'{a}ri, Benzineb, and Karetka]{fall2003automated}
C.~J. Fall, A.~T\"{o}rcsv\'{a}ri, K.~Benzineb, and G.~Karetka.
\newblock Automated categorization in the international patent classification.
\newblock \emph{SIGIR Forum}, 37\penalty0 (1):\penalty0 10–25, April 2003.
\newblock ISSN 0163-5840.
\newblock \doi{10.1145/945546.945547}.
\newblock URL \url{https://doi.org/10.1145/945546.945547}.

\bibitem[Gong and Guo(2020)]{gong2020image}
Guangjun Gong and Mingrui Guo.
\newblock Image based design patent retrieval with classification and indexing.
\newblock In \emph{2020 2nd International Conference on Information Technology and Computer Application (ITCA)}, pages 481--488, 2020.
\newblock \doi{10.1109/ITCA52113.2020.00107}.

\bibitem[Helmers et~al.(2019)Helmers, Horn, Biegler, Oppermann, and Müller]{Helmers2019}
Lea Helmers, Franziska Horn, Franziska Biegler, Tim Oppermann, and Klaus-Robert Müller.
\newblock Automating the search for a patent’s prior art with a full text similarity search.
\newblock \emph{PLOS ONE}, 14\penalty0 (3):\penalty0 e0212103, March 2019.
\newblock ISSN 1932-6203.
\newblock \doi{10.1371/journal.pone.0212103}.
\newblock URL \url{http://dx.doi.org/10.1371/journal.pone.0212103}.

\bibitem[Hofst\"{a}tter et~al.(2019)Hofst\"{a}tter, Rekabsaz, Lupu, Eickhoff, and Hanbury]{Hofstatter2019}
Sebastian Hofst\"{a}tter, Navid Rekabsaz, Mihai Lupu, Carsten Eickhoff, and Allan Hanbury.
\newblock Enriching word embeddings for patent retrieval with global context.
\newblock In \emph{Advances in Information Retrieval: 41st European Conference on IR Research, ECIR 2019, Cologne, Germany, April 14–18, 2019, Proceedings, Part I}, page 810–818, Berlin, Heidelberg, 2019. Springer-Verlag.
\newblock ISBN 978-3-030-15711-1.
\newblock \doi{10.1007/978-3-030-15712-8_57}.
\newblock URL \url{https://doi.org/10.1007/978-3-030-15712-8_57}.

\bibitem[Hu et~al.(2023)Hu, Zhou, and Lin]{Hu2023}
Zewen Hu, Xiji Zhou, and Angela Lin.
\newblock Evaluation and identification of potential high-value patents in the field of integrated circuits using a multidimensional patent indicators pre-screening strategy and machine learning approaches.
\newblock \emph{Journal of Informetrics}, 17\penalty0 (2):\penalty0 101406, 2023.
\newblock ISSN 1751-1577.
\newblock \doi{https://doi.org/10.1016/j.joi.2023.101406}.
\newblock URL \url{https://www.sciencedirect.com/science/article/pii/S1751157723000317}.

\bibitem[Ikoma and Mitamura(2025{\natexlab{a}})]{ikoma2025ai}
Hayato Ikoma and Teruko Mitamura.
\newblock Can ai examine novelty of patents?: Novelty evaluation based on the correspondence between patent claim and prior art, 2025{\natexlab{a}}.
\newblock URL \url{https://arxiv.org/abs/2502.06316}.

\bibitem[Ikoma and Mitamura(2025{\natexlab{b}})]{ikoma2025aiexamine}
Hayato Ikoma and Teruko Mitamura.
\newblock Can ai examine novelty of patents?: Novelty evaluation based on the correspondence between patent claim and prior art, 2025{\natexlab{b}}.
\newblock URL \url{https://arxiv.org/abs/2502.06316}.

\bibitem[Jang et~al.(2023)Jang, Kim, and Yoon]{Jang2023}
Hyejin Jang, Sunhye Kim, and Byungun Yoon.
\newblock An explainable ai (xai) model for text-based patent novelty analysis.
\newblock \emph{Expert Systems with Applications}, 231:\penalty0 120839, 2023.
\newblock ISSN 0957-4174.
\newblock \doi{https://doi.org/10.1016/j.eswa.2023.120839}.
\newblock URL \url{https://www.sciencedirect.com/science/article/pii/S0957417423013416}.

\bibitem[Jeon et~al.(2022)Jeon, Ahn, Kim, and Lee]{Jeon2022}
Daeseong Jeon, Joon~Mo Ahn, Juram Kim, and Changyong Lee.
\newblock A doc2vec and local outlier factor approach to measuring the novelty of patents.
\newblock \emph{Technological Forecasting and Social Change}, 174:\penalty0 121294, 2022.
\newblock ISSN 0040-1625.
\newblock \doi{https://doi.org/10.1016/j.techfore.2021.121294}.
\newblock URL \url{https://www.sciencedirect.com/science/article/pii/S0040162521007289}.

\bibitem[Jiang et~al.(2023{\natexlab{a}})Jiang, Fan, Zhang, and Zhu]{Jiang2023}
Hongxun Jiang, Shaokun Fan, Nan Zhang, and Bin Zhu.
\newblock Deep learning for predicting patent application outcome: The fusion of text and network embeddings.
\newblock \emph{Journal of Informetrics}, 17\penalty0 (2):\penalty0 101402, 2023{\natexlab{a}}.
\newblock ISSN 1751-1577.
\newblock \doi{https://doi.org/10.1016/j.joi.2023.101402}.
\newblock URL \url{https://www.sciencedirect.com/science/article/pii/S1751157723000275}.

\bibitem[Jiang et~al.(2023{\natexlab{b}})Jiang, Fan, Zhang, and Zhu]{hongxun2023deep}
Hongxun Jiang, Shaokun Fan, Nan Zhang, and Bin Zhu.
\newblock Deep learning for predicting patent application outcome: The fusion of text and network embeddings.
\newblock \emph{Journal of Informetrics}, 17\penalty0 (2):\penalty0 101402, 2023{\natexlab{b}}.
\newblock ISSN 1751-1577.
\newblock \doi{https://doi.org/10.1016/j.joi.2023.101402}.
\newblock URL \url{https://www.sciencedirect.com/science/article/pii/S1751157723000275}.

\bibitem[Jiang and Goetz(2024)]{Jiang2024}
Lekang Jiang and Stephan Goetz.
\newblock Natural language processing in patents: A survey, 2024.
\newblock URL \url{https://arxiv.org/abs/2403.04105}.

\bibitem[Jiang et~al.(2025{\natexlab{a}})Jiang, Li, and Goetz]{jiang2025enrich}
Lekang Jiang, Chengzu Li, and Stephan Goetz.
\newblock Enriching patent claim generation with european patent dataset, 2025{\natexlab{a}}.
\newblock URL \url{https://arxiv.org/abs/2505.12568}.

\bibitem[Jiang et~al.(2025{\natexlab{b}})Jiang, Scherz, and Goetz]{jiang2025patent}
Lekang Jiang, Pascal~A. Scherz, and Stefan Goetz.
\newblock Patent-{CR}: A dataset for patent claim revision.
\newblock In Luis Chiruzzo, Alan Ritter, and Lu~Wang, editors, \emph{Proceedings of the 2025 Conference of the Nations of the Americas Chapter of the Association for Computational Linguistics: Human Language Technologies (Volume 1: Long Papers)}, pages 2300--2314, Albuquerque, New Mexico, April 2025{\natexlab{b}}. Association for Computational Linguistics.
\newblock ISBN 979-8-89176-189-6.
\newblock \doi{10.18653/v1/2025.naacl-long.116}.
\newblock URL \url{https://aclanthology.org/2025.naacl-long.116/}.

\bibitem[Jiang et~al.(2025{\natexlab{c}})Jiang, Zhang, Scherz, and Goetz]{jiang2025large}
Lekang Jiang, Caiqi Zhang, Pascal~A. Scherz, and Stefan Goetz.
\newblock Can large language models generate high-quality patent claims?
\newblock In Luis Chiruzzo, Alan Ritter, and Lu~Wang, editors, \emph{Findings of the Association for Computational Linguistics: NAACL 2025}, pages 1272--1287, Albuquerque, New Mexico, April 2025{\natexlab{c}}. Association for Computational Linguistics.
\newblock ISBN 979-8-89176-195-7.
\newblock \doi{10.18653/v1/2025.findings-naacl.70}.
\newblock URL \url{https://aclanthology.org/2025.findings-naacl.70/}.

\bibitem[Krestel et~al.(2021)Krestel, Chikkamath, Hewel, and Risch]{ralf2021survey}
Ralf Krestel, Renukswamy Chikkamath, Christoph Hewel, and Julian Risch.
\newblock A survey on deep learning for patent analysis.
\newblock \emph{World Patent Information}, 65:\penalty0 102035, 2021.
\newblock ISSN 0172-2190.
\newblock \doi{https://doi.org/10.1016/j.wpi.2021.102035}.
\newblock URL \url{https://www.sciencedirect.com/science/article/pii/S017221902100017X}.

\bibitem[Kucer et~al.(2022)Kucer, Oyen, Castorena, and Wu]{kucer2022deeppatent}
Michal Kucer, Diane Oyen, Juan Castorena, and Jian Wu.
\newblock Deeppatent: Large scale patent drawing recognition and retrieval.
\newblock In \emph{Proceedings of the IEEE/CVF Winter Conference on Applications of Computer Vision (WACV)}, pages 2309--2318, January 2022.

\bibitem[Lee and Hsiang(2019)]{lee2019patent}
Jieh-Sheng Lee and Jieh Hsiang.
\newblock Patentbert: Patent classification with fine-tuning a pre-trained bert model, 2019.
\newblock URL \url{https://arxiv.org/abs/1906.02124}.

\bibitem[Lin(2004)]{lin2004rouge}
Chin-Yew Lin.
\newblock {ROUGE}: A package for automatic evaluation of summaries.
\newblock In \emph{Text Summarization Branches Out}, pages 74--81, Barcelona, Spain, July 2004. Association for Computational Linguistics.
\newblock URL \url{https://aclanthology.org/W04-1013/}.

\bibitem[Lin et~al.(2018)Lin, Wang, Du, Wu, Chang, and Chen]{Lin2018}
Hongjie Lin, Hao Wang, Dongfang Du, Han Wu, Biao Chang, and Enhong Chen.
\newblock Patent quality valuation with deep learning models.
\newblock In \emph{Database Systems for Advanced Applications: 23rd International Conference, DASFAA 2018, Gold Coast, QLD, Australia, May 21-24, 2018, Proceedings, Part II}, page 474–490, Berlin, Heidelberg, 2018. Springer-Verlag.
\newblock ISBN 978-3-319-91457-2.
\newblock \doi{10.1007/978-3-319-91458-9_29}.
\newblock URL \url{https://doi.org/10.1007/978-3-319-91458-9_29}.

\bibitem[Liu et~al.(2018)Liu, Wu, Ye, Zhao, Liu, and Du]{Liu2018}
Qi~Liu, Han Wu, Yuyang Ye, Hongke Zhao, Chuanren Liu, and Dongfang Du.
\newblock Patent litigation prediction: A convolutional tensor factorization approach.
\newblock pages 5052--5059, 07 2018.
\newblock \doi{10.24963/ijcai.2018/701}.

\bibitem[Liu et~al.(2023)Liu, Li, Cao, and Wang]{Liu2023}
Weidong Liu, Shuai Li, Yan Cao, and Yu~Wang.
\newblock Multi-task learning based high-value patent and standard-essential patent identification model.
\newblock \emph{Inf. Process. Manage.}, 60\penalty0 (3), May 2023.
\newblock ISSN 0306-4573.
\newblock \doi{10.1016/j.ipm.2023.103327}.
\newblock URL \url{https://doi.org/10.1016/j.ipm.2023.103327}.

\bibitem[Lo et~al.(2024)Lo, Chu, Hsiang, and Cho]{lo2024large}
Hao-Cheng Lo, Jung-Mei Chu, Jieh Hsiang, and Chun-Chieh Cho.
\newblock Large language model informed patent image retrieval, 2024.
\newblock URL \url{https://arxiv.org/abs/2404.19360}.

\bibitem[Moll{\'a} and Seneviratne(2018)]{molla2018overview}
Diego Moll{\'a} and Dilesha Seneviratne.
\newblock Overview of the 2018 {ALTA} shared task: Classifying patent applications.
\newblock In Sunghwan~Mac Kim and Xiuzhen~(Jenny) Zhang, editors, \emph{Proceedings of the Australasian Language Technology Association Workshop 2018}, pages 84--88, Dunedin, New Zealand, December 2018.
\newblock URL \url{https://aclanthology.org/U18-1011/}.

\bibitem[Nguyen et~al.(2024)Nguyen, Bui, Tran-Tien, Le, and Vu]{nguyen2024cinpatent}
Minh-Tien Nguyen, Nhung Bui, Manh Tran-Tien, Linh Le, and Huy-The Vu.
\newblock Cinpatent: Datasets for patent classification, 2024.
\newblock URL \url{https://arxiv.org/abs/2212.12192}.

\bibitem[OpenAI et~al.(2024)OpenAI, :, Hurst, Lerer, Goucher, Perelman, Ramesh, Clark, Ostrow, Welihinda, Hayes, Radford, Mądry, Baker-Whitcomb, Beutel, Borzunov, Carney, Chow, Kirillov, Nichol, Paino, Renzin, Passos, Kirillov, Christakis, Conneau, Kamali, Jabri, Moyer, Tam, Crookes, Tootoochian, Tootoonchian, Kumar, Vallone, Karpathy, Braunstein, Cann, Codispoti, Galu, Kondrich, Tulloch, Mishchenko, Baek, Jiang, Pelisse, Woodford, Gosalia, Dhar, Pantuliano, Nayak, Oliver, Zoph, Ghorbani, Leimberger, Rossen, Sokolowsky, Wang, Zweig, Hoover, Samic, McGrew, Spero, Giertler, Cheng, Lightcap, Walkin, Quinn, Guarraci, Hsu, Kellogg, Eastman, Lugaresi, Wainwright, Bassin, Hudson, Chu, Nelson, Li, Shern, Conger, Barette, Voss, Ding, Lu, Zhang, Beaumont, Hallacy, Koch, Gibson, Kim, Choi, McLeavey, Hesse, Fischer, Winter, Czarnecki, Jarvis, Wei, Koumouzelis, Sherburn, Kappler, Levin, Levy, Carr, Farhi, Mely, Robinson, Sasaki, Jin, Valladares, Tsipras, Li, Nguyen, Findlay, Oiwoh, Wong, Asdar, Proehl, Yang, Antonow,
  Kramer, Peterson, Sigler, Wallace, Brevdo, Mays, Khorasani, Such, Raso, Zhang, von Lohmann, Sulit, Goh, Oden, Salmon, Starace, Brockman, Salman, Bao, Hu, Wong, Wang, Schmidt, Whitney, Jun, Kirchner, de~Oliveira~Pinto, Ren, Chang, Chung, Kivlichan, O'Connell, O'Connell, Osband, Silber, Sohl, Okuyucu, Lan, Kostrikov, Sutskever, Kanitscheider, Gulrajani, Coxon, Menick, Pachocki, Aung, Betker, Crooks, Lennon, Kiros, Leike, Park, Kwon, Phang, Teplitz, Wei, Wolfe, Chen, Harris, Varavva, Lee, Shieh, Lin, Yu, Weng, Tang, Yu, Jang, Candela, Beutler, Landers, Parish, Heidecke, Schulman, Lachman, McKay, Uesato, Ward, Kim, Huizinga, Sitkin, Kraaijeveld, Gross, Kaplan, Snyder, Achiam, Jiao, Lee, Zhuang, Harriman, Fricke, Hayashi, Singhal, Shi, Karthik, Wood, Rimbach, Hsu, Nguyen, Gu-Lemberg, Button, Liu, Howe, Muthukumar, Luther, Ahmad, Kai, Itow, Workman, Pathak, Chen, Jing, Guy, Fedus, Zhou, Mamitsuka, Weng, McCallum, Held, Ouyang, Feuvrier, Zhang, Kondraciuk, Kaiser, Hewitt, Metz, Doshi, Aflak, Simens, Boyd,
  Thompson, Dukhan, Chen, Gray, Hudnall, Zhang, Aljubeh, Litwin, Zeng, Johnson, Shetty, Gupta, Shah, Yatbaz, Yang, Zhong, Glaese, Chen, Janner, Lampe, Petrov, Wu, Wang, Fradin, Pokrass, Castro, de~Castro, Pavlov, Brundage, Wang, Khan, Murati, Bavarian, Lin, Yesildal, Soto, Gimelshein, Cone, Staudacher, Summers, LaFontaine, Chowdhury, Ryder, Stathas, Turley, Tezak, Felix, Kudige, Keskar, Deutsch, Bundick, Puckett, Nachum, Okelola, Boiko, Murk, Jaffe, Watkins, Godement, Campbell-Moore, Chao, McMillan, Belov, Su, Bak, Bakkum, Deng, Dolan, Hoeschele, Welinder, Tillet, Pronin, Tillet, Dhariwal, Yuan, Dias, Lim, Arora, Troll, Lin, Lopes, Puri, Miyara, Leike, Gaubert, Zamani, Wang, Donnelly, Honsby, Smith, Sahai, Ramchandani, Huet, Carmichael, Zellers, Chen, Chen, Nigmatullin, Cheu, Jain, Altman, Schoenholz, Toizer, Miserendino, Agarwal, Culver, Ethersmith, Gray, Grove, Metzger, Hermani, Jain, Zhao, Wu, Jomoto, Wu, Shuaiqi, Xia, Phene, Papay, Narayanan, Coffey, Lee, Hall, Balaji, Broda, Stramer, Xu, Gogineni,
  Christianson, Sanders, Patwardhan, Cunninghman, Degry, Dimson, Raoux, Shadwell, Zheng, Underwood, Markov, Sherbakov, Rubin, Stasi, Kaftan, Heywood, Peterson, Walters, Eloundou, Qi, Moeller, Monaco, Kuo, Fomenko, Chang, Zheng, Zhou, Manassra, Sheu, Zaremba, Patil, Qian, Kim, Cheng, Zhang, He, Zhang, Jin, Dai, and Malkov]{openai2024gpt4ocard}
OpenAI, :, Aaron Hurst, Adam Lerer, Adam~P. Goucher, Adam Perelman, Aditya Ramesh, Aidan Clark, AJ~Ostrow, Akila Welihinda, Alan Hayes, Alec Radford, Aleksander Mądry, Alex Baker-Whitcomb, Alex Beutel, Alex Borzunov, Alex Carney, Alex Chow, Alex Kirillov, Alex Nichol, Alex Paino, Alex Renzin, Alex~Tachard Passos, Alexander Kirillov, Alexi Christakis, Alexis Conneau, Ali Kamali, Allan Jabri, Allison Moyer, Allison Tam, Amadou Crookes, Amin Tootoochian, Amin Tootoonchian, Ananya Kumar, Andrea Vallone, Andrej Karpathy, Andrew Braunstein, Andrew Cann, Andrew Codispoti, Andrew Galu, Andrew Kondrich, Andrew Tulloch, Andrey Mishchenko, Angela Baek, Angela Jiang, Antoine Pelisse, Antonia Woodford, Anuj Gosalia, Arka Dhar, Ashley Pantuliano, Avi Nayak, Avital Oliver, Barret Zoph, Behrooz Ghorbani, Ben Leimberger, Ben Rossen, Ben Sokolowsky, Ben Wang, Benjamin Zweig, Beth Hoover, Blake Samic, Bob McGrew, Bobby Spero, Bogo Giertler, Bowen Cheng, Brad Lightcap, Brandon Walkin, Brendan Quinn, Brian Guarraci, Brian Hsu,
  Bright Kellogg, Brydon Eastman, Camillo Lugaresi, Carroll Wainwright, Cary Bassin, Cary Hudson, Casey Chu, Chad Nelson, Chak Li, Chan~Jun Shern, Channing Conger, Charlotte Barette, Chelsea Voss, Chen Ding, Cheng Lu, Chong Zhang, Chris Beaumont, Chris Hallacy, Chris Koch, Christian Gibson, Christina Kim, Christine Choi, Christine McLeavey, Christopher Hesse, Claudia Fischer, Clemens Winter, Coley Czarnecki, Colin Jarvis, Colin Wei, Constantin Koumouzelis, Dane Sherburn, Daniel Kappler, Daniel Levin, Daniel Levy, David Carr, David Farhi, David Mely, David Robinson, David Sasaki, Denny Jin, Dev Valladares, Dimitris Tsipras, Doug Li, Duc~Phong Nguyen, Duncan Findlay, Edede Oiwoh, Edmund Wong, Ehsan Asdar, Elizabeth Proehl, Elizabeth Yang, Eric Antonow, Eric Kramer, Eric Peterson, Eric Sigler, Eric Wallace, Eugene Brevdo, Evan Mays, Farzad Khorasani, Felipe~Petroski Such, Filippo Raso, Francis Zhang, Fred von Lohmann, Freddie Sulit, Gabriel Goh, Gene Oden, Geoff Salmon, Giulio Starace, Greg Brockman, Hadi
  Salman, Haiming Bao, Haitang Hu, Hannah Wong, Haoyu Wang, Heather Schmidt, Heather Whitney, Heewoo Jun, Hendrik Kirchner, Henrique~Ponde de~Oliveira~Pinto, Hongyu Ren, Huiwen Chang, Hyung~Won Chung, Ian Kivlichan, Ian O'Connell, Ian O'Connell, Ian Osband, Ian Silber, Ian Sohl, Ibrahim Okuyucu, Ikai Lan, Ilya Kostrikov, Ilya Sutskever, Ingmar Kanitscheider, Ishaan Gulrajani, Jacob Coxon, Jacob Menick, Jakub Pachocki, James Aung, James Betker, James Crooks, James Lennon, Jamie Kiros, Jan Leike, Jane Park, Jason Kwon, Jason Phang, Jason Teplitz, Jason Wei, Jason Wolfe, Jay Chen, Jeff Harris, Jenia Varavva, Jessica~Gan Lee, Jessica Shieh, Ji~Lin, Jiahui Yu, Jiayi Weng, Jie Tang, Jieqi Yu, Joanne Jang, Joaquin~Quinonero Candela, Joe Beutler, Joe Landers, Joel Parish, Johannes Heidecke, John Schulman, Jonathan Lachman, Jonathan McKay, Jonathan Uesato, Jonathan Ward, Jong~Wook Kim, Joost Huizinga, Jordan Sitkin, Jos Kraaijeveld, Josh Gross, Josh Kaplan, Josh Snyder, Joshua Achiam, Joy Jiao, Joyce Lee, Juntang
  Zhuang, Justyn Harriman, Kai Fricke, Kai Hayashi, Karan Singhal, Katy Shi, Kavin Karthik, Kayla Wood, Kendra Rimbach, Kenny Hsu, Kenny Nguyen, Keren Gu-Lemberg, Kevin Button, Kevin Liu, Kiel Howe, Krithika Muthukumar, Kyle Luther, Lama Ahmad, Larry Kai, Lauren Itow, Lauren Workman, Leher Pathak, Leo Chen, Li~Jing, Lia Guy, Liam Fedus, Liang Zhou, Lien Mamitsuka, Lilian Weng, Lindsay McCallum, Lindsey Held, Long Ouyang, Louis Feuvrier, Lu~Zhang, Lukas Kondraciuk, Lukasz Kaiser, Luke Hewitt, Luke Metz, Lyric Doshi, Mada Aflak, Maddie Simens, Madelaine Boyd, Madeleine Thompson, Marat Dukhan, Mark Chen, Mark Gray, Mark Hudnall, Marvin Zhang, Marwan Aljubeh, Mateusz Litwin, Matthew Zeng, Max Johnson, Maya Shetty, Mayank Gupta, Meghan Shah, Mehmet Yatbaz, Meng~Jia Yang, Mengchao Zhong, Mia Glaese, Mianna Chen, Michael Janner, Michael Lampe, Michael Petrov, Michael Wu, Michele Wang, Michelle Fradin, Michelle Pokrass, Miguel Castro, Miguel Oom~Temudo de~Castro, Mikhail Pavlov, Miles Brundage, Miles Wang, Minal
  Khan, Mira Murati, Mo~Bavarian, Molly Lin, Murat Yesildal, Nacho Soto, Natalia Gimelshein, Natalie Cone, Natalie Staudacher, Natalie Summers, Natan LaFontaine, Neil Chowdhury, Nick Ryder, Nick Stathas, Nick Turley, Nik Tezak, Niko Felix, Nithanth Kudige, Nitish Keskar, Noah Deutsch, Noel Bundick, Nora Puckett, Ofir Nachum, Ola Okelola, Oleg Boiko, Oleg Murk, Oliver Jaffe, Olivia Watkins, Olivier Godement, Owen Campbell-Moore, Patrick Chao, Paul McMillan, Pavel Belov, Peng Su, Peter Bak, Peter Bakkum, Peter Deng, Peter Dolan, Peter Hoeschele, Peter Welinder, Phil Tillet, Philip Pronin, Philippe Tillet, Prafulla Dhariwal, Qiming Yuan, Rachel Dias, Rachel Lim, Rahul Arora, Rajan Troll, Randall Lin, Rapha~Gontijo Lopes, Raul Puri, Reah Miyara, Reimar Leike, Renaud Gaubert, Reza Zamani, Ricky Wang, Rob Donnelly, Rob Honsby, Rocky Smith, Rohan Sahai, Rohit Ramchandani, Romain Huet, Rory Carmichael, Rowan Zellers, Roy Chen, Ruby Chen, Ruslan Nigmatullin, Ryan Cheu, Saachi Jain, Sam Altman, Sam Schoenholz, Sam
  Toizer, Samuel Miserendino, Sandhini Agarwal, Sara Culver, Scott Ethersmith, Scott Gray, Sean Grove, Sean Metzger, Shamez Hermani, Shantanu Jain, Shengjia Zhao, Sherwin Wu, Shino Jomoto, Shirong Wu, Shuaiqi, Xia, Sonia Phene, Spencer Papay, Srinivas Narayanan, Steve Coffey, Steve Lee, Stewart Hall, Suchir Balaji, Tal Broda, Tal Stramer, Tao Xu, Tarun Gogineni, Taya Christianson, Ted Sanders, Tejal Patwardhan, Thomas Cunninghman, Thomas Degry, Thomas Dimson, Thomas Raoux, Thomas Shadwell, Tianhao Zheng, Todd Underwood, Todor Markov, Toki Sherbakov, Tom Rubin, Tom Stasi, Tomer Kaftan, Tristan Heywood, Troy Peterson, Tyce Walters, Tyna Eloundou, Valerie Qi, Veit Moeller, Vinnie Monaco, Vishal Kuo, Vlad Fomenko, Wayne Chang, Weiyi Zheng, Wenda Zhou, Wesam Manassra, Will Sheu, Wojciech Zaremba, Yash Patil, Yilei Qian, Yongjik Kim, Youlong Cheng, Yu~Zhang, Yuchen He, Yuchen Zhang, Yujia Jin, Yunxing Dai, and Yury Malkov.
\newblock Gpt-4o system card, 2024.
\newblock URL \url{https://arxiv.org/abs/2410.21276}.

\bibitem[Plantec et~al.(2021)Plantec, {Le Masson}, and Weil]{Plantec2021}
Quentin Plantec, Pascal {Le Masson}, and Benoît Weil.
\newblock Impact of knowledge search practices on the originality of inventions: A study in the oil \& gas industry through dynamic patent analysis.
\newblock \emph{Technological Forecasting and Social Change}, 168:\penalty0 120782, 2021.
\newblock ISSN 0040-1625.
\newblock \doi{https://doi.org/10.1016/j.techfore.2021.120782}.
\newblock URL \url{https://www.sciencedirect.com/science/article/pii/S0040162521002146}.

\bibitem[Qwen et~al.(2025)Qwen, :, Yang, Yang, Zhang, Hui, Zheng, Yu, Li, Liu, Huang, Wei, Lin, Yang, Tu, Zhang, Yang, Yang, Zhou, Lin, Dang, Lu, Bao, Yang, Yu, Li, Xue, Zhang, Zhu, Men, Lin, Li, Tang, Xia, Ren, Ren, Fan, Su, Zhang, Wan, Liu, Cui, Zhang, and Qiu]{qwen2025qwen25technicalreport}
Qwen, :, An~Yang, Baosong Yang, Beichen Zhang, Binyuan Hui, Bo~Zheng, Bowen Yu, Chengyuan Li, Dayiheng Liu, Fei Huang, Haoran Wei, Huan Lin, Jian Yang, Jianhong Tu, Jianwei Zhang, Jianxin Yang, Jiaxi Yang, Jingren Zhou, Junyang Lin, Kai Dang, Keming Lu, Keqin Bao, Kexin Yang, Le~Yu, Mei Li, Mingfeng Xue, Pei Zhang, Qin Zhu, Rui Men, Runji Lin, Tianhao Li, Tianyi Tang, Tingyu Xia, Xingzhang Ren, Xuancheng Ren, Yang Fan, Yang Su, Yichang Zhang, Yu~Wan, Yuqiong Liu, Zeyu Cui, Zhenru Zhang, and Zihan Qiu.
\newblock Qwen2.5 technical report, 2025.
\newblock URL \url{https://arxiv.org/abs/2412.15115}.

\bibitem[Research et~al.(2024)Research, An, Bae, Choi, Choi, Choi, Hong, Hwang, Jeon, Jo, Jo, Jung, Jung, Kim, Kim, Kim, Kim, Kim, Kim, Kim, Kim, Lee, Lee, Lee, Lee, Lee, Lim, Park, Park, Park, Yang, Yeen, and Yun]{research2024exaone35serieslarge}
LG~AI Research, Soyoung An, Kyunghoon Bae, Eunbi Choi, Kibong Choi, Stanley~Jungkyu Choi, Seokhee Hong, Junwon Hwang, Hyojin Jeon, Gerrard~Jeongwon Jo, Hyunjik Jo, Jiyeon Jung, Yountae Jung, Hyosang Kim, Joonkee Kim, Seonghwan Kim, Soyeon Kim, Sunkyoung Kim, Yireun Kim, Yongil Kim, Youchul Kim, Edward~Hwayoung Lee, Haeju Lee, Honglak Lee, Jinsik Lee, Kyungmin Lee, Woohyung Lim, Sangha Park, Sooyoun Park, Yongmin Park, Sihoon Yang, Heuiyeen Yeen, and Hyeongu Yun.
\newblock Exaone 3.5: Series of large language models for real-world use cases, 2024.
\newblock URL \url{https://arxiv.org/abs/2412.04862}.

\bibitem[Research et~al.(2025)Research, Bae, Choi, Choi, Choi, Choi, Hong, Hwang, Jeon, Jeon, Jo, Jo, Jung, Kim, Kim, Kim, Kim, Kim, Kim, Kim, Kim, Lee, Lee, Lee, Lee, Lee, Park, Park, Yang, Yeen, Yi, and Yun]{research2025exaonedeepreasoningenhanced}
LG~AI Research, Kyunghoon Bae, Eunbi Choi, Kibong Choi, Stanley~Jungkyu Choi, Yemuk Choi, Seokhee Hong, Junwon Hwang, Hyojin Jeon, Kijeong Jeon, Gerrard~Jeongwon Jo, Hyunjik Jo, Jiyeon Jung, Hyosang Kim, Joonkee Kim, Seonghwan Kim, Soyeon Kim, Sunkyoung Kim, Yireun Kim, Yongil Kim, Youchul Kim, Edward~Hwayoung Lee, Haeju Lee, Honglak Lee, Jinsik Lee, Kyungmin Lee, Sangha Park, Yongmin Park, Sihoon Yang, Heuiyeen Yeen, Sihyuk Yi, and Hyeongu Yun.
\newblock Exaone deep: Reasoning enhanced language models, 2025.
\newblock URL \url{https://arxiv.org/abs/2503.12524}.

\bibitem[Risch et~al.(2020)Risch, Alder, Hewel, and Krestel]{risch2020patent}
Julian Risch, Nicolas Alder, Christoph Hewel, and Ralf Krestel.
\newblock Patentmatch: A dataset for matching patent claims \& prior art, 2020.
\newblock URL \url{https://arxiv.org/abs/2012.13919}.

\bibitem[Schmitt et~al.(2023)Schmitt, Walter, and Schnittker]{Schmitt2023}
Valentin~J. Schmitt, Lothar Walter, and Frank~C. Schnittker.
\newblock Assessment of patentability by means of semantic patent analysis – a mathematical-logical approach.
\newblock \emph{World Patent Information}, 73:\penalty0 102182, 2023.
\newblock ISSN 0172-2190.
\newblock \doi{https://doi.org/10.1016/j.wpi.2023.102182}.
\newblock URL \url{https://www.sciencedirect.com/science/article/pii/S0172219023000121}.

\bibitem[Sellam et~al.(2020)Sellam, Das, and Parikh]{sellam2020bleurtlearningrobustmetrics}
Thibault Sellam, Dipanjan Das, and Ankur~P. Parikh.
\newblock Bleurt: Learning robust metrics for text generation, 2020.
\newblock URL \url{https://arxiv.org/abs/2004.04696}.

\bibitem[Sharma et~al.(2019)Sharma, Li, and Wang]{sharma2019bigpatent}
Eva Sharma, Chen Li, and Lu~Wang.
\newblock {BIGPATENT}: A large-scale dataset for abstractive and coherent summarization.
\newblock In Anna Korhonen, David Traum, and Llu{\'i}s M{\`a}rquez, editors, \emph{Proceedings of the 57th Annual Meeting of the Association for Computational Linguistics}, pages 2204--2213, Florence, Italy, July 2019. Association for Computational Linguistics.
\newblock \doi{10.18653/v1/P19-1212}.
\newblock URL \url{https://aclanthology.org/P19-1212/}.

\bibitem[Shibayama et~al.(2021)Shibayama, Yin, and Matsumoto]{Shibayama2021}
Sotaro Shibayama, Deyun Yin, and Kuniko Matsumoto.
\newblock Measuring novelty in science with word embedding.
\newblock \emph{PLOS ONE}, 16\penalty0 (7):\penalty0 1--16, 07 2021.
\newblock \doi{10.1371/journal.pone.0254034}.
\newblock URL \url{https://doi.org/10.1371/journal.pone.0254034}.

\bibitem[Shomee et~al.(2024)Shomee, Wang, Medya, and Ravi]{shomee2024advance}
Homaira~Huda Shomee, Zhu Wang, Sourav Medya, and Sathya~N. Ravi.
\newblock Impact: A large-scale integrated multimodal patent analysis and creation dataset for design patents.
\newblock In A.~Globerson, L.~Mackey, D.~Belgrave, A.~Fan, U.~Paquet, J.~Tomczak, and C.~Zhang, editors, \emph{Advances in Neural Information Processing Systems}, volume~37, pages 125520--125546. Curran Associates, Inc., 2024.
\newblock URL \url{https://proceedings.neurips.cc/paper_files/paper/2024/file/e3301977b92f28e32639ec99eb08f4a1-Paper-Datasets_and_Benchmarks_Track.pdf}.

\bibitem[Siddharth et~al.(2020)Siddharth, N., and Chakrabarti]{Siddharth2020}
L.~Siddharth, Madhusudanan N., and Amaresh Chakrabarti.
\newblock Toward automatically assessing the novelty of engineering design solutions.
\newblock \emph{Journal of Computing and Information Science in Engineering}, page~1, 03 2020.
\newblock \doi{10.1115/1.4044318}.

\bibitem[Suzgun et~al.(2022)Suzgun, Melas-Kyriazi, Sarkar, Kominers, and Shieber]{Suzgun2022}
Mirac Suzgun, Luke Melas-Kyriazi, Suproteem~K. Sarkar, Scott~Duke Kominers, and Stuart~M. Shieber.
\newblock The harvard uspto patent dataset: A large-scale, well-structured, and multi-purpose corpus of patent applications, 2022.
\newblock URL \url{https://arxiv.org/abs/2207.04043}.

\bibitem[Team et~al.(2024)Team, Mesnard, Hardin, Dadashi, Bhupatiraju, Pathak, Sifre, Rivière, Kale, Love, Tafti, Hussenot, Sessa, Chowdhery, Roberts, Barua, Botev, Castro-Ros, Slone, Héliou, Tacchetti, Bulanova, Paterson, Tsai, Shahriari, Lan, Choquette-Choo, Crepy, Cer, Ippolito, Reid, Buchatskaya, Ni, Noland, Yan, Tucker, Muraru, Rozhdestvenskiy, Michalewski, Tenney, Grishchenko, Austin, Keeling, Labanowski, Lespiau, Stanway, Brennan, Chen, Ferret, Chiu, Mao-Jones, Lee, Yu, Millican, Sjoesund, Lee, Dixon, Reid, Mikuła, Wirth, Sharman, Chinaev, Thain, Bachem, Chang, Wahltinez, Bailey, Michel, Yotov, Chaabouni, Comanescu, Jana, Anil, McIlroy, Liu, Mullins, Smith, Borgeaud, Girgin, Douglas, Pandya, Shakeri, De, Klimenko, Hennigan, Feinberg, Stokowiec, hui Chen, Ahmed, Gong, Warkentin, Peran, Giang, Farabet, Vinyals, Dean, Kavukcuoglu, Hassabis, Ghahramani, Eck, Barral, Pereira, Collins, Joulin, Fiedel, Senter, Andreev, and Kenealy]{gemmateam2024gemmaopenmodelsbased}
Gemma Team, Thomas Mesnard, Cassidy Hardin, Robert Dadashi, Surya Bhupatiraju, Shreya Pathak, Laurent Sifre, Morgane Rivière, Mihir~Sanjay Kale, Juliette Love, Pouya Tafti, Léonard Hussenot, Pier~Giuseppe Sessa, Aakanksha Chowdhery, Adam Roberts, Aditya Barua, Alex Botev, Alex Castro-Ros, Ambrose Slone, Amélie Héliou, Andrea Tacchetti, Anna Bulanova, Antonia Paterson, Beth Tsai, Bobak Shahriari, Charline~Le Lan, Christopher~A. Choquette-Choo, Clément Crepy, Daniel Cer, Daphne Ippolito, David Reid, Elena Buchatskaya, Eric Ni, Eric Noland, Geng Yan, George Tucker, George-Christian Muraru, Grigory Rozhdestvenskiy, Henryk Michalewski, Ian Tenney, Ivan Grishchenko, Jacob Austin, James Keeling, Jane Labanowski, Jean-Baptiste Lespiau, Jeff Stanway, Jenny Brennan, Jeremy Chen, Johan Ferret, Justin Chiu, Justin Mao-Jones, Katherine Lee, Kathy Yu, Katie Millican, Lars~Lowe Sjoesund, Lisa Lee, Lucas Dixon, Machel Reid, Maciej Mikuła, Mateo Wirth, Michael Sharman, Nikolai Chinaev, Nithum Thain, Olivier Bachem,
  Oscar Chang, Oscar Wahltinez, Paige Bailey, Paul Michel, Petko Yotov, Rahma Chaabouni, Ramona Comanescu, Reena Jana, Rohan Anil, Ross McIlroy, Ruibo Liu, Ryan Mullins, Samuel~L Smith, Sebastian Borgeaud, Sertan Girgin, Sholto Douglas, Shree Pandya, Siamak Shakeri, Soham De, Ted Klimenko, Tom Hennigan, Vlad Feinberg, Wojciech Stokowiec, Yu~hui Chen, Zafarali Ahmed, Zhitao Gong, Tris Warkentin, Ludovic Peran, Minh Giang, Clément Farabet, Oriol Vinyals, Jeff Dean, Koray Kavukcuoglu, Demis Hassabis, Zoubin Ghahramani, Douglas Eck, Joelle Barral, Fernando Pereira, Eli Collins, Armand Joulin, Noah Fiedel, Evan Senter, Alek Andreev, and Kathleen Kenealy.
\newblock Gemma: Open models based on gemini research and technology, 2024.
\newblock URL \url{https://arxiv.org/abs/2403.08295}.

\bibitem[Touvron et~al.(2023)Touvron, Martin, Stone, Albert, Almahairi, Babaei, Bashlykov, Batra, Bhargava, Bhosale, Bikel, Blecher, Ferrer, Chen, Cucurull, Esiobu, Fernandes, Fu, Fu, Fuller, Gao, Goswami, Goyal, Hartshorn, Hosseini, Hou, Inan, Kardas, Kerkez, Khabsa, Kloumann, Korenev, Koura, Lachaux, Lavril, Lee, Liskovich, Lu, Mao, Martinet, Mihaylov, Mishra, Molybog, Nie, Poulton, Reizenstein, Rungta, Saladi, Schelten, Silva, Smith, Subramanian, Tan, Tang, Taylor, Williams, Kuan, Xu, Yan, Zarov, Zhang, Fan, Kambadur, Narang, Rodriguez, Stojnic, Edunov, and Scialom]{touvron2023llama2openfoundation}
Hugo Touvron, Louis Martin, Kevin Stone, Peter Albert, Amjad Almahairi, Yasmine Babaei, Nikolay Bashlykov, Soumya Batra, Prajjwal Bhargava, Shruti Bhosale, Dan Bikel, Lukas Blecher, Cristian~Canton Ferrer, Moya Chen, Guillem Cucurull, David Esiobu, Jude Fernandes, Jeremy Fu, Wenyin Fu, Brian Fuller, Cynthia Gao, Vedanuj Goswami, Naman Goyal, Anthony Hartshorn, Saghar Hosseini, Rui Hou, Hakan Inan, Marcin Kardas, Viktor Kerkez, Madian Khabsa, Isabel Kloumann, Artem Korenev, Punit~Singh Koura, Marie-Anne Lachaux, Thibaut Lavril, Jenya Lee, Diana Liskovich, Yinghai Lu, Yuning Mao, Xavier Martinet, Todor Mihaylov, Pushkar Mishra, Igor Molybog, Yixin Nie, Andrew Poulton, Jeremy Reizenstein, Rashi Rungta, Kalyan Saladi, Alan Schelten, Ruan Silva, Eric~Michael Smith, Ranjan Subramanian, Xiaoqing~Ellen Tan, Binh Tang, Ross Taylor, Adina Williams, Jian~Xiang Kuan, Puxin Xu, Zheng Yan, Iliyan Zarov, Yuchen Zhang, Angela Fan, Melanie Kambadur, Sharan Narang, Aurelien Rodriguez, Robert Stojnic, Sergey Edunov, and Thomas
  Scialom.
\newblock Llama 2: Open foundation and fine-tuned chat models, 2023.
\newblock URL \url{https://arxiv.org/abs/2307.09288}.

\bibitem[Trappey et~al.(2021)Trappey, Trappey, and Hsieh]{Trappey2021}
Amy Trappey, Charles~V. Trappey, and Alex Hsieh.
\newblock {An intelligent patent recommender adopting machine learning approach for natural language processing: A case study for smart machinery technology mining}.
\newblock \emph{Technological Forecasting and Social Change}, 164\penalty0 (C), 2021.
\newblock \doi{10.1016/j.techfore.2020.1}.
\newblock URL \url{https://ideas.repec.org/a/eee/tefoso/v164y2021ics0040162520313378.html}.

\bibitem[Verhoeven et~al.(2016)Verhoeven, Bakker, and Veugelers]{Verhoeven2016}
Dennis Verhoeven, Jurriën Bakker, and Reinhilde Veugelers.
\newblock Measuring technological novelty with patent-based indicators.
\newblock \emph{Research Policy}, 45\penalty0 (3):\penalty0 707--723, 2016.
\newblock ISSN 0048-7333.
\newblock \doi{https://doi.org/10.1016/j.respol.2015.11.010}.
\newblock URL \url{https://www.sciencedirect.com/science/article/pii/S0048733315001857}.

\bibitem[Vowinckel and Hähnke(2023)]{Vowinckel2023}
Konrad Vowinckel and Volker~D. Hähnke.
\newblock Searchformer: Semantic patent embeddings by siamese transformers for prior art search.
\newblock \emph{World Patent Information}, 73:\penalty0 102192, 2023.
\newblock ISSN 0172-2190.
\newblock \doi{https://doi.org/10.1016/j.wpi.2023.102192}.
\newblock URL \url{https://www.sciencedirect.com/science/article/pii/S0172219023000224}.

\bibitem[Wang and Zhang(2024)]{wang2024learning}
Hongsong Wang and Yuqi Zhang.
\newblock Learning efficient representations for image-based patent retrieval.
\newblock In Qingshan Liu, Hanzi Wang, Zhanyu Ma, Weishi Zheng, Hongbin Zha, Xilin Chen, Liang Wang, and Rongrong Ji, editors, \emph{Pattern Recognition and Computer Vision}, pages 15--26, Singapore, 2024. Springer Nature Singapore.
\newblock ISBN 978-981-99-8540-1.

\bibitem[Wang and Chen(2019)]{Wang2019}
Juite Wang and Yi-Jing Chen.
\newblock A novelty detection patent mining approach for analyzing technological opportunities.
\newblock \emph{Advanced Engineering Informatics}, 42:\penalty0 100941, 2019.
\newblock ISSN 1474-0346.
\newblock \doi{https://doi.org/10.1016/j.aei.2019.100941}.
\newblock URL \url{https://www.sciencedirect.com/science/article/pii/S147403461830421X}.

\bibitem[Wang et~al.(2024)Wang, Yang, Huang, Yang, Majumder, and Wei]{wang2024multilinguale5textembeddings}
Liang Wang, Nan Yang, Xiaolong Huang, Linjun Yang, Rangan Majumder, and Furu Wei.
\newblock Multilingual e5 text embeddings: A technical report, 2024.
\newblock URL \url{https://arxiv.org/abs/2402.05672}.

\bibitem[Wu et~al.(2023)Wu, Zhu, Liu, Zhu, Wang, Zhao, Liu, Chen, and Xiong]{Wu2023}
Han Wu, Guanqi Zhu, Qi~Liu, Hengshu Zhu, Hao Wang, Hongke Zhao, Chuanren Liu, Enhong Chen, and Hui Xiong.
\newblock A multi-aspect neural tensor factorization framework for patent litigation prediction.
\newblock \emph{IEEE Transactions on Big Data}, PP:\penalty0 1--18, 01 2023.
\newblock \doi{10.1109/TBDATA.2023.3313030}.

\bibitem[Wu et~al.(2024)Wu, Zhu, Liu, Zhu, Wang, Zhao, Liu, Chen, and Xiong]{wu2024multi}
Han Wu, Guanqi Zhu, Qi~Liu, Hengshu Zhu, Hao Wang, Hongke Zhao, Chuanren Liu, Enhong Chen, and Hui Xiong.
\newblock A multi-aspect neural tensor factorization framework for patent litigation prediction.
\newblock \emph{IEEE Transactions on Big Data}, 10\penalty0 (1):\penalty0 35--54, 2024.
\newblock \doi{10.1109/TBDATA.2023.3313030}.

\bibitem[Zanella et~al.(2023)Zanella, Liu, and Choo]{Zanella2023}
Gianluca Zanella, Charles~Zhechao Liu, and Kim-Kwang~Raymond Choo.
\newblock Understanding the trends in blockchain domain through an unsupervised systematic patent analysis.
\newblock \emph{IEEE Transactions on Engineering Management}, 70\penalty0 (6):\penalty0 1991--2005, 2023.
\newblock \doi{10.1109/TEM.2021.3074310}.

\bibitem[Zhang et~al.(2020)Zhang, Kishore, Wu, Weinberger, and Artzi]{zhang2020bertscoreevaluatingtextgeneration}
Tianyi Zhang, Varsha Kishore, Felix Wu, Kilian~Q. Weinberger, and Yoav Artzi.
\newblock Bertscore: Evaluating text generation with bert, 2020.
\newblock URL \url{https://arxiv.org/abs/1904.09675}.

\end{thebibliography}

\section*{Checklist}

\begin{enumerate}

\item {\bf Claims}
    \item[] Question: Do the main claims made in the abstract and introduction accurately reflect the paper's contributions and scope?
    \item[] Answer: \answerYes{} 
    \item[] Justification: Our claims match our experimental results.
    \item[] Guidelines:
    \begin{itemize}
        \item The answer NA means that the abstract and introduction do not include the claims made in the paper.
        \item The abstract and/or introduction should clearly state the claims made, including the contributions made in the paper and important assumptions and limitations. A No or NA answer to this question will not be perceived well by the reviewers. 
        \item The claims made should match theoretical and experimental results, and reflect how much the results can be expected to generalize to other settings. 
        \item It is fine to include aspirational goals as motivation as long as it is clear that these goals are not attained by the paper. 
    \end{itemize}

\item {\bf Limitations}
    \item[] Question: Does the paper discuss the limitations of the work performed by the authors?
    \item[] Answer: \answerYes{}
    \item[] Justification: We discuss the limitations of our experiment in Section 4.
    \item[] Guidelines:
    \begin{itemize}
        \item The answer NA means that the paper has no limitation while the answer No means that the paper has limitations, but those are not discussed in the paper. 
        \item The authors are encouraged to create a separate "Limitations" section in their paper.
        \item The paper should point out any strong assumptions and how robust the results are to violations of these assumptions (e.g., independence assumptions, noiseless settings, model well-specification, asymptotic approximations only holding locally). The authors should reflect on how these assumptions might be violated in practice and what the implications would be.
        \item The authors should reflect on the scope of the claims made, e.g., if the approach was only tested on a few datasets or with a few runs. In general, empirical results often depend on implicit assumptions, which should be articulated.
        \item The authors should reflect on the factors that influence the performance of the approach. For example, a facial recognition algorithm may perform poorly when image resolution is low or images are taken in low lighting. Or a speech-to-text system might not be used reliably to provide closed captions for online lectures because it fails to handle technical jargon.
        \item The authors should discuss the computational efficiency of the proposed algorithms and how they scale with dataset size.
        \item If applicable, the authors should discuss possible limitations of their approach to address problems of privacy and fairness.
        \item While the authors might fear that complete honesty about limitations might be used by reviewers as grounds for rejection, a worse outcome might be that reviewers discover limitations that aren't acknowledged in the paper. The authors should use their best judgment and recognize that individual actions in favor of transparency play an important role in developing norms that preserve the integrity of the community. Reviewers will be specifically instructed to not penalize honesty concerning limitations.
    \end{itemize}

\item {\bf Theory assumptions and proofs}
    \item[] Question: For each theoretical result, does the paper provide the full set of assumptions and a complete (and correct) proof?
    \item[] Answer: \answerNA{}
    \item[] Justification: This paper focuses on dataset construction and does not include theoretical proofs.
    \item[] Guidelines:
    \begin{itemize}
        \item The answer NA means that the paper does not include theoretical results. 
        \item All the theorems, formulas, and proofs in the paper should be numbered and cross-referenced.
        \item All assumptions should be clearly stated or referenced in the statement of any theorems.
        \item The proofs can either appear in the main paper or the supplemental material, but if they appear in the supplemental material, the authors are encouraged to provide a short proof sketch to provide intuition. 
        \item Inversely, any informal proof provided in the core of the paper should be complemented by formal proofs provided in appendix or supplemental material.
        \item Theorems and Lemmas that the proof relies upon should be properly referenced. 
    \end{itemize}

    \item {\bf Experimental result reproducibility}
    \item[] Question: Does the paper fully disclose all the information needed to reproduce the main experimental results of the paper to the extent that it affects the main claims and/or conclusions of the paper (regardless of whether the code and data are provided or not)?
    \item[] Answer: \answerYes{}
    \item[] Justification: We provide all the details to the best of our capacity. Some of the details have been shared in the appendix.
    \item[] Guidelines:
    \begin{itemize}
        \item The answer NA means that the paper does not include experiments.
        \item If the paper includes experiments, a No answer to this question will not be perceived well by the reviewers: Making the paper reproducible is important, regardless of whether the code and data are provided or not.
        \item If the contribution is a dataset and/or model, the authors should describe the steps taken to make their results reproducible or verifiable. 
        \item Depending on the contribution, reproducibility can be accomplished in various ways. For example, if the contribution is a novel architecture, describing the architecture fully might suffice, or if the contribution is a specific model and empirical evaluation, it may be necessary to either make it possible for others to replicate the model with the same dataset, or provide access to the model. In general. releasing code and data is often one good way to accomplish this, but reproducibility can also be provided via detailed instructions for how to replicate the results, access to a hosted model (e.g., in the case of a large language model), releasing of a model checkpoint, or other means that are appropriate to the research performed.
        \item While NeurIPS does not require releasing code, the conference does require all submissions to provide some reasonable avenue for reproducibility, which may depend on the nature of the contribution. For example
        \begin{enumerate}
            \item If the contribution is primarily a new algorithm, the paper should make it clear how to reproduce that algorithm.
            \item If the contribution is primarily a new model architecture, the paper should describe the architecture clearly and fully.
            \item If the contribution is a new model (e.g., a large language model), then there should either be a way to access this model for reproducing the results or a way to reproduce the model (e.g., with an open-source dataset or instructions for how to construct the dataset).
            \item We recognize that reproducibility may be tricky in some cases, in which case authors are welcome to describe the particular way they provide for reproducibility. In the case of closed-source models, it may be that access to the model is limited in some way (e.g., to registered users), but it should be possible for other researchers to have some path to reproducing or verifying the results.
        \end{enumerate}
    \end{itemize}

\item {\bf Open access to data and code}
    \item[] Question: Does the paper provide open access to the data and code, with sufficient instructions to faithfully reproduce the main experimental results, as described in supplemental material?
    \item[] Answer: \answerYes{}
    \item[] Justification: We shared our data and code, see the appendix for details.
    \item[] Guidelines:
    \begin{itemize}
        \item The answer NA means that paper does not include experiments requiring code.
        \item Please see the NeurIPS code and data submission guidelines (\url{https://nips.cc/public/guides/CodeSubmissionPolicy}) for more details.
        \item While we encourage the release of code and data, we understand that this might not be possible, so “No” is an acceptable answer. Papers cannot be rejected simply for not including code, unless this is central to the contribution (e.g., for a new open-source benchmark).
        \item The instructions should contain the exact command and environment needed to run to reproduce the results. See the NeurIPS code and data submission guidelines (\url{https://nips.cc/public/guides/CodeSubmissionPolicy}) for more details.
        \item The authors should provide instructions on data access and preparation, including how to access the raw data, preprocessed data, intermediate data, and generated data, etc.
        \item The authors should provide scripts to reproduce all experimental results for the new proposed method and baselines. If only a subset of experiments are reproducible, they should state which ones are omitted from the script and why.
        \item At submission time, to preserve anonymity, the authors should release anonymized versions (if applicable).
        \item Providing as much information as possible in supplemental material (appended to the paper) is recommended, but including URLs to data and code is permitted.
    \end{itemize}

\item {\bf Experimental setting/details}
    \item[] Question: Does the paper specify all the training and test details (e.g., data splits, hyperparameters, how they were chosen, type of optimizer, etc.) necessary to understand the results?
    \item[] Answer: \answerYes{} 
    \item[] Justification: Details in the Appendix.
    \item[] Guidelines:
    \begin{itemize}
        \item The answer NA means that the paper does not include experiments.
        \item The experimental setting should be presented in the core of the paper to a level of detail that is necessary to appreciate the results and make sense of them.
        \item The full details can be provided either with the code, in appendix, or as supplemental material.
    \end{itemize}

\item {\bf Experiment statistical significance}
    \item[] Question: Does the paper report error bars suitably and correctly defined or other appropriate information about the statistical significance of the experiments?
    \item[] Answer: \answerYes{} 
    \item[] Justification: We report on our data collection attempts and errors in the Appendix.
    \item[] Guidelines:
    \begin{itemize}
        \item The answer NA means that the paper does not include experiments.
        \item The authors should answer "Yes" if the results are accompanied by error bars, confidence intervals, or statistical significance tests, at least for the experiments that support the main claims of the paper.
        \item The factors of variability that the error bars are capturing should be clearly stated (for example, train/test split, initialization, random drawing of some parameter, or overall run with given experimental conditions).
        \item The method for calculating the error bars should be explained (closed form formula, call to a library function, bootstrap, etc.)
        \item The assumptions made should be given (e.g., Normally distributed errors).
        \item It should be clear whether the error bar is the standard deviation or the standard error of the mean.
        \item It is OK to report 1-sigma error bars, but one should state it. The authors should preferably report a 2-sigma error bar than state that they have a 96\% CI, if the hypothesis of Normality of errors is not verified.
        \item For asymmetric distributions, the authors should be careful not to show in tables or figures symmetric error bars that would yield results that are out of range (e.g. negative error rates).
        \item If error bars are reported in tables or plots, The authors should explain in the text how they were calculated and reference the corresponding figures or tables in the text.
    \end{itemize}

\item {\bf Experiments compute resources}
    \item[] Question: For each experiment, does the paper provide sufficient information on the computer resources (type of compute workers, memory, time of execution) needed to reproduce the experiments?
    \item[] Answer: \answerYes{}
    \item[] Justification: We documented this in the Appendix.
    \item[] Guidelines:
    \begin{itemize}
        \item The answer NA means that the paper does not include experiments.
        \item The paper should indicate the type of compute workers CPU or GPU, internal cluster, or cloud provider, including relevant memory and storage.
        \item The paper should provide the amount of compute required for each of the individual experimental runs as well as estimate the total compute. 
        \item The paper should disclose whether the full research project required more compute than the experiments reported in the paper (e.g., preliminary or failed experiments that didn't make it into the paper). 
    \end{itemize}
    
\item {\bf Code of ethics}
    \item[] Question: Does the research conducted in the paper conform, in every respect, with the NeurIPS Code of Ethics \url{https://neurips.cc/public/EthicsGuidelines}?
    \item[] Answer: \answerYes{} 
    \item[] Justification: We carefully reviewed the NeurIPS Code of Ethics.
    \item[] Guidelines:
    \begin{itemize}
        \item The answer NA means that the authors have not reviewed the NeurIPS Code of Ethics.
        \item If the authors answer No, they should explain the special circumstances that require a deviation from the Code of Ethics.
        \item The authors should make sure to preserve anonymity (e.g., if there is a special consideration due to laws or regulations in their jurisdiction).
    \end{itemize}

\item {\bf Broader impacts}
    \item[] Question: Does the paper discuss both potential positive societal impacts and negative societal impacts of the work performed?
    \item[] Answer: \answerYes{}
    \item[] Justification: We described the potential social impact of this research in the Appendix.
    \item[] Guidelines:
    \begin{itemize}
        \item The answer NA means that there is no societal impact of the work performed.
        \item If the authors answer NA or No, they should explain why their work has no societal impact or why the paper does not address societal impact.
        \item Examples of negative societal impacts include potential malicious or unintended uses (e.g., disinformation, generating fake profiles, surveillance), fairness considerations (e.g., deployment of technologies that could make decisions that unfairly impact specific groups), privacy considerations, and security considerations.
        \item The conference expects that many papers will be foundational research and not tied to particular applications, let alone deployments. However, if there is a direct path to any negative applications, the authors should point it out. For example, it is legitimate to point out that an improvement in the quality of generative models could be used to generate deepfakes for disinformation. On the other hand, it is not needed to point out that a generic algorithm for optimizing neural networks could enable people to train models that generate Deepfakes faster.
        \item The authors should consider possible harms that could arise when the technology is being used as intended and functioning correctly, harms that could arise when the technology is being used as intended but gives incorrect results, and harms following from (intentional or unintentional) misuse of the technology.
        \item If there are negative societal impacts, the authors could also discuss possible mitigation strategies (e.g., gated release of models, providing defenses in addition to attacks, mechanisms for monitoring misuse, mechanisms to monitor how a system learns from feedback over time, improving the efficiency and accessibility of ML).
    \end{itemize}
    
\item {\bf Safeguards}
    \item[] Question: Does the paper describe safeguards that have been put in place for the responsible release of data or models that have a high risk for misuse (e.g., pretrained language models, image generators, or scraped datasets)?
    \item[] Answer: \answerNo{}
    \item[] Justification: Since our data is based on patent documents that are already publicly available, no separate safeguard was considered.
    \item[] Guidelines:
    \begin{itemize}
        \item The answer NA means that the paper poses no such risks.
        \item Released models that have a high risk for misuse or dual-use should be released with necessary safeguards to allow for controlled use of the model, for example by requiring that users adhere to usage guidelines or restrictions to access the model or implementing safety filters. 
        \item Datasets that have been scraped from the Internet could pose safety risks. The authors should describe how they avoided releasing unsafe images.
        \item We recognize that providing effective safeguards is challenging, and many papers do not require this, but we encourage authors to take this into account and make a best faith effort.
    \end{itemize}

\item {\bf Licenses for existing assets}
    \item[] Question: Are the creators or original owners of assets (e.g., code, data, models), used in the paper, properly credited and are the license and terms of use explicitly mentioned and properly respected?
    \item[] Answer: \answerYes{}
    \item[] Justification: The USPTO data we used is all public data.
    \item[] Guidelines:
    \begin{itemize}
        \item The answer NA means that the paper does not use existing assets.
        \item The authors should cite the original paper that produced the code package or dataset.
        \item The authors should state which version of the asset is used and, if possible, include a URL.
        \item The name of the license (e.g., CC-BY 4.0) should be included for each asset.
        \item For scraped data from a particular source (e.g., website), the copyright and terms of service of that source should be provided.
        \item If assets are released, the license, copyright information, and terms of use in the package should be provided. For popular datasets, \url{paperswithcode.com/datasets} has curated licenses for some datasets. Their licensing guide can help determine the license of a dataset.
        \item For existing datasets that are re-packaged, both the original license and the license of the derived asset (if it has changed) should be provided.
        \item If this information is not available online, the authors are encouraged to reach out to the asset's creators.
    \end{itemize}

\item {\bf New assets}
    \item[] Question: Are new assets introduced in the paper well documented and is the documentation provided alongside the assets?
    \item[] Answer: \answerYes{}
    \item[] Justification: We shared details of the dataset generation process and the dataset created to the best of our capabilities. 
    \item[] Guidelines:
    \begin{itemize}
        \item The answer NA means that the paper does not release new assets.
        \item Researchers should communicate the details of the dataset/code/model as part of their submissions via structured templates. This includes details about training, license, limitations, etc. 
        \item The paper should discuss whether and how consent was obtained from people whose asset is used.
        \item At submission time, remember to anonymize your assets (if applicable). You can either create an anonymized URL or include an anonymized zip file.
    \end{itemize}

\item {\bf Crowdsourcing and research with human subjects}
    \item[] Question: For crowdsourcing experiments and research with human subjects, does the paper include the full text of instructions given to participants and screenshots, if applicable, as well as details about compensation (if any)? 
    \item[] Answer: \answerNA{} 
    \item[] Justification: This research does not involve human participants during the dataset construction.
    \item[] Guidelines:
    \begin{itemize}
        \item The answer NA means that the paper does not involve crowdsourcing nor research with human subjects.
        \item Including this information in the supplemental material is fine, but if the main contribution of the paper involves human subjects, then as much detail as possible should be included in the main paper. 
        \item According to the NeurIPS Code of Ethics, workers involved in data collection, curation, or other labor should be paid at least the minimum wage in the country of the data collector. 
    \end{itemize}

\item {\bf Institutional review board (IRB) approvals or equivalent for research with human subjects}
    \item[] Question: Does the paper describe potential risks incurred by study participants, whether such risks were disclosed to the subjects, and whether Institutional Review Board (IRB) approvals (or an equivalent approval/review based on the requirements of your country or institution) were obtained?
    \item[] Answer: \answerNA{}
    \item[] Justification: This research does not involve a study with human participants.
    \item[] Guidelines:
    \begin{itemize}
        \item The answer NA means that the paper does not involve crowdsourcing nor research with human subjects.
        \item Depending on the country in which research is conducted, IRB approval (or equivalent) may be required for any human subjects research. If you obtained IRB approval, you should clearly state this in the paper. 
        \item We recognize that the procedures for this may vary significantly between institutions and locations, and we expect authors to adhere to the NeurIPS Code of Ethics and the guidelines for their institution. 
        \item For initial submissions, do not include any information that would break anonymity (if applicable), such as the institution conducting the review.
    \end{itemize}

\item {\bf Declaration of LLM usage}
    \item[] Question: Does the paper describe the usage of LLMs if it is an important, original, or non-standard component of the core methods in this research? Note that if the LLM is used only for writing, editing, or formatting purposes and does not impact the core methodology, scientific rigorousness, or originality of the research, declaration is not required.
    \item[] Answer: \answerYes{} 
    \item[] Justification: We describe in detail how we used LLM in the data parsing process in section\ref{sec:parsing} and the appendix.
    \item[] Guidelines:
    \begin{itemize}
        \item The answer NA means that the core method development in this research does not involve LLMs as any important, original, or non-standard components.
        \item Please refer to our LLM policy (\url{https://neurips.cc/Conferences/2025/LLM}) for what should or should not be described.
    \end{itemize}

\end{enumerate}

\appendix
\section*{Appendix}

\section{Dataset Details}

\subsection{Data Availability}\label{data_availability}
The PANORAMA dataset and related code are publicly available:
\begin{itemize}
\item \textbf{Task and Code}: \url{https://github.com/LGAI-Research/PANORAMA}
\item \textbf{Hugging Face Dataset}: \url{https://huggingface.co/datasets/LG-AI-Research/PANORAMA}
\end{itemize}

Sample test code and instructions for reproducibility are provided in the Task GitHub repository above.

\subsection{Dataset Structure}

Below, we provide a representative JSON snippet illustrating the structure of an individual PANORAMA dataset record. The dataset distinguishes between \textbf{initial claims}, defined as the claims presented immediately before receiving the first Office Action (OA), and \textbf{final claims}, defined as the claims immediately before receiving the Notice of Allowance (NOA). Here, \textbf{Office Action (OA)} refers to official correspondence from a patent examiner evaluating patentability, which can be either a \textbf{Non-Final Office Action (CTNF)}—a preliminary evaluation identifying issues that require amendment or response—or a \textbf{Final Office Action (CTFR)}—the examiner's definitive determination regarding patentability unless appealed or further amended. The \textbf{Notice of Allowance (NOA)} indicates that the examiner has deemed the claims allowable, signifying readiness for patent issuance upon payment of associated fees. This example includes key fields such as application number, abstract, initial claims, final claims, CTNF and NOA document texts, and cited patents.

\begin{lstlisting}[language=json, numbers=none]
{
    "record_id": 3,
    "applicationNumber": 14735126,
    "title": "Gas distributors used in wafer carriers",
    "abstract": "ABSTRACT The present invention relates to gas distributors used in wafer carriers. The gas distributors comprise a body with an interior space, a separator configured at the front side of the body in the interior space, and an air inlet connected with the body. One edge of the separator and the front side of the body together form a passage. ...",
    "initialClaims": [
        "1. A gas distributor used in wafer carriers, comprising: a body having an interior space; and a separator configured in the interior space, wherein one edge of the separator and the front side of the body form a passage, and wherein the separator divides the interior space into: a first room connected with an air inlet, wherein the air inlet and passage are configured at adjacent sides of the body; and a second room connected with at least one air outlet, wherein the at least one air outlet is configured on the rear side of the body.",
        "2. The gas distributor used in wafer carriers as claimed in claim 1, wherein the separator further comprises an extended member, and wherein the extended member extends into the first room.",
        "3. The gas distributor used in wafer carriers as claimed in claim 2, wherein the extended member has a plane facing the front side of the body, and wherein the width of the plane is greater at the end away from the air inlet.",
        ...
        ],
    "finalClaims": [
        "1. (Cancelled)",
        "2. (Cancelled)",
        "3. A gas distributor used in wafer carriers, comprising: a body having an interior space, wherein the interior space comprises a front side and a back side; a separator being configured in the interior space, wherein the separator comprises a first edge and a second edge, the first edge is connected with the front side of the interior space and the second edge reaches the back side of the interior space, forming a passage; wherein the separator divides the interior space into: a first room connected with an air inlet, wherein the air inlet is configured in a bottom of the first room; and a second room connected with at least one air outlet, wherein the at least one air outlet is configured on the front side of the interior space; wherein the separator further comprises an extended member which is connected with the second edge of the separator, and the extended member is parallel to the back side of an interior space in the first room; The-gas- distriuter used in wafer cer claimed in claim 1, wherein the extended member has a first end and a second end; wherein the first end is located at the bottom of the first room, and the second end is located on top of the first room; wherein a first width of the extended member between the first end of the extended member and the second edge of the separator is smaller than a second width of the extended member between the second end of the extended member and the second edge of the separator.",
        ...
    ],
    "CTNFDocumentIdentifier": "IWGK9PBYRXEAPX1",
    "CTNFBodyText": "",
    "NOABodyText": "",
    "patentsCitedByExaminer": [
        {
            "referenceIdentifier": "6758876",
            "title": "Substrate transport apparatus, pod and method",
            "abstract": "A method of using a substrate transport pod suitable for manufacturing semiconductor devices...",
            "claims": [
                "1. A method for transporting substrates between a plurality of processes, comprising: loading the substrates into a pod...",
                ...
            ]
        }
    ],
    "firstInventorToFileIndicator": "Y",
    "applicationStatusCode": 150,
    "applicationTypeCode": "UTL",
    "entityStatusData": {
        "smallEntityStatusIndicator": false,
        "businessEntityStatusCategory": "Small"
    },
    "filingDate": "2015-06-09",
    "class/subclass": "137/573",
    "nationalStageIndicator": false,
    "firstInventorName": "CHIN-MING LIN",
    "cpcClassificationBag": [
        "H01L21/67393",
        "Y10T137/86212"
    ],
    "effectiveFilingDate": "2015-06-09",
    "publicationDateBag": [
        "2015-12-10"
    ],
    "publicationSequenceNumberBag": [
        "0357218"
    ],
    "earliestPublicationDate": "2015-12-10",
    "applicationTypeLabelName": "Utility",
    "applicationStatusDate": "2018-01-31",
    "class": "137",
    "applicationTypeCategory": "REGULAR",
    "applicationStatusDescriptionText": "Patented Case",
    "patentNumber": "9899246",
    "grantDate": "2018-02-20",
    "applicantBag": [
        {
            "applicantNameText": "GUDENG PRECISION INDUSTRIAL CO., LTD",
            "correspondenceAddressBag": [
                {
                    "cityName": "New Taipei City",
                    "countryCode": "TW",
                    "nameLineOneText": "GUDENG PRECISION INDUSTRIAL CO., LTD",
                    "countryName": "TAIWAN",
                    "postalAddressCategory": "postal"
                }
            ]
        }
    ],
    "firstApplicantName": "GUDENG PRECISION INDUSTRIAL CO., LTD",
    "customerNumber": 88174,
    "groupArtUnitNumber": "3753",
    "earliestPublicationNumber": "US20150357218A1",
    "inventionTitle": "GAS DISTRIBUTOR USED IN WAFER CARRIERS",
    "applicationConfirmationNumber": 3190,
    "examinerNameText": "HICKS, ANGELISA",
    "subclass": "573",
    "publicationCategoryBag": [
        "Granted/Issued",
        "Pre-Grant Publications - PGPub"
    ],
    "docketNumber": "KS-00041"
}
\end{lstlisting}

\subsection{Parsed CTNF structure}
The following section describes the structured JSON representation obtained by parsing Non-Final Office Action (CTNF) documents. Each CTNF document is converted into a structured JSON format containing detailed information for every individual claim examined within the action. Each claim is represented with a set of clearly defined attributes. The attribute \texttt{claimNumber} specifies the numerical identifier of the claim, and \texttt{parentClaim} indicates the claim it depends on, with a value of $-1$ signifying an independent claim. The boolean attribute \texttt{isReject} records whether the claim has been rejected by the examiner. If the claim has been rejected, the detailed reasoning behind the rejection is provided under the \texttt{reasons} field. Each rejection reason entry includes a U.S.C. section code (\texttt{sectionCode}) to specify the legal grounds (e.g., §102 for anticipation or §103 for obviousness), along with explicit evidence from cited patents. Such evidence includes the cited patent number (\texttt{patentNum}), relevant paragraph numbers (\texttt{text}), and associated figure reference numerals (\texttt{img}). Additionally, the examiner's detailed textual explanation for the rejection is documented in the \texttt{reason} field. This structured representation provides comprehensive and logically organized information from CTNF, facilitating accurate computational analysis and efficient retrieval of patent examination details for subsequent benchmarking tasks.
\begin{lstlisting}[language=json, numbers=none]
{
    "claims": [
        {
            "claimNumber": 1,
            "parentClaim": -1,
            "isReject": true,
            "reasons": [
                {
                    "sectionCode": 102,
                    "citedPatents": [
                        {
                            "patentNum": "US 20070159740",
                            "text": [112, 113, 114, 115, 116, 117, 118, 119],
                            "img": ["12"]
                        }
                    ],
                    "reason": "Regarding claim 1, Williams teaches a power cord..."
                }
            ]
        },
        {
            "claimNumber": 2,
            "parentClaim": 1,
            "isReject": true,
            "reasons": [
                {
                    "sectionCode": 102,
                    "citedPatents": [
                        {
                            "patentNum": "US 20070159740",
                            "text": [120, 121],
                            "img": ["12"]
                        }
                    ],
                    "reason": "Regarding claim 2, Williams teaches the power cord with leakage current detection and interruption device of claim 1,..."
                }
            ]
        },
        // ... additional items with same structure
    ]
}
\end{lstlisting}

\subsection{Parsed Specification structure}
The following section describes the structured JSON representation obtained by parsing patent specification documents. Each parsed specification is represented as a JSON object containing an array named \texttt{items}, which stores entries corresponding to individual paragraphs within the original document. Each entry consists of two fields: a four-digit numerical key (\texttt{key}) and the associated textual content (\texttt{content}). These numerical keys directly reflect the paragraph numbering present in the original patent specification. This structured JSON representation maintains the sequential integrity of the original specification, facilitating efficient computational access and precise information retrieval for downstream benchmarking tasks.

\begin{lstlisting}[language=json, numbers=none]
{
  "id": "spec_txt_20020051537",
  "items": [
    {
      "key": "0001",
      "content": "This application hereby claims priority 35 U.S.C. section 119 to U.S. Provisional Patent Application..."
    },
    {
      "key": "0002",
      "content": "1. Field of the Invention"
    },
    {
      "key": "0003",
      "content": "The present invention relates generally to..."
    },
    {
      "key": "0004",
      "content": "2. Related Art"
    },
    // ... additional items with same structure
    {
      "key": "0107",
      "content": "The foregoing descriptions of embodiments of..."
    }
  ]
}
\end{lstlisting}

\subsection{Statistical Overview}
The PANORAMA dataset provides comprehensive insights into patent prosecution patterns through various statistical distributions. We analyzed key metrics including claim counts, rejection codes, technological categories, citation patterns, and rejection rates across the dataset.

\begin{figure}
    \centering
    \includegraphics[width=1\linewidth]{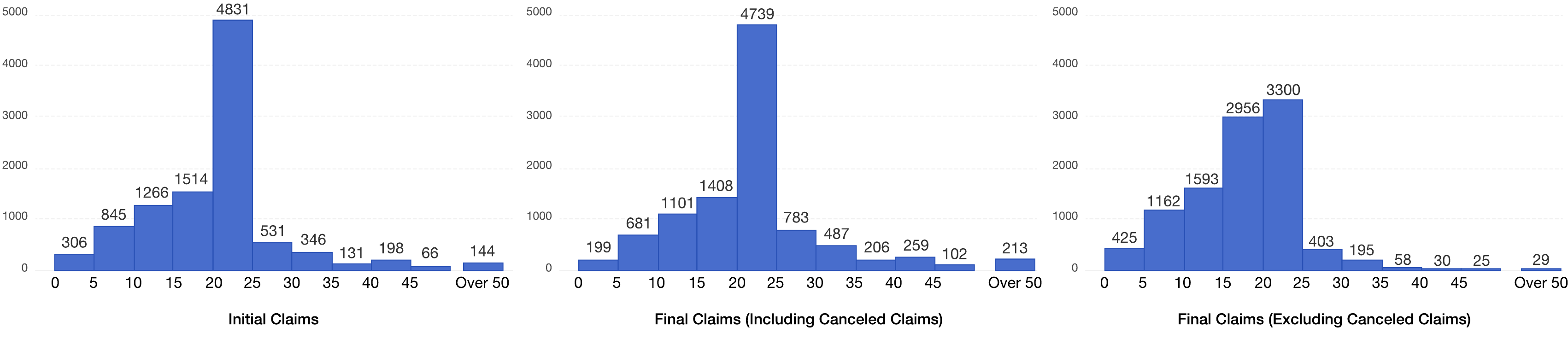}
    \caption{Distribution of Claims Count in USPTO Patent Applications. The three panels show the frequency distribution of: (Left) initial claims count in original patent applications, (Center) final claims count including canceled claims during prosecution, and (Right) final claims count excluding canceled claims.}
    \label{fig:charts-claims-excluding-canceled}
\end{figure}

Figure \ref{fig:charts-claims-excluding-canceled} illustrates the distribution of patent claims across three perspectives. Initial applications typically contain 20 claims, with a notable peak at 20-25 claims reflecting the USPTO's base filing fee structure that covers up to 20 claims. During prosecution, many claims are modified or canceled, resulting in different final claim distributions. When excluding canceled claims, the majority of patents maintain 15-20 active claims, with fewer applications exceeding 25 claims.

\begin{figure}
    \centering
    \includegraphics[width=1\linewidth]{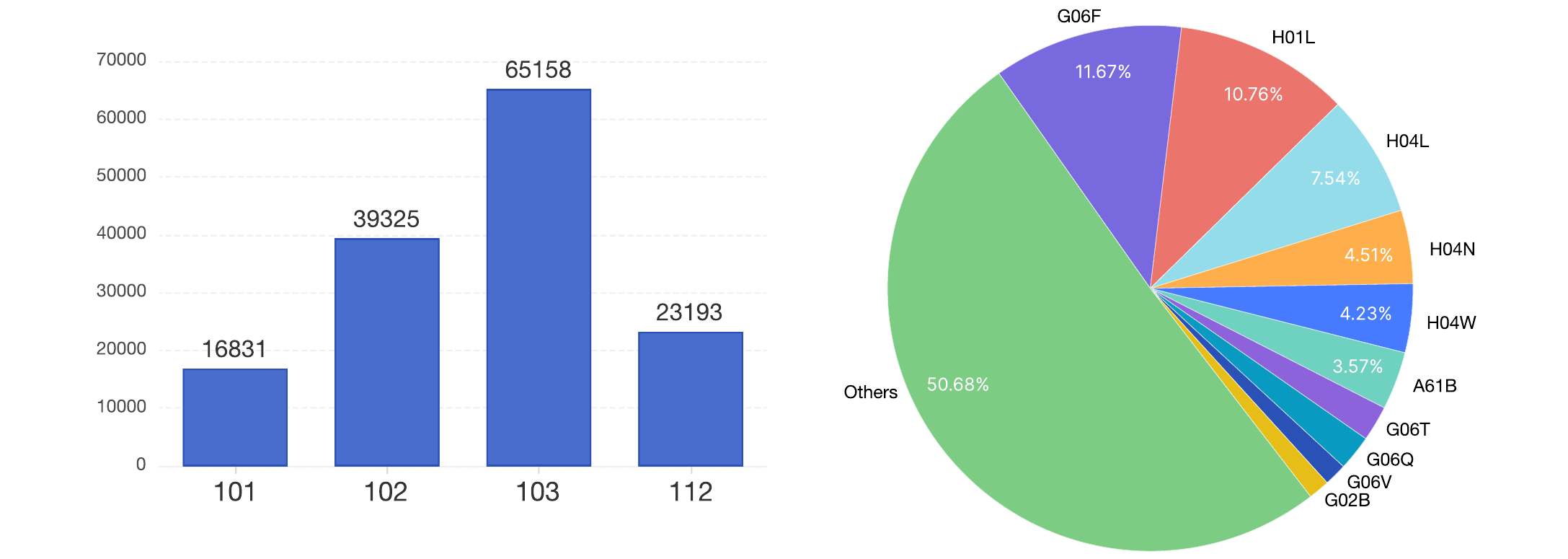}
    \caption{Distribution of rejections by section code (left) and IPC categories (right) in the PANORAMA dataset.}
    \label{fig:charts-section-ipc}
\end{figure}

Figure \ref{fig:charts-section-ipc} presents rejection patterns and technological distributions within PANORAMA. The left panel shows that §103 (non-obviousness) dominates with 65,158 rejections, followed by §102 (novelty) with 39,325 rejections. §112 (specification) and §101 (subject matter eligibility) account for 23,193 and 16,831 rejections, respectively. The right panel reveals that over half of the patents fall into diverse technical fields ("Others" at 50.68\%), while major categories include G06F (Digital Data Processing, 11.67\%), H01L (Semiconductor Devices, 10.76\%), and H04L (Digital Information Transmission, 7.54\%).

\begin{figure}
    \centering
    \includegraphics[width=1\linewidth]{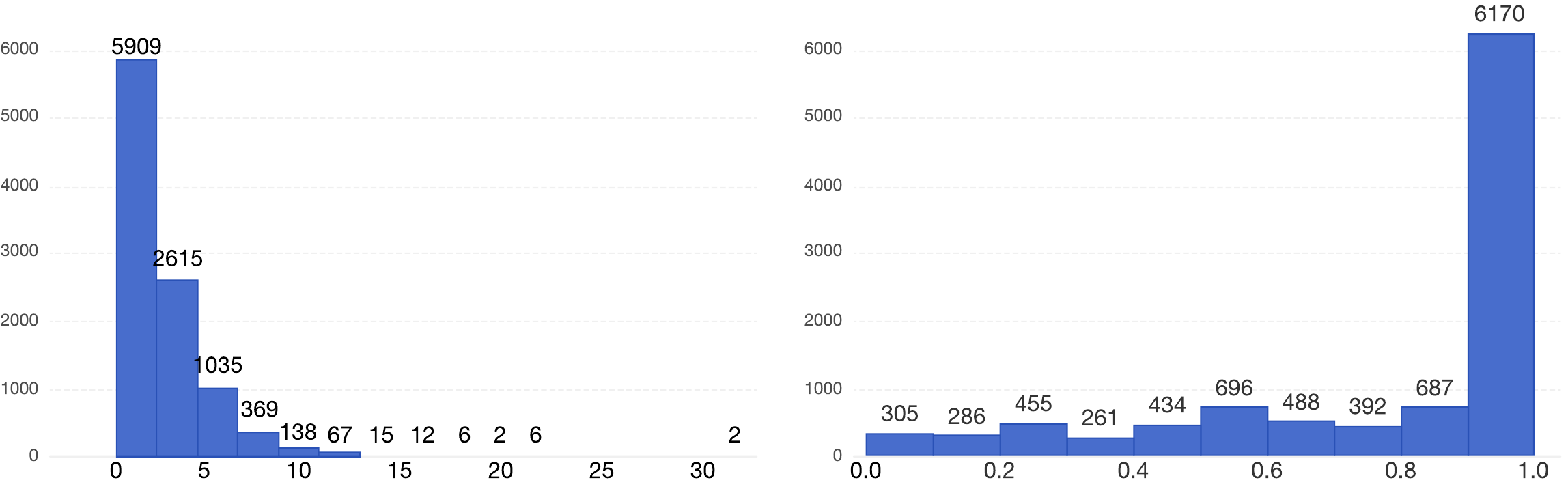}
    \caption{Distribution of cited patents count (left) and rejection rates (right) in the PANORAMA dataset.}
    \label{fig:charts-cited-rejection}
\end{figure}

Figure \ref{fig:charts-cited-rejection} examines citation patterns and rejection outcomes. The left panel demonstrates that most patent applications cite minimal prior art, with 5,909 applications citing 0-2 patents and the majority citing fewer than five references. The right panel reveals a heavily skewed distribution of rejection rates, with an overwhelming concentration at 1.0, where 6,170 applications experienced complete rejection. This extreme skewness—contrasting with only 305 applications at 0.0 rejection rate—indicates that the vast majority of patent applications face complete initial rejection of all claims, rather than partial acceptance.

Table~\ref{tab:patent-component-stats} presents the statistics of patent components in our dataset. Initial claims have a mean count of 17.81 per patent application with considerable variance (SD = 6.92), ranging from applications with no claims to those with as many as 118 claims. The mean textual length is 434.88 characters per claim. Final claims show similar patterns with a slightly lower mean count (16.78) but greater variability in length, with some claims extending to 17,733 characters. This indicates a refinement process where claims tend to become more detailed and specific during prosecution. The specification components display significant complexity with an average of 132.81 items per document and substantial variability (SD = 173.34). The mean content length of specification items is 534.99 characters, with some reaching up to 176,266 characters, highlighting the detailed technical descriptions contained in patent applications.

Table~\ref{tab:doc-length-stats} provides statistics on the length of different patent document types. The substantial difference in document lengths reflects their distinct purposes in the patent prosecution process. Patent abstracts are concise (mean length = 849.65 characters), while Non-Final Office Actions (CTNF) are the most verbose (mean = 13,414.35 characters), containing detailed examiner feedback. Notices of Allowance (NOA) fall in between (mean = 4,982.83 characters), focusing on the rationale for patent approval.

\begin{table}[t]
\renewcommand{\arraystretch}{1.2}
\centering
\begin{tabular}{llrrr}
\toprule
\textbf{Component} & \textbf{Metric} & \textbf{Mean $\pm$ SD} & \textbf{Min} & \textbf{Max} \\
\midrule
\multirow{2}{*}{\textbf{Initial Claims}} & Count & 17.81 $\pm$ 6.92 & 0 & 118 \\
 & Length (chars) & 434.88 $\pm$ 307.87 & 49 & 12,323 \\
\midrule
\multirow{2}{*}{\textbf{Final Claims}} & Count & 16.78 $\pm$ 7.12 & 0 & 134 \\
 & Length (chars) & 505.33 $\pm$ 395.75 & 74 & 17,733 \\
\midrule
\multirow{2}{*}{\textbf{Specification}} & Item Count & 132.81 $\pm$ 173.34 & 6 & 8144 \\
 & Content Length (chars) & 534.99 $\pm$ 541.46 & 0 & 176,266 \\
\bottomrule
\end{tabular}
\vspace{0.2cm}
\caption{Statistics of Patent Claims and Specification Components}\label{tab:patent-component-stats}
\renewcommand{\arraystretch}{1.0}
\end{table}

\begin{table}[t]
\renewcommand{\arraystretch}{1.2}
\centering
\begin{tabular}{lrrr}
\toprule
\textbf{Document Type} & \textbf{Mean $\pm$ SD} & \textbf{Min} & \textbf{Max} \\
\midrule
Abstract & 849.65 $\pm$ 909.18 & 64 & 75,767 \\
CTNF & 13,414.35 $\pm$ 10,874.00 & 4 & 133,119 \\
NOA & 4,982.83 $\pm$ 3,624.40 & 266 & 53,917 \\
\bottomrule
\end{tabular}
\vspace{0.2cm}
\caption{Statistics of Patent Document Lengths (in characters)}\label{tab:doc-length-stats}
\renewcommand{\arraystretch}{1.0}
\end{table}

\subsection{Licensing Information}\label{license}
The dataset is released under the Creative Commons Attribution-NonCommercial 4.0 International License.

\section{Curation Details of Dataset}
This section describes how the PANORAMA dataset was curated using publicly accessible APIs provided by the USPTO. 

\subsection{Data Collection and Initial Filtering}\label{data_collection}

The PANORAMA dataset was systematically curated using publicly accessible APIs provided by the USPTO. The detailed data collection and initial filtering procedures are outlined below:

\begin{enumerate} \item \textbf{Retrieval of Non-Final Rejection Documents:}
Non-Final Rejection documents were retrieved via the USPTO OA Text Retrieval API (\url{https://developer.uspto.gov/ds-api/oa\_actions/v1/records}). Specifically, we targeted Non-Final Rejection documents issued between January 1, 2015, and December 31, 2024, utilizing query criteria based on legacy document code identifiers and grant date ranges.

\item \textbf{Filtering for First Non-Final Rejection Documents:}
Retrieved Non-Final Rejection  documents were filtered to include only the earliest-issued document per the patent application, excluding any subsequent documents for the same application.

\item \textbf{Final Rejection and Allowance Status Verification:}
Applications were filtered based on examination outcomes, to include only those that ultimately received a Notice of Allowance (NOA) without encountering a Final Rejection (CTFR).

\item \textbf{Document Consistency Check (Non-Final Rejection and Notice of Allowance):}
Applications with inconsistencies, including modifications in abstracts or specifications between the initial Non-Final Rejection  document and the subsequent Notice of Allowance document, were excluded to maintain data consistency and validity.

\item \textbf{Detailed Document Collection via Patent File Wrapper API:}
For each valid application, additional documentation was systematically collected using the Patent File Wrapper API (\url{https://beta-api.uspto.gov/api/v1/patent/applications/\{applicationNumber\}/documents}). Key documents retrieved included application specifications, abstracts, initial claims (claims presented immediately before the first Office Action), final claims (claims presented immediately before the Notice of Allowance), and associated drawings. XML parsing was extensively used to extract structured content from these documents.

\item \textbf{Citation Data Extraction:}
Patents cited by examiners within the Non-Final Rejection documents were comprehensively identified and collected. For each cited patent, relevant information—including abstracts, claims, specifications, and drawings—was systematically extracted using the Enriched Cited Reference Metadata API and the \texttt{patent\_client} Python library.

\item \textbf{Structured Data Compilation:}
All collected and extracted data were meticulously structured into standardized JSON format, consolidating application metadata, claim details, examiner-cited patent information, and textual content of associated documents.
\end{enumerate}

In total, we initially retrieved 206,767 patent records, which were then subjected to our rigorous filtering criteria. After applying all filtering steps, only 12,839 records (6.21\%) were retained for further processing. The detailed statistics of this filtering process are presented in Table~\ref{tab:error_analysis}.
\begin{table}[h]
\centering
\renewcommand{\arraystretch}{1.2}
\begin{tabular}{llc}
\toprule
\textbf{Details} & \textbf{Count} \\ 
\midrule
Document was not the first CTNF for the application & 64,646 \\
application received Final Rejection (CTFR) & 4 \\
NOA document XML format missing & 2 \\
NOA XML parsing errors & 41,586 \\
Specification retrieval or parsing failure & 18,476 \\
Abstract retrieval or parsing failure & 9,186 \\
Drawing retrieval or parsing failure & 593 \\
Initial claims retrieval or parsing failure & 10,609 \\
Final claims retrieval or parsing failure & 4,988 \\
Cited patent missing claims information & 43,838 \\
\bottomrule
\end{tabular}
\vspace{0.2cm}
\caption{Detailed summary of data exclusion errors} \label{tab:error_analysis} \end{table}

\subsection{Generation of Parsed Non-Final Rejection Documents}\label{parsed_ctnf}

Following the initial data collection and filtering, we transformed the retained patent records into structured Non-Final Rejection documents suitable for further validation and analysis. To automate this parsing procedure, we employed the GPT-4o model, leveraging its capabilities to systematically extract claim rejection details from raw Non-Final Rejection Documents texts. The detailed steps are described below:

\begin{enumerate}
    \item \textbf{Organization of Input Data:}
    The retained records were stored as JSON files in a standardized naming convention (\texttt{rec\_rXXXXX\_\{applicationNumber\}.json}) within a dedicated directory, facilitating automated processing and traceability.

    \item \textbf{Automated Parsing Process:}
    Each record file underwent the following automated parsing steps:
    \begin{itemize}
        \item Application-specific data such as the application number, initial claims, and the raw CTNF body text (\texttt{CTNFBodyText}) were extracted.
        \item These data points, combined with the structured parsing prompt provided in Section~\ref{subsubsec:parsing_prompt}, were input into the GPT-4o model via the OpenAI API.
        \item The model's responses, structured as JSON, were retrieved and rigorously validated to ensure format consistency and correctness.
    \end{itemize}

    \item \textbf{Structured Output Generation:}
    The validated, structured outputs from GPT-4o were saved as new JSON files named \texttt{pC\_rXXXXX\_\{applicationNumber\}.json}. This structured storage ensured easy access, further validation, and reproducibility in subsequent steps.

    \item \textbf{Robust Error Management:}
    Throughout this automated parsing stage, rigorous error handling was implemented. All encountered issues—including missing input files, invalid JSON outputs from GPT-4o, or parsing inconsistencies—were systematically logged and reviewed, thereby maintaining the integrity of the generated parsed Non-Final Rejection documents.
\end{enumerate}

\subsubsection{Parsing Prompt}\label{subsubsec:parsing_prompt}
The following prompt was used to instruct the GPT-4o model on extracting structured information from the Non-Final Rejection documents:

\begin{lstlisting}[extendedchars=true, literate={§}{{\S}}1]
IMPORTANT: You must ONLY return the requested JSON structure. Do not include ANY additional text, explanations, or comments before or after the JSON.

Task Overview:
The task you need to perform is to parse the CTNF txt document into JSON. The corresponding Claims(in array) are provided as a reference, but you are not parsing them.
The final output should look like the following JSON structure.

OUTPUT FORMAT:
{
  "claims": [
    {
        "claimNumber": <integer>,
        "parentClaim": <integer>,
        "isReject": <boolean>,
        "reasons": [{
          "sectionCode": <integer>,
          "citedPatents": <array of strings>,
          "reason": <string>
        }]
    }
  ]
}

Parsing Instructions:
1. Initial Analysis:
- Read CTNF document and corresponding Claims carefully.
- Identify all claim numbers mentioned anywhere in CTNF.
- Look for Common range formats, such as:
  * "Claims 1-19"
  * "Claims 1, 2, and 4-7"
  * "Claims 1-10 and 15"
- For claim ranges:
  * Expand ranges to include all individual claim numbers.
    - For example, "Claims 1, 2, and 4-7" should be processed as claims 1, 2, 4, 5, 6, 7.
 * Even if the CTNF bundles multiple claims together, the JSON output should list them as individual claims.

2. Extraction Guidelines for Each Field:
For claimNumber:
- Return claim number as an integer.
- Include ALL claims that appears in the corresponding Claims.
- Exclude CANCELED claims among the corresponding claims. 

For parentClaim:
- Determine if the claim is independent or dependent by checking the corresponding Claims.
  * Dependent claims begin with expressions like "The crossover of claim 7...", "The method of claim 13...", etc.
- If the claim is dependent, return the number of the referenced claim.
- If the claim is independent, return -1.
- Note: Claim 1 is always an independent claim.

For isReject:
- Look for sections titled "Claim Rejections" or similar in CTNF.
- Return true if the claim is rejected under any U.S.C. section.
- Return false if:
  * The claim is only objected to.
  * The claim is indicated as allowable.
  * The claim is not mentioned in the CTNF document.
  * The claim is allowed but depends on a rejected claim (e.g., "Claims 4 and 17-18 are objected to as being dependent upon a rejected base claim, but would be allowable if rewritten in independent form including all of the limitations of the base claim and any intervening claims.").

For reasons:
- Find all U.S.C. section codes and corresponding reasons under which the claim is rejected.
- For each section code, add an object to the reasons array containing sectionCode, citedPatents, and reason.
- If the claim is not rejected, return an empty array [].
- In most cases, there will be one reject reason for one section code, but if there are multiple rejections for the same claim with different cited patents, there may be multiple json children with the same section code.

Subfields of reasons:
For sectionCode:
- Return the numerical section code under which the claim was rejected.
- Common formats include "35 U.S.C. § 102a1", "35 U.S.C. § 103", "35 U.S.C. § 101", "35 U.S.C. § 112".
- Extract only the numerical parts (e.g., 102, 103).

For citedPatents:
- Return array of all relevant patent citations for each claim.
- Identify citations in formats such as:
  * US patent applications: "US 20150048242", "US 2015/0048242".
  * US patents: "US 9495285", "9,495,285".
  * Foreign patents: "EP 1234567 ", "JP 2015-123456".
- Citation locations:
  * Usually found in the rejection heading (e.g., "rejected under 35 U.S.C. 103 as being unpatentable over Remillard et al (US 20150048242)").
  * May appear in combination formats (e.g., "Remillard et al (US 20150048242) in view of HSU et al (US 9495285)").
  * May be referenced later by author name only (e.g., "Remillard et al").
- For different rejection types:
  * 102 rejections: Include the single cited reference.
  * 103 rejections:
    - Include all references.
    - Maintain citation order (primary reference first).
    - Include references after "in view of ", "and further in view of", etc.
- Special cases:
  * If only the author name is mentioned, look for the full citation earlier in the document.
  * Standardize formatting variations to a simplified format.
  * If no patent/publication number is found but the reference is clearly cited, omit that citation.
- Return as array of strings in standardized format:
  * Include the country code (e.g., "JP 2015-123456").
  * US patent applications: "US" + space + 11-digit number (e.g., "US 20150048242")
  * US patents: "US" + space + 7 or 8-digit number (e.g., "US 9495285")
  * Return an empty array [] if no citations are found.

For reason:
- Extract detailed technical reasoning from the rejection explanation.
- IMPORTANT: Do not summarize. Keep the rationale sentences from the original document intact.
- The reason must start with "Regarding Claim #" followed by the single corresponding claimNumber.
- Include specific elements and their relationships mentioned in the rejection.
- Do not include the phrase "claim _ rejected under 35 U.S.C § _" in the reason field.
- Even if a reason is written for multiple claims in the original document, the reason should only pertain to the corresponding claimNumber.
- Add references to specific cited patents by replacing author citations with full patent numbers:
  * Original text: "Walberg et al. teaches an electrosurgical device..."
  * Should become: "Walberg et al. (US 20150151601) teaches an electrosurgical device..."
  * Always include the full patent number in parentheses after the author name
  * Do this for all author citations in the reason text
- Look for patterns such as:
  * Component definitions with reference numbers (e.g., "first electrode (16)").
  * Paragraph references (e.g., "paragraph 0013").
  * Figure references (e.g., "Figure 11B").
  * Technical relationships between components.
  * Material specifications.
  * Functional descriptions.
- When multiple components are described:
  * Include their structural relationships (e.g., "disposed between ", "disposed on").
  * Include their functional relationships.
  * Include reference numbers and paragraph citations.
- For claim dependencies:
  * Include which parent claim is being referenced.
  * Include any specific limitations being added.
- Return the reason as a string, maintaining technical detail while being concise.
- For claims in a range, also indicate the range reference.

3. Example Scenarios:
Claim Range with Multiple Rejections:
CTNF Text:
"Claims 1-5 are rejected under 35 U.S.C. 112... Claims 1-5 are rejected under 35 U.S.C. 103..."

Example JSON for Claim 1:
{
    "claimNumber": 1,
    "parentClaim": -1,
    "isReject": true,
    "reasons": [
      {
        "sectionCode": 112,
        "citedPatents": [],
        "reason": "Regarding Claim 1, the phrase or other laydown area required for transfer cask operations renders the claim(s) indefinite because the claim(s) include(s) elements not actually disclosed (those encompassed by or other laydown area), thereby rendering the scope of the claim(s) unascertainable."
      },
      {
        "sectionCode": 103,
        "citedPatents": ["US 20150048242", "US 9495285"],
        "reason": "Regarding Claim 1, US 20150048242 discloses a method, the method comprising: loading a container of spent fuel (waste canister) into a cavity of a transfer cask (transporter cask) (pg. 5.57, paras [0001]-[0002]); placing a shielding sleeve around the transfer cask (pg. 5.57, para [0002]); simultaneously lifting the transfer cask and the shielding sleeve over a storage cask (dry well) (pg. 5.57, paras [0001]-[0002], [0004]); and transferring the container of spent fuel from the transfer cask to the storage cask (pg. 5.57, para [0004]-pg. 5.58, para [0000]). US 20150048242 does not specifically disclose the method is for transferring spent fuel from wet storage to dry storage; however, US 9495285 discloses transferring spent fuel from wet storage to dry storage (Abstract, para [0022]). It would have been obvious to a person skilled in the art to modify the method of Rasmussen in accordance with the teachings of 340 such that the method is for transferring spent fuel from wet storage to dry storage, since it would allow the fuel to be placed in long term or off site storage (see 340 para [0002])."
      }
    ]
}

Allowed Claims Range:
CTNF Text:
"Claims 1-19 are allowed..."
Example JSON for Claim 6:
{
    "claimNumber": 6,
    "parentClaim": 1,
    "isReject": false,
    "reasons": []
}

4. Special Instructions:
- For claims not mentioned in the CTNF document, use the following defaults:
  * isReject: false
  * reasons: []
- Maintain claim number order in the output array.
- If rejected for multiple reasons, they are arranged in the following order: 101, 112, 102, 103, etc.
- Be precise in extracting section codes and patent numbers.
- Return ONLY the JSON structure with NO additional text.

Now, analyze the provided CTNF document and extract the requested information into ONLY the JSON structure. Return nothing but the JSON.
\end{lstlisting}

We performed additional parsing to extract specific paragraphs or particular drawing elements of the cited patent mentioned in the Non-Final Rejection document. (Although our experiment carried out parsing twice, the procedure could be consolidated into a single parsing step.) The prompt is as follows:

\begin{lstlisting}[extendedchars=true, literate={§}{{\S}}1]
IMPORTANT: You must ONLY return the requested JSON structure. Do not include any additional text, explanations, or comments before or after the JSON.

Task:
Your job is to read the original CTNF document and enrich the citedPatents part of the CTNF data parsed into JSON. (If the original CTNF document is not provided, refer to the reason text to perform this task.)

1. For every claim in the JSON, check if `isReject` is `true` and any `reason` has a `sectionCode` of `102` or `103`.

2. For those claims, look for paragraph references in the original CTNF document of the form `[NNNN]` (4 digits), including ranges such as `[000N]-[00NN]`. 
   - Convert each reference to an integer and, if a range is found, expand it. For example, `[0011]-[0016]` -> `[11, 12, 13, 14, 15, 16]`.

3. For those same claims, also look for figure references in the original CTNF document of the form `(fig.XYZ)`, `(Fig.12)`, etc.
   - Normalize them to strings without "fig" or "Fig.". For example, `(fig.3A)` -> `"3A"`, `(Fig.12)` -> `"12"`.
   - If the same figure reference appears multiple times for the same patent, list it only once (no duplicates).

IMPORTANT: Only include paragraph numbers in 'text' and 'img' if they specifically apply to the same patent number mentioned in the reason text. If paragraph references are associated with a different patent, do not include them. If the same paragraph number or range is mentioned multiple times for the same patent, list it only once (no duplicates).

4. For each item in the `"citedPatents"` array (currently a list of strings like `["US 20070159740"]`), transform it into an array of objects with the following structure:
   ```json
   [
     {
       "patentNum": "<the existing patent number string>",
       "text": [<unique paragraph numbers>],
       "img": [<unique figure labels>]
     }
   ]

Apply these "text" and "img" references to every cited patent in the same reason if the references specifically belong to that patent.

5. If no paragraph or figure references are found for a particular patent within a reason, the "text" or "img" field should be an empty array ([]).

6. Return ONLY the final transformed JSON, with no extra commentary. Keep the existing JSON structure (claims, reasons, etc.) and replace "citedPatents": ["..."] with "citedPatents": [{"patentNum": "...", "text": [...], "img": [...]}] as specified.
\end{lstlisting}

\subsection{Analysis of Post Data Validation}\label{validation_procedure}

\begin{table}[h]
\centering
\renewcommand{\arraystretch}{1.2}
\begin{tabular}{llc}
\toprule
\textbf{Details} & \textbf{Count} \\ 
\midrule
Invalid format parsing by GPT & 1,763 \\
Patents only in parsed CTNF & 1,462 \\ 
CTNF-record Claims count mismatch & 1,055 \\ 
Multiple CTNF files found & 131 \\ 
\bottomrule
\end{tabular}
\vspace{0.2cm}
\caption{Detailed summary of validation errors (duplicate instances may occur)}
\label{tab:validation_error_analysis}
\end{table}







After initial data collection and parsing, a rigorous validation process was applied to ensure dataset accuracy and quality. This validation consisted of the following steps:

\begin{itemize} \item \textbf{Claims Consistency Check:} Confirms that the total number of claims in each record and its corresponding Non-Final Rejection file are identical.
\item \textbf{Cited Patents Consistency Check:} Checks whether the set of cited patents in the Non-Final Rejection matches those listed in the collected records. Suppose a patent document does not exist in the USPTO database (commonly because it originates from other jurisdictions such as Europe, Japan, or Korea). In that case, it may appear as missing or unavailable.
\item \textbf{Delete duplicated parsed Non-Final Rejection files:} 
When multiple parsed Non-Final Rejection files possess an identical application number, the file with the most substantial size is retained, while all others are eliminated.
\end{itemize}

Table~\ref{tab:validation_error_analysis} summarizes the statistics for validation-related failures. And, the Table~\ref{tab:data_retention} summarizes the initial data collection attempts to the final data collected.

\begin{table}[h]
\centering
\renewcommand{\arraystretch}{1.4}
\begin{tabular}{lccc}
\hline
Data Processing Stage & Records Attempted & Records Retained & Retention Rate \\ 
\hline
Data Collection and Initial Filtering & 206,767 & 12,839 & 6.21\% \\ 
Post Data Validation & 12,839 & 8,143 & 63.42\% \\ 
\hline
\textbf{Final Dataset} & \textbf{206,767} & \textbf{8,143} & \textbf{3.94\%} \\ 
\hline
\end{tabular}
\vspace{0.2cm}
\caption{Data retention through curation steps}
\label{tab:data_retention}
\end{table}

\subsection{Details of Expert Validation}

\subsubsection{Validation System and Procedure}\label{expert_evaluation_system}
We designed a structured expert evaluation procedure to validate the performance and accuracy of our GPT-4o-based parsing system for extracting claim rejection details from USPTO Non-Final Rejection documents. Experienced USPTO patent researchers and agents were recruited through Upwork to conduct this evaluation using a customized web-based platform explicitly developed for this study.

The evaluation involved 100 randomly selected patent applications from the PANORAMA dataset, evenly distributed across five technical domains: Circuit-Signal, Device-Hardware, IT-Data Processing, Manufacturing-Mechanics, and Chemistry-Bio. We selected 20 patent applications from each domain containing 10–26 claims each, which reflects a range within one standard deviation of the dataset's average claim count. These applications were further grouped into evaluation bundles of five applications each, resulting in four bundles per domain and 20 evaluation bundles across all domains.

\begin{figure}[htbp]
    \includegraphics[width=1\linewidth]{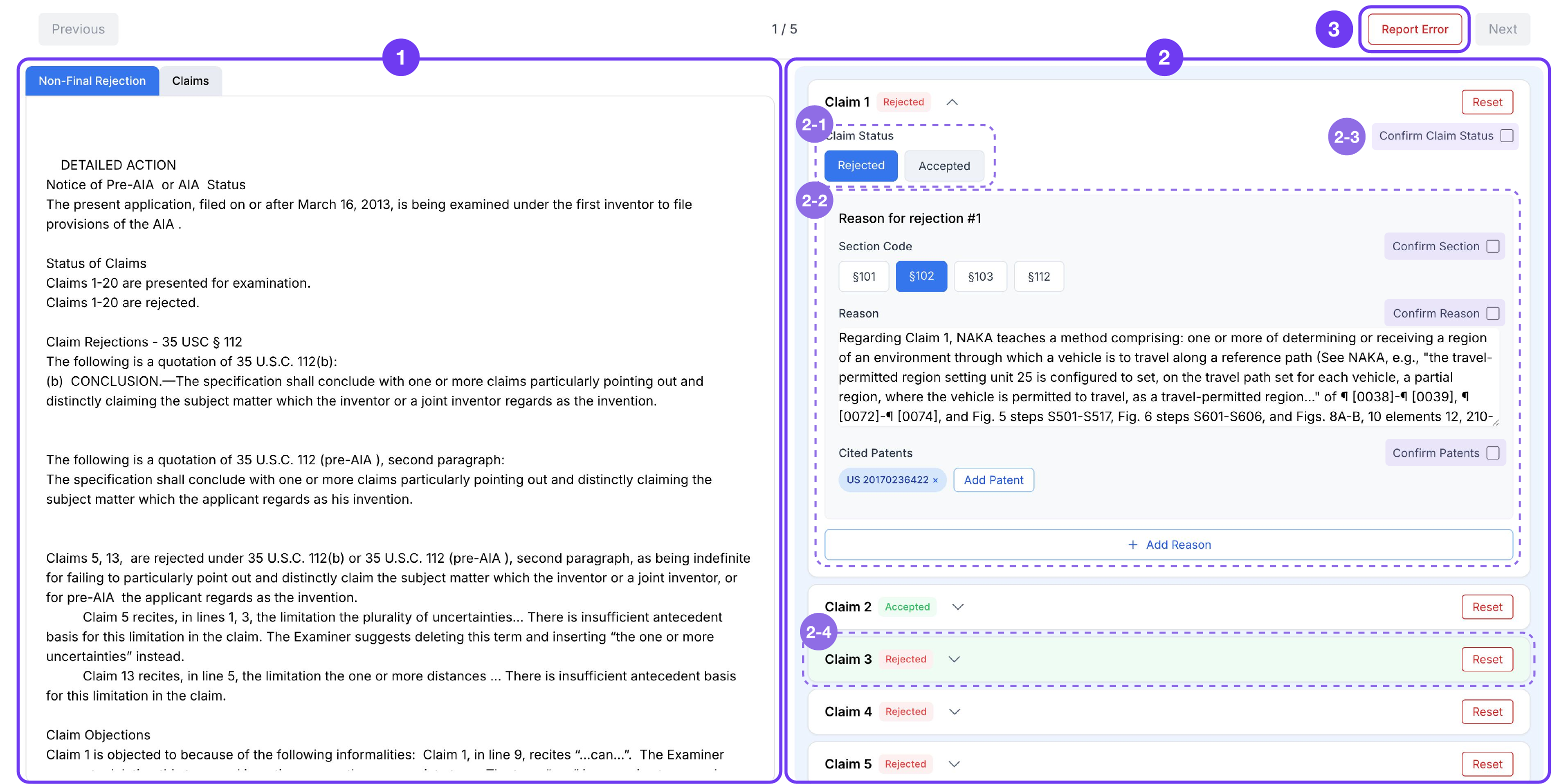}
    \caption{Bundle Evaluation Interface. The left panel (B-1) displays the original Non-Final Rejection and Claims documents in a tabbed view, with the default tab showing the rejection document. The right panel (B-2) shows the GPT-parsed claim information, including rejection status, legal codes, cited patents, and editable fields.}
    \label{fig:validation_UI}
\end{figure}

Upon selecting an evaluation bundle aligned with their expertise, experts accessed the detailed evaluation interface shown in Figure~\ref{fig:validation_UI}. This interface was specifically designed to facilitate systematic and accurate validation of GPT-4o's parsed claim rejection data against original USPTO Non-Final Rejection documents.

The evaluation interface consists of two main panels:

\textbf{Original Documents Panel (1).}
Panel (1) provides experts with the original Non-Final Rejection documents authored by USPTO patent examiners. It is implemented as a tabbed view containing two separate tabs: the default tab, "Non-Final Rejection," which shows the complete CTNF text, and the "Claims" tab, listing all the claims from the patent application under evaluation. Experts primarily reference the "Non-Final Rejection" tab content to verify the accuracy of parsed claims.

\textbf{Parsed Claims Evaluation Panel (2).}
Panel (2) presents structured parsing results generated by the GPT-4o model. Parsed claims are displayed in an accordion-style list format, where each claim can be individually expanded for detailed review. Experts evaluate each claim systematically by comparing the parsed output against the original rejection document in Panel (1).

Within this panel, the following key components are highlighted:

\textbf{Claim Status Buttons (2-1).}
These buttons indicate the parsed rejection or acceptance status of each claim. Selecting `Rejected' reveals further detailed information about the rejection reasons.

\textbf{Rejection Reasons Section (2-2).}
When a claim is marked as `Rejected,' the interface displays specific legal section codes (e.g., §101, §102, §103, §112) and corresponding detailed textual explanations parsed by GPT-4o. A single claim may contain multiple rejection reasons; thus, this section supports displaying multiple reasons clearly. If experts identify a missing reason, they can manually add it using the "Add Reason" button. Additionally, cited patents related to the rejection reasons can be reviewed and modified directly through the interface.

\textbf{Confirmation Checkboxes (2-3).}
Experts are required to carefully verify each parsed element—claim status, rejection reasons, section codes, and cited patents. If no discrepancies are found or after making necessary corrections, experts confirm their review by checking these confirmation boxes. Once all checkboxes for a claim are confirmed, the claim’s accordion header changes its background color to green (2-4), visually indicating that the claim’s evaluation is complete.

\textbf{Report Error Button (3).}
If experts encounter issues that cannot be corrected directly within the provided editing interface, such as severe parsing errors or critical system issues, they utilize the "Report Error" button positioned at the top-right corner of the interface. Upon clicking this button, experts specify the error type and provide detailed descriptions, allowing the research team to separately review and address these issues.

Upon completion of the review for all claims within a document, experts proceed to the next document in the bundle. After evaluating all five documents in the bundle, a final submission step records the experts' corrections and confirmations. All evaluated data, including corrections and confirmations, are systematically saved to the server in JSON format and subsequently analyzed to assess discrepancies with GPT-4o’s original parsing output.

\subsubsection{Result of Expert Validation}\label{expert_validation_result}

\begin{table}[ht]
\centering
\renewcommand{\arraystretch}{1.2}
\begin{tabular}{lcccc}
\toprule
 & \textbf{Fleiss' $\kappa$} & \textbf{3 agree (\%)} & \textbf{2 agree (\%)} & \textbf{All different (\%)} \\
\midrule
isReject               & 0.787 & 91.14 & 8.86 & 0.00 \\
sectionCode            & 0.795 & 75.35 & 23.11 & 1.55 \\
citedPatents           & 0.771 & 70.38 & 22.68 & 6.94 \\
reasons\_length        & 0.740 & 79.56 & 19.80 & 0.64 \\
Composite (all fields) & 0.751 & 67.45 & 25.45 & 7.10 \\
\bottomrule
\end{tabular}
\vspace{0.2cm}
\caption{Consensus statistics across three experts}
\label{expert_consensus}
\renewcommand{\arraystretch}{1.0}
\end{table}
We validated the results of parsing 100 patent documents from seven experts. We compared the LLM-parsed Non-Final Rejection documents to those corrected by the experts, analyzing the differences across four fields: isReject, sectionCode, citedPatent, and reason. 
We first assessed the agreement among three experts correcting a single Non-Final Rejection (Table \ref{expert_consensus}). For all fields, Fleiss' kappa values exceeded 0.7 for all fields, indicating substantial agreement. However, the overall agreement was 67.45\%, reflecting discrepancies in some fields. These differences largely stem from the inherent ambiguity of Non-Final Rejection documents, which often lack explicitly filled fields. Notably, it was rare for all three experts to disagree simultaneously, and there were no cases of disagreement in the isReject field.
\begin{table}[ht]
\centering
\renewcommand{\arraystretch}{1.2}
\begin{tabular}{lccc}
\toprule
 & \textbf{Cohen's $\kappa$} & \textbf{Exact match (\%)} & \textbf{Unmatched \#}  \\
\midrule
isReject               & 0.981 & 99.5 & 10 \\
sectionCode            & 0.910 & 92.8 & 135 \\
citedPatents           & 0.987 & 98.8 & 23 \\
reasons\_length        & 0.982 & 99.1 & 17 \\
Composite (all fields) & 0.922 & 92.5 & 141 \\
\bottomrule
\end{tabular}
\vspace{0.2cm}
\caption{Similarity between model output and expert validation result.}
\label{validation_result}
\renewcommand{\arraystretch}{1.0}
\end{table}

Given the differences among the experts, we compared the majority consensus to the LLM-parsed data (Table\ref{validation_result}). We found a very high agreement rate, with a Cohen's Kappa of over 0.9, indicating that most of our data was parsed accurately. However, some exceptions were identified, notably 135 cases where the sectionCode was incorrect.

\section{Details of Benchmark Tasks}

\subsection{Data Availability}\label{data_availability_bench}
All benchmarks are publicly available:
\begin{itemize}
\item \textbf{PAR4PC}: 
\url{https://huggingface.co/datasets/DxD-Lab/PANORAMA-PAR4PC-Bench}
\item \textbf{PI4PC}: 
\url{https://huggingface.co/datasets/DxD-Lab/PANORAMA-PI4PC-Bench}
\item \textbf{NOC4PC}: 
\url{https://huggingface.co/datasets/DxD-Lab/PANORAMA-NOC4PC-Bench}
\end{itemize}

\subsection{Prior Art Retrieval for Patent Claims (PAR4PC)}

\subsubsection{Task Description}

Table \ref{tab:par4pc_summary} provides a statistical overview of the PAR4PC benchmark dataset, showing token length distributions across training, validation, and test splits using the o200k\_base tokenizer. The dataset consists of 59,953 total samples with consistently long sequences averaging over 16,000 tokens, positioning it as a demanding benchmark for evaluating long-context understanding capabilities.

\begin{table}[htbp]
\centering
\begin{tabular}{llccccc}
\toprule
\textbf{Dataset} & \textbf{Mode} & \textbf{Samples} & \textbf{Mean $\pm$ Std Dev} \\
\midrule
\multirow{2}{*}{Test} 
& Zero-Shot & \multirow{2}{*}{2,896} & $16,060.44 \pm 3,140.35$  \\
& CoT & & $16,111.44 \pm 3,140.35$  \\
\midrule
\multirow{2}{*}{Validation} 
& Zero-Shot & \multirow{2}{*}{3,029} & $16,267.18 \pm 3,170.92$ \\
& CoT & & $16,318.18 \pm 3,170.92$ \\
\midrule
\multirow{2}{*}{Train} 
& Zero-Shot & \multirow{2}{*}{54,028} & $16,027.33 \pm 3,696.42$ \\
& CoT & & $16,078.33 \pm 3,696.42$ \\
\bottomrule
\end{tabular}
\vspace{0.2cm}
\caption{Token Length Statistics for PAR4PC Benchmark Dataset using o200k\_base Tokenizer}\label{tab:par4pc_summary}
\end{table}

The Task consists of individual JSON files, each representing a single question designed to evaluate the model's ability to identify relevant cited patents. Each JSON file contains structured information about the patent application under examination (the `context') and several candidate patents presented as options (A-H). The structure includes the application's title, abstract, and claims, as well as similar details for each candidate patent option. Additionally, it provides the ground truth labels (`gold', `silver', `negative') indicating the relevance of each option as a cited patent for a specific claim of the application under examination. Table \ref{tab:par4pc_input} details the key fields and their data types within a typical Task sample file.

\begin{table}[htbp]
\renewcommand{\arraystretch}{1.3}
\footnotesize
\begin{tabularx}{\linewidth}{c|c| >{\RaggedRight}X | >{\RaggedRight}X}
\hline
\multicolumn{2}{c|}{} & Description & Example \\
\hline
\multicolumn{2}{c|}{\makecell{Question}} & Natural language question asking which patents from the given options were cited against the specified claim & Question: Based only on the provided context and options, which patent(s) (A-H) were cited against claim X? \\
\hline
\multirow{8}{*}{Patent Application}
& title & Title of the invention in the application under examination & Cylinder Liner, Block Manufacturing Method... \\ \cline{2-4}
& abstract & Abstract of the invention in the application under examination & A cylinder liner that is casted in a block... includes: a cylindrical liner body; a projection part...; and a bore adjacent part... \\ \cline{2-4}
& claim & List of all claims in the patent application under examination & ["1. A cylinder liner that is casted in a block...", "2. The cylinder liner according to claim 1...", ...] \\
\hline
\multirow{8}{*}{Prior Art}
& key & Represents an individual option ('A', 'B', etc.) & A \\ \cline{2-4}
& title & Title of the invention of the option patent & ENCRYPTION AND AUTHENTICATION OF DATA... \\ \cline{2-4}
& abstract & Abstract of the option patent & Techniques for encryption and authentication of data...  \\ \cline{2-4}
& claim & List of claims of the option patent & ["1. Method for encryption and authentication...", "2. A method according to claim 1...", ...] \\
\hline
\multicolumn{2}{c|}{\makecell{Gold Answer}} & List of 'Gold' (correct) cited patent option keys & ["E", "F"] \\ \cline{1-4}
\multicolumn{2}{c|}{\makecell{Silver Answer}} & List of 'Silver' (partially correct) cited patent option keys & ["B"] \\ \cline{1-4}
\multicolumn{2}{c|}{\makecell{Negative Answer}} & List of 'Negative' (incorrect) cited patent option keys & ["A", "C", "D", "G", "H"] \\
\hline
\end{tabularx}
\vspace{0.2cm}
\caption{Input Structure for the PAR4PC Task} 
\label{tab:par4pc_input} 
\renewcommand{\arraystretch}{1.0}
\end{table}

\subsubsection{Details of Task Construction}\label{PAR4PC_construction}

We selected patent application records that passed validity verification. Each application record has a unique identifier (rec\_num) and application number (app\_num), which were used to connect related files. To ensure data quality, we applied two primary filtering criteria. First, we selected only applications where the number of patents cited by examiners ranged from 1 to 5. This filtering criterion was implemented to maintain data consistency by excluding overly complex cases with numerous citations or cases without any citations, which would not provide meaningful evaluation scenarios. Second, we restricted our task to applications containing only rejections under 35 U.S.C. \S 102 or \S 103. This focus on core patent law provisions helped eliminate complexity arising from other technical rejection reasons and ensured that the task represented the most common and critical examination scenarios.

For each application that passed the filtering process, we identified claims rejected under sections \S 102 or \S 103. For each claim, we classified the patents cited by examiners into three categories: (1) Gold citations: core citations directly used in the rejection of the specific claim, (2) Silver citations: citations used in the rejection of other claims within the same application, and (3) Negative citations: citations not used in the examination of the application. Negative citations were extracted from other applications with the same patent classification (class) and similar filing dates, excluding those already classified as Gold or Silver.

Finally, we generated a task instance for each claim. Each instance includes the application number, claim number, application context (title, abstract, claim text), eight options, and answer classifications (Gold, Silver, Negative). This structure allows for evaluating how accurately models can predict an examiner's citation decisions during the patent examination process.

\subsubsection{Data Processing and Evaluation}

The task set was split into training, validation, and test sets to prevent data leakage by ensuring patent applications did not appear across multiple splits. Specifically, patent applications were grouped based on their unique identifiers, and these groups were randomly shuffled with a fixed random seed for reproducibility. Following predefined ratios (90\% train, 5\% validation, 5\% test), each group was allocated to one of the three splits. Aggregated datasets for each split were saved in JSONL and Parquet formats.

LLM predictions were obtained using zero-shot or Chain-of-Thought (CoT) prompting strategies. Predictions were compared against gold standard labels using Exact Match Accuracy and a Custom Score calculated as follows:

\begin{equation}
\text{Avg Score \%} = \left( \frac{\sum_{i=1}^{N} \max(0, \text{raw\_score}_i)}{\sum_{i=1}^{N} (2 \times |G_i|)} \right) \times 100
\label{eq:final_avg_custom_score}
\end{equation}
where:
\begin{itemize}
    \item \( \text{Avg Score \%} \) is the Average Custom Score Percentage over the dataset.
    \item \( N \) is the total number of valid questions evaluated.
    \item \( i \) is the index for an individual question.
    \item \( \text{raw\_score}_i \) is the initial score calculated for question \( i \) before clipping, defined as: \\
          \( \text{raw\_score}_i = (2 \times |P_i \cap G_i|) - (1 \times |P_i \setminus (G_i \cup S_i)|) - (1 \times |G_i \setminus P_i|) \).
    \item \( \max(0, \text{raw\_score}_i) \) represents the score for question \( i \) after clipping any negative value to zero.
    \item \( P_i \) is the set of answer letters predicted by the model for question \( i \).
    \item \( G_i \) is the set of correct `Gold' answer letters for question \( i \).
    \item \( S_i \) is the set of `Silver' answer letters for question \( i \).
    \item \( |G_i| \) is the number of Gold answers for question \( i \).
    \item \( (2 \times |G_i|) \) represents the maximum possible score for question \( i \), used in the denominator sum.
\end{itemize}

\subsubsection{Test Prompts}\label{PAR4PC_prompt}

The evaluation utilized dynamically generated prompts based on the data and the selected prompting strategy (\texttt{zero-shot} or \texttt{cot}). Each prompt includes a common context section, followed by mode-specific instructions populated with actual data during runtime.

The common introductory part of the prompt, shared by both modes, is presented below:

\begin{lstlisting}
You are a patent expert tasked with identifying cited patents for a specific claim rejection based *only* on the provided context and options.

**Context:**
*   **Application Number:** {app_number}
*   **Title:** {context.get("title", "N/A")}
*   **Abstract:** {context.get("abstract", "N/A")}
*   **Initial Claims:**
    {context_claims_json}

**Target Claim for Analysis:** Claim {claim_number}

**Options (Potential Cited Patents):**
{options_formatted_string}
% The {options_formatted_string} placeholder is expanded as follows for each option (A-H):
% A: Patent ID: {patent_id}
%    Title: {title}
%    Abstract: {abstract}
%    Claims: {claims_str}
% ... (Repeated for B through H) ...
\end{lstlisting}

Following the common section, mode-specific instructions are appended.

\paragraph{Zero-Shot Prompt:} This prompt directly asks the LLM to identify the cited patent(s) based on the provided information and specifies the required JSON output format.

\begin{lstlisting}
Based only on the provided context and options, which patent(s) (A-H) were cited?
Answer format (JSON only):
```json
{{"answer": "A"}}
```
If multiple patents are cited
```json
{{"answer": ["A","C","F"]}}
```
\end{lstlisting}

\paragraph{Chain-of-Thought (CoT) Prompt:} This prompt guides the LLM through a structured reasoning process before providing the final answer. It explicitly asks the model to identify the claim focus, match it against the prior art options, and then select the cited patent(s), returning both the reasoning and the answer in the specified JSON format.

\begin{lstlisting}
Based only on the provided context and options, which patent(s) (A-H) were cited?
Think through the steps required to evaluate this, craft the supporting rationale accordingly, and then deliver your answer based on that rationale.
Always write the "reason" **first** and then write the "answer".

Answer format (JSON only):
```json
{{"reason":"", "answer": "A"}}
```
If multiple patents are cited
```json
{{"reason":"", "answer": ["A","C","F"]}}
```
\end{lstlisting}

\subsubsection{Detailed LLM Performance Analysis}\label{task1_result_detail}
Table \ref{tab:par4pc_performance} presents model performance on the PAR4PC dataset, showing both custom score and exact match accuracy metrics with a breakdown by rejection types (§102 only and §103 only). The evaluation excludes the 28 cases with both §102 and §103 rejections, allowing for cleaner analysis of how models perform on distinct patent rejection categories. 

\begin{table}[htbp]
\centering
\resizebox{\textwidth}{!}{%
\begin{tabular}{llrrrrrr}
\toprule
& & \multicolumn{3}{c}{\textbf{Custom Score}} & \multicolumn{3}{c}{\textbf{(Exact Match) Accuracy}} \\
\cmidrule(lr){3-5} \cmidrule(lr){6-8}
\textbf{Model} & \textbf{Mode} & \textbf{All} & \textbf{§102 only} & \textbf{§103 only} & \textbf{All} & \textbf{§102 only} & \textbf{§103 only} \\
\midrule
baseline &  & 5.63 & 3.76 & 6.94 & 0.54 & 1.45 & 0.74 \\
\midrule
\multirow{2}{*}{GPT-4o} & ZS & 47.34 & 82.41 & 33.52 & \textbf{48.69} & \textbf{79.8} & 26.63 \\
                        & CoT & 56.95 & \textbf{86.37} & 45.11 & \textbf{51.04} & 73.6 & \textbf{34.81} \\
\midrule
\multirow{2}{*}{Claude-3.7-Sonnet} & ZS & 40.12 & 75.33 & 26.46 & 45.48 & 75.94 & 23.94 \\
                       & CoT & 40.29 & 75.21 & 26.75 & 46.31 & 75.86 & 25.43 \\
\midrule
\multirow{2}{*}{Gemini-2.0-Flash} & ZS & 37.56 & 65.61 & 26.51 & 38.88 & 62.28 & 22.21 \\
                       & CoT & 43.61 & 58.11 & 33.14 & 34.50 & 51.89 & 21.96 \\
\midrule
\multirow{2}{*}{\makecell{Llama-3.1\\-8B-Instruct}} & ZS & 13.45 & 16.09 & 11.38 & 0.00 & 0.00 & 0.00 \\
                                       & CoT & 37.99 & 47.15 & 31.37 & 0.00 & 0.00 & 0.00 \\
\midrule
\multirow{2}{*}{\makecell{Qwen2.5-7B\\-Instruct}} & ZS & 66.11 & 64.25 & \textbf{66.93} & 33.05 & 56.83 & 16.06 \\
                                          & CoT & 67.42 & 66.14 & 67.97 & 34.43 & 58.34 & 17.37 \\
\midrule
\multirow{2}{*}{\makecell{EXAONE-3.5\\-7.8B-Instruct}} & ZS & 0.00 & 0.00 & 0.00 & 5.42 & 7.63 & 3.70 \\
                                          & CoT & 22.52 & 29.46 & 17.50 & 0.00 & 0.00 & 0.00 \\
\midrule
\multirow{2}{*}{Gemma-3-12B-Instruct} & ZS & 56.47 & 54.61 & 57.37 & 29.49 & 48.62 & 15.76 \\
                         & CoT & \textbf{77.30} & 75.27 & \textbf{78.42} & 30.73 & 44.43 & 20.54 \\
\midrule
\multirow{2}{*}{\makecell{Qwen2.5\\-32B-Instruct}} & ZS & \textbf{68.94} & \textbf{85.08} & 57.48 & 47.2 & 58.93 & \textbf{38.75} \\
                                          & CoT & 55.05 & 78.44 & 38.54 & 46.41 & \textbf{75.94} & 25.49 \\
\midrule
\multirow{2}{*}{\makecell{EXAONE-3.5\\-32B-Instruct}} & ZS & 51.46 & 61.3 & 44.26 & 31.66 & 39.56 & 26.09 \\
                                          & CoT & 44.93 & 66.47 & 29.61 & 36.74 & 62.20 & 18.57 \\
\midrule
\multirow{2}{*}{Gemma-3-27B-Instruct} & ZS & 50.19 & 71.02 & 35.48 & 42.30 & 70.33 & 22.45 \\
                         & CoT & 55.36 & 75.44 & 41.21 & 44.85 & 70.49 & 26.69 \\
\midrule
QWQ-32B & CoT & 59.03 & 81.80 & 42.93 & 48.33 & 75.10 & 29.36 \\
\midrule
EXAONE-Deep-32B & CoT & 42.59 & 62.35 & 28.52 & 36.86 & 61.15 & 19.58 \\
\bottomrule
\end{tabular}
}
\vspace{0.2cm}
\caption{Model Performance Summary with Rejection Type Breakdown (Count: §102=1193, §103=1675). Cases with both §102 and §103 rejections (28 cases) were excluded from scoring.}
\label{tab:par4pc_performance}
\end{table}

Table~\ref{tab:par4pc_additional_eval} presents a detailed analysis of performance differences across technology domains. Each model exhibited strengths in different areas; for instance, GPT demonstrated higher performance in the Physics domain, whereas Qwen achieved better results in the Electricity domain.

\begin{table*}[t]
\centering
\resizebox{\textwidth}{!}{
\begin{tabular}{l l c c c c c c c c c c}
\toprule
\textbf{Model} & \textbf{Prompt} & \textbf{A} & \textbf{B} & \textbf{C} & \textbf{D} & \textbf{E} & \textbf{F} & \textbf{G} & \textbf{H} & \textbf{Y} & \textbf{Total} \\
\midrule
\textbf{\# samples} &  & 228 & 242 & 35 & 6 & 75 & 111 & 963 & 1,200 & 16 & 2,876 \\
\midrule
Baseline & --- & 3.02 & 2.15 & 7.33 & 0.91 & 2.48 & 2.79 & 2.20 & 2.89 & 3.33 & 5.63 \\
GPT-4o & ZS & 42.66 & 58.76 & 36.00 & 31.82 & 36.75 & 39.70 & 63.78 & 38.67 & 5.55 & 47.34 \\
GPT-4o & CoT & 59.70 & 72.05 & 41.33 & 31.82 & 55.98 & 52.42 & 68.86 & 48.12 & 52.78 & 56.95 \\
Claude-3.7-Sonnet & ZS & 36.70 & 51.06 & 27.33 & 31.82 & 31.20 & 36.67 & 54.25 & 32.00 & 8.33 & 40.12 \\
Claude-3.7-Sonnet & CoT & 36.70 & 52.57 & 26.67 & 31.82 & 31.20 & 38.48 & 53.65 & 32.34 & 8.33 & 40.29 \\
Qwen 2.5-7B-Instruct & ZS & 66.23 & 69.83 & 74.29 & \textbf{100.00} & 80.66 & 79.73 & 60.18 & 68.46 & 71.88 & 66.11 \\
Qwen 2.5-7B-Instruct & CoT & 73.25 & 72.93 & \textbf{85.71} & 33.33 & 80.00 & 68.02 & 65.58 & 66.38 & 71.88 & 67.42 \\
Gemma-3-12B-Instruct & ZS & 56.58 & 64.88 & 74.29 & 91.67 & 64.67 & 63.51 & 56.44 & 52.13 & 62.50 & 56.47 \\
Gemma-3-12B-Instruct & CoT & \textbf{85.53} & \textbf{72.93} & 80.00 & 33.33 & \textbf{83.33} & \textbf{81.08} & \textbf{77.36} & \textbf{76.75} & \textbf{90.63} & \textbf{77.30} \\
\bottomrule
\end{tabular}
}
\caption{Performance comparison across CPC sections. (
A—Human Necessities; B—Operations, Transport; C—Chemistry, Metallurgy; D—Textiles, Paper; 
E—Fixed Constructions; F—Mechanical Engineering; G—Physics; H—Electricity; Y—Other.)}
\label{tab:par4pc_additional_eval}
\end{table*}

\subsection{Paragraph Identification for Patent Claims (PI4PC) Task}

\subsubsection{Task Description}

Table \ref{tab:pi4pc_summary} presents the token length statistics for the PI4PC benchmark dataset, comprising 71,549 total samples distributed across training (64,210), validation (3,937), and test (3,402) splits. With average token lengths ranging from approximately 12,900 to 13,700 tokens and notably higher standard deviations (around 6,000 tokens) compared to PAR4PC, PI4PC exhibits greater variability in sequence lengths while maintaining substantial context requirements. 

\begin{table}[htbp]
\centering
\begin{tabular}{llccccc}
\toprule
\textbf{Dataset} & \textbf{Mode} & \textbf{Samples} & \textbf{Mean $\pm$ Std Dev} \\
\midrule
\multirow{2}{*}{Test} 
& Zero-Shot & \multirow{2}{*}{3,402} & $12,936.78 \pm 6,061.09$ \\
& CoT & & $12,964.78 \pm 6,061.09$ \\
\midrule
\multirow{2}{*}{Validation} 
& Zero-Shot & \multirow{2}{*}{3,937} & $13,658.95 \pm 6,238.58$ \\
& CoT & & $13,686.95 \pm 6,238.58$ \\
\midrule
\multirow{2}{*}{Train} 
& Zero-Shot & \multirow{2}{*}{64,210} & $13,275.66 \pm 6,040.43$ \\
& CoT & & $13,303.66 \pm 6,040.43$ \\
\bottomrule
\end{tabular}
\vspace{0.2cm}
\caption{Token Length Statistics for PI4PC Benchmark Dataset using o200k\_base Tokenizer}\label{tab:pi4pc_summary}
\end{table}

Each PI4PC Task instance is structured as a JSON file, encapsulating the target patent application's context, details of the cited prior art (including the full specification), five paragraph options from the prior art, and ground-truth labels identifying the Gold-standard paragraph. The detailed structure is provided in Table \ref{tab:pi4pc_input}.

\begin{table}[htbp]
\centering
\renewcommand{\arraystretch}{1.3}
\footnotesize
\begin{tabularx}{\linewidth}{c|c| >{\RaggedRight}X | >{\RaggedRight\arraybackslash}X}
\hline
\multicolumn{2}{c|}{} & Description & Example \\
\hline
\multicolumn{2}{c|}{\makecell{Question}} & Natural language question asking which single paragraph from the cited prior art specification are relevant for rejecting the specified claim & Question: Based on the provided context (claim X and the cited prior art specification), which paragraph is the most relevant? \\
\hline
\multirow{5}{*}{Patent Application}
& title & Title of the invention in the target application & "Biometric monitoring system" \\ \cline{2-4}
& abstract & Abstract of the invention in the target application & "ABSTRACT A system for monitoring biometric signals for..." \\ \cline{2-4}
& claims & List of claims in the target application & ["1. A biometric monitoring system, comprising...", ...] \\
\hline
\multirow{10}{*}{Prior Art} 
& <KEY> & Represents an individual paragraph option (key: number, value: text) from the specification above & 39 \\ \cline{2-4} 
& title & Title of the cited prior art document & "Physiological status..." \\ \cline{2-4}
& abstract & Abstract of the cited prior art document & "A physiological status..." \\ \cline{2-4}
& claims & Claims of the cited prior art document & ["1. A physiological...", ...] \\ \cline{2-4}
& specification & Full text specification of the cited prior art document & "BACKGROUND [0001] As..." \\ 
\hline
\multicolumn{2}{c|}{\makecell{Gold Answer}} & Single correct ('Gold') paragraph key & [39] \\ \cline{1-4}
\multicolumn{2}{c|}{\makecell{Silver Answer}} & Optional ('Silver') partially relevant paragraph key (0 or 1 element) & [10] \\ \cline{1-4}
\multicolumn{2}{c|}{\makecell{Negative Answer}} & Irrelevant ('Negative') paragraph keys (3 or 4 elements) & [20, 48, 65] \\
\hline
\end{tabularx}
\vspace{0.2cm}
\caption{Input Structure for the PI4PC Task} 
\label{tab:pi4pc_input} 
\renewcommand{\arraystretch}{1.0} 
\end{table}

\subsubsection{Details of Task Construction}\label{PI4PC_construction}

We began by identifying patent applications with clear rejection documentation, specifically focusing on Office Actions that contained paragraph-specific citations to prior art. Each application in our task includes a unique identifier and application number that allowed us to link the rejection documents with their corresponding specification files and cited prior art documents. To ensure data quality and consistency, we filtered the applications to include only those with well-structured paragraph citations and complete specification texts for both the application under examination and the cited prior art.

For each selected application, we extracted claims that were rejected under 35 U.S.C. \S 102 (novelty) or \S 103 (non-obviousness), along with the specific prior art documents cited by examiners. We then identified the exact paragraphs within these cited documents that examiners referenced in their rejections. These paragraphs were classified as "Gold" citations – the specific text portions that directly supported the rejection of the claim in question.

To create a challenging evaluation scenario, we supplemented each Gold paragraph with several distractor paragraphs from the same cited document. These distractors were carefully selected to include: (1) paragraphs adjacent to the Gold paragraph, which might contain related but less relevant information; (2) paragraphs containing similar terminology but discussing different aspects of the invention; and (3) randomly selected paragraphs from elsewhere in the document to provide clear negative examples. This approach ensures that models must demonstrate genuine understanding of both the claim's technical content and the cited document's disclosure, rather than relying on superficial keyword matching.

Each task instance consists of the target patent application's context (including title, abstract, and the specific claim text), the cited prior art document's full text divided into numbered paragraphs, and the correct answer indicating which paragraph(s) the examiner actually cited as grounds for rejection. By structuring the task in this manner, we create a realistic scenario that mirrors the precise analytical work performed by patent examiners when they must identify specific disclosures within lengthy technical documents.

\subsubsection{Data Processing and Evaluation}

Similar to PAR4PC, the PI4PC dataset was partitioned into training, validation, and test subsets at the patent application level to prevent data leakage, using a 90\% (training), 5\% (validation), and 5\% (test) split. Language model predictions were obtained using either zero-shot or Chain-of-Thought (CoT) prompting strategies.

Models were evaluated based on their ability to select the single correct Gold paragraph from the five provided options. The primary evaluation metrics were:

\begin{itemize}
    \item \textbf{Exact Match Accuracy}: Percentage of predictions exactly matching the Gold paragraph.
    
    \item \textbf{Average Custom Score Percentage}: Partial credit was granted when Silver paragraphs were correctly identified:
    \begin{itemize}
        \item 2 points for selecting a Gold paragraph.
        \item 1 point for selecting a Silver paragraph.
        \item 0 points otherwise (Negative or invalid selections).
    \end{itemize}
    The metric was computed as:
    \begin{equation}
    \text{Avg Score \%} = \left( \frac{\sum_{i=1}^{N} \text{score}_i}{2 \times N} \right) \times 100
    \label{eq:pi4pc_avg_custom_score}
    \end{equation}
    where \( \text{score}_i \) is the individual question score (0, 1, or 2), and \( N \) is the total number of evaluated questions.
\end{itemize}

\subsubsection{Test Prompts}\label{PI4PC_prompt}

Prompts were dynamically generated using a template file, incorporating details from the specific Task instance. Both zero-shot and CoT strategies shared a common introductory section.

The common section provided context about the task, the target application, the specific claim under analysis, details of the cited prior art document (including its full specification), and the list of paragraph options.

\begin{lstlisting}[extendedchars=true, literate={§}{{\S}}1 {¶}{{\P}}1]
You are an expert patent examiner reviewing a patent application.
Your task is to identify the **single most relevant paragraph** from the provided Prior Art Specification that is cited to reject Claim {claim_num} of the Target Application ({app_num}).

**Target Application Context:**
*   **Title:** {target_title}
*   **Abstract:** {target_abstract}
*   **Claim {claim_num}:** {target_claim_text}

**Prior Art Specification Context:**
*   **Patent ID:** {prior_art_patent_id}
*   **Title:** {prior_art_title}
*   **Abstract:** {prior_art_abstract}
*   **Full Specification Text:**
{prior_art_spec_text}

*   **Cited Paragraph Options (Excerpts from the Full Specification above):**
Review the following paragraphs, which are excerpts from the full specification provided above. Choose the **one paragraph key (integer)** from the list below that best supports the rejection of Claim {claim_num}.

Options:
{key_1}: {paragraph_text_1}
{key_2}: {paragraph_text_2}
{key_3}: {paragraph_text_3}
{key_4}: {paragraph_text_4}
{key_5}: {paragraph_text_5}

**CRITICAL INSTRUCTION:**  
Based on your analysis of the **Full Specification Text** and the **Target Application Claim {claim_num}**, you **MUST** select **EXACTLY ONE** integer key from the **5 options** provided above.  
Under no circumstances should you choose a key not present in the options or provide multiple keys, ranges, reasoning, or explanations.
\end{lstlisting}

Following this common block, mode-specific instructions were added.

\paragraph{Zero-Shot Prompt:} This prompt directly asks for the single most relevant paragraph key.

\begin{lstlisting}
Answer format (JSON only)
Return ONLY this JSON object - DO NOT INCLUDE ANY REASON IN THE ANSWER.
```json
{{"answer": ##}}
```
\end{lstlisting}

\paragraph{Chain-of-Thought (CoT) Prompt:} This prompt guides the model through a detailed analysis before requesting the answer, including reasoning.

\begin{lstlisting}
Think through the steps required to evaluate this, craft the supporting rationale accordingly, and then deliver your answer based on that rationale.
Always write the "reason" **first** and then write the "answer".

Answer format (JSON only):
```json
{{"reason":"...","answer": ##}}
```
\end{lstlisting}

\subsubsection{Detailed LLM Performance Analysis}\label{task2_result_detail}
Table \ref{tab:pi4pc_performance} presents a comparative analysis of Large Language Model performance on the PI4PC benchmark task across different USPTO rejection grounds. The benchmark focuses explicitly on rejections under 35 U.S.C. \S 102, rejections under 35 U.S.C. \S 103, and cases where applications received concurrent rejections under both \S 102 and \S 103. GPT-4o demonstrates superior performance on \S 102 rejections with a 64.93\% accuracy in zero-shot settings, while Gemini 2.0 Flash achieves the highest performance (51.67\%) on the challenging cases involving both \S 102 and \S 103 rejections using CoT reasoning. Notably, all models show a significant performance drop when handling applications with concurrent \S 102 and \S 103 rejections, suggesting that prior art analysis becomes substantially more complex when both novelty and non-obviousness issues must be addressed simultaneously.

\begin{table}[htbp]
\centering
\resizebox{\textwidth}{!}{%
\begin{tabular}{llrrrrrr}
\toprule
& & \multicolumn{3}{c}{\textbf{Custom Score}} & \multicolumn{3}{c}{\textbf{(Exact Match) Accuracy}} \\
\cmidrule(lr){3-5} \cmidrule(lr){6-8}
\textbf{Model} & \textbf{Mode} & \textbf{All} & \textbf{§102 only} & \textbf{§103 only} & \textbf{All} & \textbf{§102 only} & \textbf{§103 only} \\
\midrule
baseline &  & 27.10 & 28.46 & 26.49 & 19.83 & 19.84 & 19.82 \\
\midrule
\multirow{2}{*}{GPT-4o} & ZS & \textbf{63.33} & \textbf{64.93} & \textbf{62.83} & \textbf{56.06} & 56.68 & \textbf{55.96} \\
                        & CoT & \textbf{62.62} & \textbf{63.41} & \textbf{62.47} & \textbf{55.73} & \textbf{55.61} & \textbf{55.96} \\
\midrule
\multirow{2}{*}{Claude-3.7-Sonnet} & ZS & 57.33 & 60.00 & 56.20 & 51.59 & 53.72 & 50.69 \\
                       & CoT & 60.55 & 62.78 & 59.59 & 54.09 & 55.70 & 53.39 \\
\midrule
\multirow{2}{*}{Gemini-2.0-Flash} & ZS & 61.96 & 63.86 & 61.19 & 55.61 & \textbf{57.31} & 54.90 \\
                       & CoT & 61.72 & 62.83 & 61.30 & 54.67 & 55.52 & 54.32 \\
\midrule
\multirow{2}{*}{\makecell{Llama-3.1\\-8B-Instruct}} & ZS & 9.61 & 10.13 & 9.48 & 7.88 & 8.61 & 7.62 \\
                                       & CoT & 0.00 & 0.00 & 0.00 & 0.00 & 0.00 & 0.00 \\
\midrule
\multirow{2}{*}{\makecell{Qwen2.5-7B\\-Instruct}} & ZS & 29.25 & 32.51 & 27.76 & 23.96 & 26.37 & 22.86 \\
                         & CoT & 48.41 & 50.72 & 47.30 & 39.71 & 40.00 & 39.52 \\
\midrule
\multirow{2}{*}{{\makecell{EXAONE-3.5\\-7.8B-Instruct}}} & ZS & 44.55 & 46.73 & 43.64 & 35.98 & 36.05 & 36.02 \\
                                          & CoT & 41.34 & 44.71 & 39.85 & 34.16 & 35.78 & 33.45 \\
\midrule
\multirow{2}{*}{Gemma-3-12B-Instruct} & ZS & 44.34 & 45.34 & 44.09 & 36.74 & 36.59 & 37.00 \\
                                          & CoT & 31.11 & 31.75 & 31.08 & 26.16 & 26.55 & 26.33 \\
\midrule
\multirow{2}{*}{\makecell{Qwen2.5\\-32B-Instruct}} & ZS & 60.55 & 61.79 & 60.06 & 53.29 & 53.81 & 53.08 \\
                                          & CoT & 59.94 & 61.97 & 59.13 & 52.44 & 53.90 & 51.84 \\
\midrule
\multirow{2}{*}{\makecell{EXAONE-3.5\\-32B-Instruct}} & ZS & 49.40 & 50.99 & 48.74 & 41.21 & 41.17 & 41.29 \\
                                          & CoT & 51.06 & 53.36 & 50.09 & 43.09 & 43.77 & 42.84 \\
\midrule
\multirow{2}{*}{Gemma-3-27B-Instruct} & ZS & 54.66 & 56.82 & 53.79 & 46.53 & 47.35 & 46.26 \\
                                      & CoT & 56.22 & 58.43 & 55.29 & 48.94 & 50.85 & 48.12 \\
\midrule
QWQ-32B & CoT & 58.98 & 60.11 & 58.71 & 52.66 & 53.26 & 52.61 \\
\midrule
EXAONE-Deep-32B & CoT & 35.80 & 36.17 & 35.74 & 30.99 & 30.48 & 31.33 \\
\bottomrule
\end{tabular}
}
\vspace{0.2cm}
\caption{Model Performance Summary with Rejection Type Breakdown (Count: §102=1115, §103=2257). Cases with both §102 and §103 rejections(30 cases) were excluded from scoring.}
\label{tab:pi4pc_performance}
\end{table}

\subsection{Novelty and Non-obviousness Classification for Patent Claims (NOC4PC) Task}

The NOC4PC Task evaluates a language model's capability to determine whether a specific patent claim is allowable or should be rejected based on novelty and non-obviousness criteria, specifically under 35 U.S.C. §102 or §103. This task directly reflects critical decisions made by patent examiners during patent prosecution.

\subsubsection{Task Description}

Table \ref{tab:noc4pc_summary} summarizes the token length distribution for the NOC4PC benchmark dataset, the largest in our benchmark suite with 146,487 total samples (136,211 training, 7,392 validation, and 2,884 test samples). NOC4PC features substantially shorter sequences than PAR4PC and PI4PC, with mean token lengths around 6,800 for test and validation sets, and approximately 6,900 for the training set.

\begin{table}[htbp]
\centering
\begin{tabular}{llccccc}
\toprule
\textbf{Dataset} & \textbf{Mode} & \textbf{Samples} & \textbf{Mean $\pm$ Std Dev} \\
\midrule
\multirow{2}{*}{Test} 
& Zero-Shot & \multirow{2}{*}{2,884} & $6,820.89 \pm 4,598.93$ \\
& CoT & & $6,863.89 \pm 4,598.93$ \\
\midrule
\multirow{2}{*}{Validation} 
& Zero-Shot & \multirow{2}{*}{7,392} & $6,533.49 \pm 3,985.46$ \\
& CoT & & $6,575.49 \pm 3,985.46$ \\
\midrule
\multirow{2}{*}{Train} 
& Zero-Shot & \multirow{2}{*}{136,211} & $6,906.63 \pm 4,811.03$ \\
& CoT & & $6,949.63 \pm 4,811.03$ \\
\bottomrule
\end{tabular}
\vspace{0.2cm}
\caption{Token Length Statistics for NOC4PC Benchmark Dataset using o200k\_base Tokenizer}\label{tab:noc4pc_summary}
\end{table}

The NOC4PC Task consists of structured JSON files, each representing an individual question designed to assess the model’s capability to determine the rejection type and rationale for a specified claim using provided patent application and cited prior art data. Each JSON input file contains detailed context about the patent application under consideration, including the title, abstract, claims, and patent ID. Similarly, it includes comprehensive details about a cited prior art document, such as its title, abstract, claims, specific cited paragraph identifiers, and corresponding paragraph content.

Additionally, each JSON file includes an explicit `Answer Code' indicating the rejection type (e.g., \S 102 or \S 103) and an `Answer Reason,' representing the examiner's rationale behind rejecting the specified claim based on the cited prior art. Table \ref{tab:noc4pc_input} describes the key fields and provides concrete examples from a representative Task sample file.

\begin{table}[htbp]
\centering
\renewcommand{\arraystretch}{1.3}
\footnotesize
\begin{tabularx}{\linewidth}{c|c| >{\RaggedRight}X | >{\RaggedRight\arraybackslash}X}
\hline
\multicolumn{2}{c|}{} & Description & Example \\
\hline
\multicolumn{2}{c|}{\makecell{Question}} & Natural language question asking for the rejection type (code) and rationale for the specified claim based on the provided context and prior art & Question: Based on the patent application and the cited prior art, what is the rejection code (e.g., 102, 103) and the reason for rejecting claim X? \\
\hline
\multirow{3}{*}{Patent Application}
& title & Title of the invention in the target application & "Biometric monitoring system" \\ \cline{2-4}
& abstract & Abstract of the invention in the target application & "ABSTRACT A system for..." \\ \cline{2-4}
& claims & List of claims in the target application & ["1. A biometric...", ...] \\
\hline
\multirow{6}{*}{Prior Art} 
& patent\_id & Patent/publication ID of a cited prior art document & "US 20070159740" \\ \cline{2-4}
& title & Title of the cited prior art document & "Physiological status..." \\ \cline{2-4}
& abstract & Abstract of the cited prior art document & "A physiological status..." \\ \cline{2-4}
& claims & Claims of the cited prior art document & ["1. A physiological...", ...] \\ \cline{2-4}
& paragraphs.key & Identifier of a paragraph cited from this prior art & 39 \\ \cline{2-4} 
& paragraphs.content & Text content of the cited paragraph & "Having thus described..." \\ 
\hline
\multicolumn{2}{c|}{\makecell{Answer Code}} & Examiner's rejection decision code (e.g., 102, 103) & "102" \\ \cline{1-4}
\multicolumn{2}{c|}{\makecell{Answer Reason}} & Examiner’s rationale for the rejection & "Regarding claim 1, Williams teaches..." \\
\hline
\end{tabularx}
\vspace{0.2cm}
\caption{Input Structure for the NOC4PC Task} 
\label{tab:noc4pc_input} 
\renewcommand{\arraystretch}{1.0} 
\end{table}

\subsubsection{Task Construction}\label{NOC4PC_construction}

We processed patent application records with complete documentation, each containing a unique identifier and application number. From these records, we extracted essential contextual information including titles, abstracts, and initial claims. Rigorous validation checks were implemented throughout the processing pipeline to filter out records with missing critical information such as application numbers or claims.

For valid application records, we located and processed the corresponding Office Action documents. These documents contain detailed information about the examiner's evaluation of each claim, including rejection status, legal basis for rejection (35 U.S.C. \S 102 or \S 103), and prior art references cited to support the rejection. This information was carefully extracted while maintaining the relationship between claims and their respective rejection reasons.

Aggregating and processing the prior art specifications cited by examiners formed a critical component of our task construction. For each cited patent, we collected comprehensive information including patent numbers, titles, abstracts, claims, and specific paragraphs referenced by examiners. This process required locating and parsing specification files for each cited patent and extracting the exact paragraphs that examiners used to support their rejection decisions. Robust error handling was implemented to manage cases where specification files were unavailable or paragraphs could not be located.

Each task instance created for individual claims includes the application context, a comprehensive collection of prior art specifications with relevant paragraphs, the examiner's decision, and for rejected claims, the specific legal basis and detailed reasoning. This structure mirrors the actual patent examination process, requiring models to analyze claim language against prior art, understand legal standards for patentability, and make determinations aligned with examiners' decisions.

\subsubsection{Data Processing and Evaluation}

To prevent data leakage, the Task dataset was partitioned into training, validation, and test sets at the patent application level, ensuring all claims from the same patent application belonged exclusively to one split. Specifically, unique application numbers were randomly shuffled and then divided into three subsets with ratios of 93\% for training, 5\% for validation, and 2\% for testing.

During evaluation, the language models' predictions were assessed based on their classification accuracy against the ground truth labels. Predictions with processing errors were identified and excluded from metric computations to maintain the integrity of the evaluation.

The primary evaluation metrics were:

\begin{itemize}
    \item \textbf{Accuracy}: Measured the percentage of correctly classified instances, providing an overall assessment of the model's ability to determine the correct patentability decision (i.e., Allowable, \S 102 rejection, or \S 103 rejection).

    \item \textbf{Custom Score}: Calculated as the Macro F1-score (the unweighted mean of F1-scores for each class) multiplied by 100, providing an intuitive percentage-like scale for comparison. This metric is particularly important for our task due to class imbalance, as it gives equal importance to the performance on each rejection type (Allowable, \S 102, and \S 103) regardless of their frequency in the dataset. The Custom Score was used for each rejection code separately as well as for overall performance evaluation.

    \item \textbf{Confusion Matrix}: Generated to visually analyze the distribution of model predictions against the true labels, facilitating identification of common misclassification patterns, such as confusion between \S 102 and \S 103 rejections.
\end{itemize}

For baseline analysis, multiple runs were evaluated and aggregated to calculate statistical measures (mean, standard deviation, minimum, and maximum) of accuracy and Custom Score across all runs, providing a robust assessment of model performance and stability.

While the current evaluation emphasizes classification accuracy and Custom Score, future analyses could extend evaluation to include quantitative comparisons between model-generated rationales (from Chain-of-Thought prompting) and examiner-provided rationales using standard NLP metrics such as ROUGE, BERTScore, or other semantic similarity measures.

\subsubsection{Test Prompts}\label{NOC4PC_prompt}

Both prompts start with a common section defining the role, task, and presenting the target application and prior art data. This common base structure is generated as follows (representing Python f-string construction):

\begin{lstlisting}[extendedchars=true, literate={§}{{\S}}1 {¶}{{\P}}1]
You are an expert AI acting as a U.S. Patent Examiner.
Your task is to analyze **Target Claim {claim_num}** of the **Target Patent Application** in view of the provided **Prior Art Specifications**.

Determine if **Target Claim {claim_num}** is allowable or should be rejected under 35 U.S.C. § 102 (lack of novelty) or 35 U.S.C. § 103 (obviousness).

**Target Patent Application Information:**
*   Application Number: {app_num}
*   Target Claim Number: {claim_num}
*   Title: {target_title}
*   Abstract: {target_abstract}
*   Target Claim {claim_num} Text to Analyze:    ```
    {target_claim_text}    ```

**Prior Art Specifications (Cited as Basis for Potential Rejection):**
The following prior art documents and specific paragraphs were cited as potentially relevant for the rejection of the target claim. Analyze the target claim against the information presented in these specific paragraphs **and the claims** of the prior art.
--- Prior Art: {pa_patent_id} ---
Title: {title}
Abstract:
{abstract}

Claims of {pa_patent_id}:
{formatted_pa_claims} /* Each claim indented */

Cited Paragraphs from {pa_patent_id}:
{formatted_pa_paragraphs} /* Each paragraph key and content indented */
--- End Prior Art: {pa_patent_id} ---
/* (Repeated for each prior art document) */
---
\end{lstlisting}

\paragraph{Zero-Shot Prompt:} These simpler instructions are appended instead, requesting only the final code.

\begin{lstlisting}
**Select Conclusion Code**
Choose one: "ALLOW", "102", or "103".
    *   `"ALLOW"`: If your reasoning concluded the claim is novel and non-obvious over the cited art.
    *   `"102"` (Rejected - Novelty): If your reasoning concluded the claim is anticipated by a single cited reference.
    *   `"103"` (Rejected - Obviousness): If your reasoning concluded the claim is obvious over the cited art.

**Answer Format (JSON only)**
Return ONLY this JSON object - DO NOT INCLUDE ANY REASON IN THE ANSWER.

```json
{{"code": "102"}}
```
\end{lstlisting}

\paragraph{Chain-of-Thought (CoT) Prompt:} The following instructions guiding the reasoning process and specifying the output format are appended.

\begin{lstlisting}
**Select Conclusion Code**
Choose one: "ALLOW", "102", or "103".
    *   `"ALLOW"`: If your reasoning concluded the claim is novel and non-obvious over the cited art.
    *   `"102"` (Rejected - Novelty): If your reasoning concluded the claim is anticipated by a single cited reference.
    *   `"103"` (Rejected - Obviousness): If your reasoning concluded the claim is obvious over the cited art.

### OUTPUT (JSON only)
Think through the steps required to evaluate this, craft the supporting rationale accordingly, and then deliver your answer based on that rationale.
Always write the "reason" first and then write the "answer".
Return exactly one JSON object:

```json
{{
  "reason": "...",
  "code": "102" | "103" | "ALLOW"
}}
```
\end{lstlisting}

\subsubsection{Detailed LLM Performance Analysis}\label{task3_result_detail}
For semantic evaluation, we utilize three metrics: cosine similarity (CS), derived from embeddings generated by \citet{wang2024multilinguale5textembeddings}, BERTScore (BS) \cite{zhang2020bertscoreevaluatingtextgeneration}, which compares token-wise contextual embeddings, BLEURT \cite{sellam2020bleurtlearningrobustmetrics}, which is fine-tuned to reflect human judgment, and lexical- level metrics ROUGE \cite{lin2004rouge}. Table \ref{tab:noc4pc_accuracy} presents performance metrics (custom score and accuracy) for various LLMs across different patent rejection categories (§102, §103, and allowed cases). Table \ref{tab:noc4pc_similarity} provides a deeper analysis of semantic similarity metrics (CS, ROUGE, BS, and BLEURT) for each model broken down by rejection type.

\begin{table}[htbp]
\centering
\resizebox{\textwidth}{!}{%
\begin{tabular}{llrrrrrrrr}
\toprule
& & \multicolumn{4}{c}{Custom Score} & \multicolumn{4}{c}{Accuracy} \\
\cmidrule(lr){3-6} \cmidrule(lr){7-10}
Model & Mode & Overall & §102 & §103 & Allow & Overall & §102 & §103 & Allow \\
\midrule
Baseline &  & 32.33 & 16.8 & 16.68 & 16.65 & 33.46 & 33.71 & 33.39 & 33.3 \\
\midrule
\multirow{2}{*}{GPT-4o} & ZS & 34.69 & 13.21 & 28.45 & 5.72 & 46.60 & 24.70 & 74.47 & 9.38 \\
                        & CoT & 32.19 & 11.67 & 12.88 & \textbf{27.99} & 33.18 & 21.22 & 23.95 & 72.36 \\
\midrule
\multirow{2}{*}{Claude-3.7-Sonnet} & ZS & \textbf{35.84} & 27.01 & 16.84 & \textbf{8.90} & 39.91 & 68.11 & 33.79 & \textbf{15.41} \\
                           & CoT & \textbf{45.40} & 21.83 & \textbf{23.25} & 17.23 & \textbf{48.27} & 48.68 & \textbf{53.54} & 34.84 \\
\midrule
\multirow{2}{*}{Gemini-2.0-Flash} & ZS & 21.06 & 31.96 & 4.89 & 1.63 & 31.14 & 92.09 & 7.91 & 2.51 \\
                                 & CoT & 31.79 & 22.67 & 16.59 & 8.83 & 34.80 & 51.52 & 33.12 & 15.27 \\
\midrule
\multirow{2}{*}{\makecell{Llama-3.1\\-8B-Instruct}} & ZS & 15.71 & 49.79 & 1.16 & 0.00 & 29.26 & 99.16 & 1.17 & 0.00 \\
                               & CoT & 19.56 & 29.67 & 19.81 & 0.08 & 35.68 & 80.22 & 24.71 & 0.17 \\
\midrule
\multirow{2}{*}{\makecell{Qwen2.5-7B\\-Instruct}} & ZS & 28.92 & 100.00 & 0.00 & 0.00 & 14.95 & 100.00 & 0.00 & 0.00 \\
                         & CoT & 20.31 & 10.92 & 2.89 & 22.69 & 28.36 & 27.94 & 6.13 & \textbf{83.08} \\
\midrule
\multirow{2}{*}{\makecell{EXAONE-3.5\\-7.8B-Instruct}} & ZS & 15.00 & \textbf{100.00} & 0.70 & 0.00 & 28.95 & \textbf{100.00} & 0.70 & 0.00 \\
                           & CoT & 24.99 & 12.87 & 14.11 & 10.04 & 35.02 & 34.65 & 39.30 & 25.13 \\
\midrule
\multirow{2}{*}{\makecell{Gemma-3\\-12B-Instruct}} & ZS & 32.54 & 23.63 & 22.66 & 1.63 & 42.34 & 54.92 & 51.48 & 2.51 \\
                                          & CoT & 17.67 & 22.89 & 6.50 & 1.42 & 32.39 & 84.41 & 14.93 & 2.18 \\
\midrule
\multirow{2}{*}{\makecell{Qwen2.5\\-32B-Instruct}} & ZS & 26.88 & 3.34 & \textbf{30.86} & 3.69 & \textbf{46.15} & 5.28 & \textbf{86.27} & 5.86 \\
                                          & CoT & 33.85 & 20.29 & 10.48 & 23.78 & 33.53 & 43.76 & 18.65 & 55.44 \\
\midrule
\multirow{2}{*}{\makecell{EXAONE-3.5\\-32B-Instruct}} & ZS & 23.05 & 15.47 & 21.59 & 0.00 & 32.87 & 30.22 & 47.90 & 0.00 \\
                                          & CoT & 28.47 & 25.42 & 18.15 & 1.00 & 37.03 & 61.64 & 37.41 & 1.53 \\
\midrule
\multirow{2}{*}{\makecell{Gemma-3\\-27B-Instruct}} & ZS & 24.00 & 24.61 & 14.53 & 0.88 & 31.24 & 58.51 & 27.87 & 1.34 \\
                                          & CoT & 22.45 & \textbf{32.05} & 6.20 & 1.63 & 32.45 & \textbf{92.57} & 10.25 & 2.51 \\
\midrule
QWQ-32B & CoT & 34.73 & 24.06 & 9.14 & 22.52 & 34.90 & 56.47 & 15.90 & 51.01 \\
\midrule
EXAONE-Deep-32B & CoT & 21.23 & 22.47 & 8.03 & 3.03 & 35.89 & 82.02 & 19.59 & 6.49 \\
\bottomrule
\end{tabular}
}
\vspace{0.2cm}
\caption{NOC4PC Performance Metrics (N: §102=834, §103=1453, allow=597). Allow cases have no comparison text data. Baseline represents the average of 20 random selection trials.}
\label{tab:noc4pc_accuracy}
\end{table}

\begin{table}[htbp]
\centering
\footnotesize
\begin{tabular}{l|c|c|rrrr}
\toprule
\textbf{Model} & \textbf{Rejection} & \textbf{Custom Score} & \textbf{CS} & \textbf{ROUGE} & \textbf{BS} & \textbf{BLEURT} \\
\midrule
GPT-4o                     & \multirow{12}{*}{Overall} & 32.19 & 0.6042 & 0.2144 & 0.3598 & -0.5403 \\
Claude 3.7 Sonnet          & & \textbf{45.40} & \textbf{0.6196} & 0.1880 & 0.3500 & -0.6146 \\
Gemini 2.0 Flash           & & 31.79 & 0.6120 & \textbf{0.2626} & \textbf{0.3848} & -0.5480 \\
Llama-3.1-8B-Instr.        & & 19.56 & 0.5620 & 0.2391 & 0.3494 & -0.5043 \\
Qwen2.5-7B-Instr.          & & 20.31 & 0.6015 & 0.2407 & 0.3607 & -0.5217 \\
EXAONE-3.5-7.8B-Instr.     & & 24.99 & 0.5649 & 0.1657 & 0.2931 & -0.6240 \\
Gemma-3-12B-Instr.         & & 17.67 & 0.5965 & 0.2399 & 0.3768 & -0.5376 \\
Qwen2.5-32B-Instr.         & & 33.85 & 0.5701 & 0.2386 & 0.3713 & \textbf{-0.4927} \\
EXAONE-3.5-32B-Instr.      & & 28.47 & 0.5838 & 0.1989 & 0.3380 & -0.5302 \\
Gemma-3-27B-Instr.         & & 22.45 & 0.6051 & 0.2174 & 0.3652 & -0.5314 \\
QWQ-32B                    & & 34.73 & 0.5633 & 0.2037 & 0.3482 & -0.5086 \\
EXAONE-Deep-32B                & & 21.23 & 0.5453 & 0.1833 & 0.3294 & -0.5043 \\
\midrule
GPT-4o                     & \multirow{12}{*}{§102} & 11.67 & 0.6266 & 0.2145 & 0.3670 & -0.5559 \\
Claude 3.7 Sonnet          &  & 21.83 & \textbf{0.6400} & 0.1927 & 0.3625 & -0.6129 \\
Gemini 2.0 Flash           &  & 22.67 & 0.6305 & \textbf{0.2764} & \textbf{0.4044} & -0.5154 \\
Llama-3.1-8B-Instr.        &  & \textbf{29.67} & 0.5778 & 0.2523 & 0.3689 & \textbf{-0.4668} \\
Qwen2.5-7B-Instr.          &  & 10.92 & 0.6261 & 0.2493 & 0.3714 & -0.5089 \\
EXAONE-3.5-7.8B-Instr.     &  & 12.87 & 0.5817 & 0.1556 & 0.2983 & -0.6511 \\
Gemma-3-12B-Instr.         &  & 22.89 & 0.6179 & 0.2399 & 0.3768 & -0.5376 \\
Qwen2.5-32B-Instr.         &  & 20.29 & 0.5798 & 0.2398 & 0.3818 & -0.4857 \\
EXAONE-3.5-32B-Instr.      &  & 25.42 & 0.6036 & 0.2017 & 0.3464 & -0.5324 \\
Gemma-3-27B-Instr.         & & 32.05 & 0.6245 & 0.2189 & 0.3769 & -0.5267 \\
QWQ-32B                    &  & 24.06 & 0.5741 & 0.2088 & 0.3603 & -0.4937 \\
EXAONE-Deep-32B                &  & 22.47 & 0.5531 & 0.1887 & 0.3460 & -0.4773 \\
\midrule
GPT-4o                     & \multirow{12}{*}{§103} & 12.88 & 0.5914 & 0.2144 & 0.3557 & -0.5313 \\
Claude 3.7 Sonnet          &   & \textbf{23.25} & \textbf{0.6078} & 0.1853 & 0.3428 & -0.6156 \\
Gemini 2.0 Flash           &   & 16.59 & 0.6012 & \textbf{0.2546} & 0.3734 & -0.5668 \\
Llama-3.1-8B-Instr.        &   & 19.81 & 0.5529 & 0.2314 & 0.3382 & -0.5234 \\
Qwen2.5-7B-Instr.          &   & 2.89  & 0.5872 & 0.2358 & 0.3546 & -0.5292 \\
EXAONE-3.5-7.8B-Instr.     &   & 14.11 & 0.5552 & 0.1572 & 0.2901 & -0.6085 \\
Gemma-3-12B-Instr.         &   & 6.50  & 0.5842 & 0.2486 & \textbf{0.3962} & -0.5106 \\
Qwen2.5-32B-Instr.         &   & 10.48 & 0.5645 & 0.2379 & 0.3653 & \textbf{-0.4967} \\
EXAONE-3.5-32B-Instr.      &   & 18.15 & 0.5725 & 0.1973 & 0.3331 & -0.5290 \\
Gemma-3-27B-Instr.         & & 6.20 & 0.5939 & 0.2165 & 0.3585 & -0.5341 \\
QWQ-32B                    &   & 9.14  & 0.5571 & 0.2008 & 0.3412 & -0.5171 \\
EXAONE-Deep-32B               &   & 8.03  & 0.5529 & 0.2314 & 0.3382 & -0.5234 \\
\bottomrule
\end{tabular}
\vspace{0.2cm}
\caption{NOC4PC Similarity Metrics for Zero-shot Mode. CS indicates cosin similarity, ROGUE indicates the RougeL-F1 score, BS indicates the BertScore-F1, and BLEURT indicates the BLEURT score.}
\label{tab:noc4pc_similarity}
\end{table}

\subsection{Chain-of-Thought Prompting}\label{app:chain-of-thought}
\begin{table}[htbp]
\centering
\footnotesize
\renewcommand{\arraystretch}{1.2}
\begin{tabularx}{\textwidth}{lcccccc}
\toprule
\multirow{2}{*}{\textbf{Model}} & \multicolumn{2}{c}{\textbf{PAR4PC}} & \multicolumn{2}{c}{\textbf{PI4PC}} & \multicolumn{2}{c}{\textbf{NOC4PC}} \\
\cmidrule(lr){2-3} \cmidrule(lr){4-5} \cmidrule(lr){6-7}
 & CoT & CoT & CoT & CoT & CoT & CoT \\
  & (Base) & (Custom Guide) & (Base) & (Custom Guide) & (Base) & (Custom Guide) \\
\midrule
GPT-4o & \textbf{56.95} & \textbf{48.11} & \textbf{62.62} & 59.91 & 32.19 & 38.71 \\
Claude-3.7-Sonnet & 40.29 & 42.18 & 60.55 & 56.57 & \textbf{45.40} & \textbf{42.95}  \\
Gemini-2.0-flash & 43.61 & 39.17 & 61.72 & \textbf{60.91} & 31.79 & 30.42 \\
\bottomrule
\end{tabularx}
\vspace{0.2cm}
\caption{Performance comparison of three LLMs across three tasks (PAR4PC, PI4PC, NOC4PC) between two CoT prompts. CoT (base) is a prompt that asks the LLMs to generate their own chain of thought, whereas CoT (custom guide) is a prompt that directs the LLMs to reason step by step according to USPTO patent-examination guidelines.} \label{tab:task_results_cot}
\renewcommand{\arraystretch}{1.0}
\end{table}

We conducted preliminary targeted experiments to investigate whether explicitly embedding step-by-step instructions in chain-of-thought (CoT) prompts improves performance. Two prompt variants were compared. We designed step-by-step custom instructions in a prompt based on the USPTO guidelines for examination of applications\footnote{https://www.uspto.gov/web/offices/pac/mpep/s2107.html} with the Broadest Reasonable Interpretation (BRI) standard\footnote{https://www.uspto.gov/web/offices/pac/mpep/s2111.html}, which are the guidelines for interpreting claims. After roughly ten rounds of prompt engineering, we settled on the best-performing version of this template. The second variant, our baseline, provided only high-level task instructions and allowed the model to generate its own reasoning.

For each of the three tasks—PAR4PC, PI4PC, and NOC4PC—we prepared both prompt types and evaluated them on GPT-4o, Claude-3.7-Sonnet, and Gemini-2.0-Flash with the temperature fixed at 0. The results are summarized in Table \ref{tab:task_results_cot}. Overall, prompts allowing the LLM to generate its own chain of thought consistently outperformed those with manually crafted USPTO-based instructions.

We speculate that the low score results from the current prompt being too simplistic to capture the complex and subtle nature of the patent examination, as it was not created by patent examiners or experts. Therefore, in this experiment, we adopted the method of having LLM generate reasoning directly, and found the need to explore a prompting technique that includes experts to improve performance in the future.

\subsubsection{Custom Guide Chain-of-Thought Prompt of PAR4PC}
\begin{lstlisting}[extendedchars=true, literate={§}{{\S}}1 {¶}{{\P}}1 {≤}{{$\le$}}1 {≥}{{$\ge$}}1]
**Step-by-Step Instrcution**  
*Apply the Broadest Reasonable Interpretation (BRI) standard.*

1. **BRI Claim Charting**  
   - Decompose Claim {claim_number} into numbered limitations [L1]-[Ln] and record element : function/relationship.

2. **Core Inventive Concept & Problem**  
   - Summarise in ≤ 20 words the inventive concept + technical problem.

3. **Single-Reference Screening (§102)**  
   - For each option (A-H) rate coverage:  
     | Opt | Maps limits | Term/synonym | Field match | Score* |  
     |-----|------------|---------------|-------------|--------|  
     *Score: 0 = no key feature, 1 = partial, 2 = full anticipation.*

4. **Multi-Reference Analysis (§103)**  
   a. Pick options with Score ≥ 1.  
   b. Build coverage matrix to find smallest combo covering all limits.  
   c. For each viable combo, supply a motivation-to-combine (same field, complementary function, predictable substitution, etc.).  
   d. Rank: full coverage -> clear motivation -> earliest primary art.

5. **Consistency & Inherency Check**  
   - Reject art that contradicts any limitation; accept inherent feature only if necessarily present.

6. **Output (JSON only)**  
    Always write the "reason" **first** and then write the "answer".
   - "reason" MUST include:  
     Step1 <claim focus>; Step2 <mapping & motivation> ; Step3 <§102 or §103>.  
   - Keep "reason" ≤ 200 words.
   - "answer" = single letter **or** list of letters.

```json
{{"reason":"Step1 ... ; Step2 ... ; Step3 ...", "answer":"A"}}] 
```
If multiple patents are cited":
```json
{{"reason":"Step1 ... ; Step2 ... ; Step3 ...", "answer": ["A","C","F"]}}]
```
\end{lstlisting}

\subsubsection{Custom Guide Chain-of-Thought Prompt of PI4PC}
\begin{lstlisting}[extendedchars=true, literate={§}{{\S}}1 {¶}{{\P}}1 {=>}{{=>}}1 {≤}{{$\le$}}1 {≥}{{$\ge$}}1]
**Step-by-Step Method**
*Use the Broadest-Reasonable-Interpretation (BRI) standard throughout.*

1.  **BRI Claim Deconstruction**
    - Break the claim {claim_num} into **numbered limitations** (e.g., [1A]-[1F]).
    - Write each limitation in examiner-style "element : function / relationship" form.
    - Try to include as much of the claim as possible.

2.  **Key Distinguishing Feature(s)**
    - Identify which limitation(s) the applicant asserts as novel / non-obvious.
    - List all the features that should be considered when evaluating novelty and non-obviousness.

3.  **Prior-Art Mapping Table (one table per option paragraph)**
    - For each of the five option paragraphs, provide a detailed mapping to Claim {claim_num} elements.
    - Use the table format to score the degree of overlap between each option paragraph and the claim limitations.
    - IMPORTANT: **Do not skip any options** - evaluate all five paragraphs.
    | Opt# | Maps to elements | Exact term / BRI synonym | Col-Line (or ¶) | Match score* |
    |------|---------------|-----------------------|---------------|-----------|
    *Scoring: 0 = missing, 1 = mention, 2 = partial, 3 = full & explicit.*

4.  **Select the Most Relevant Paragraph for Patentability Evaluation**
    - Your goal is to identify exactly ONE paragraph most relevant to evaluate the novelty/non-obviousness of the applicant's claimed invention.
    - Select exactly one paragraph based on its relevance to the novelty or non-obviousness of the Key Distinguishing Features (KD-x) in Claim {claim_num}.
    - Do not select multiple keys or provide general reasoning.
    - Focus on technical relevance, improvements, and system integration when selecting your paragraph.
    Selection Criteria:
    - Consider paragraphs that scored ≥ 1 points in Step 3.
    - Technical Objectives: Does the paragraph directly support the technical objectives of the Claim? Does it provide a solution to the problem presented by the Claim?
    - Prior Art Improvements: Does the paragraph present innovative improvements to existing systems or technologies?
    - System Integration: Does the paragraph explain how elements of the system described in the Claim interact or integrate with each other?
    - Motivation to Combine: Does the paragraph offer a motivational context for combining features, particularly for a §103 rejection?

    Output Requirements:
    - Clearly indicate your final selection as the Primary Reference (PR).
    - Provide a concise reason for your selection based strictly on the criteria above.

5. **Consistency & Inherency Check**
   - Verify the selected paragraph does not contradict any claim limitation.

6. **Output (JSON only)**
    Always write the "reason" first and then write the "answer".
   - 'reason' MUST list Step1-Step6 in order, each separated by ';'.
     - Step1 <statutory/context> ;
     - Step2 <limits> ;
     - Step3 <key feature> ;
     - Step4 <mapping & score> ;
     - Step5 <rank/tie-break> ;
     - Step6 <consistency/inherency & §102 or §103 result>.
   - Keep "reason" ≤ 1000 words.
   - "answer" = single paragraph key (int).

```json
{{"reason":"Step1 ... ; Step2 ... ; Step3 ... ; Step4 ...; Step5 ...; Step6 ...","answer": 17}}
```
\end{lstlisting}

\subsubsection{Custom Guide Chain-of-Thought Prompt of NOC4PC}
\begin{lstlisting}[extendedchars=true, literate={§}{{\S}}1 {¶}{{\P}}1 {≤}{{$\le$}}1]
**Analysis Task and Response Instructions:**

Perform your internal reasoning first, then draft the *Office-Action-style* text.
---

### INTERNAL REASONING (not shown to applicant)
1. Apply the Broadest-Reasonable-Interpretation (BRI) to Claim {claim_num}; chart every limitation [L1]-[Ln].
   1-a. *Statutory check* - confirm Claim{claim_num} fits a statutory class (process, machine, manufacture, composition).
   1-b. *Limitation numbering*-  break the claim into [L1]-[Ln] and record in "element : function / relationship" form.
   1-c. *Key-feature flag* - mark limitations asserted (or apparent) as novel / non-obvious.
2. Compare each limitation to the teachings (claims + cited paragraphs) of every prior-art reference.
3. Decide:
   - §102 anticipation if a single reference explicitly, implicitly, or inherently discloses each and every limitation.
        Under BRI, interpret broadly: functional equivalence or conventional components (processors, databases, modules, memory, standard network elements, known protocols, etc.) count as implicit disclosures.
    - §103 obviousness if any of the following apply:
        (a) A primary reference discloses at least 70% of the limitations explicitly or implicitly, and remaining limitations constitute routine modifications, predictable optimizations (e.g., efficiency, speed, cost reduction, miniaturization), or standard practices known to a person of ordinary skill in the field.
        (b) A combination of references collectively covers all limitations and demonstrates a clear, implicit or explicit KSR rationale, such as addressing the same technical problem, improving performance, enhancing usability, or following common industry practices.
        (c) The limitations not explicitly disclosed are obvious through common general knowledge or widely recognized industry standards or textbooks in the field.
    - ALLOW Only if:
    No single reference or combination of references, even considering implicit disclosures and general knowledge, discloses or renders obvious specific, detailed implementation aspects (unique structures, algorithmic specifics, or non-trivial process steps), AND
    No reasonable motivation or rationale (performance improvement, standard practice, or known solution) can be objectively articulated to bridge these gaps.

---

### DRAFT OA LANGUAGE (will be revealed)
Write the *reason* paragraph exactly like an Office Action:

* Start: **"Regarding Claim {claim_num}, ..."**
* Use examiner diction:
  - "Reference X (Col. Y, lines Z) discloses ..."
  - "Therefore, Claim {claim_num} is rejected under 35 U.S.C. § 102(a) as being anticipated by Reference X."
  - or "It would have been obvious to one of ordinary skill to modify X with Y (same field, predictable results) ... => § 103 rejection."
  - or "The cited references do not teach or render obvious limitation [Lk] ... Claim {claim_num} is allowable."

* If § 103, list **all** references in the combination (e.g., "in view of Smith '123").
* Cite at least one column-line or paragraph for each matched limitation.
* Keep length ≤ 200 words.

---

### OUTPUT (JSON only)
Always write the "reason" first and then write the "answer".
Return exactly one JSON object:

```json
{{
  "reason": "<OA-style paragraph above>",
  "code": "102" | "103" | "ALLOW"
}}
```
\end{lstlisting}



\subsection{Experimental Details}\label{resource}
We describe the detailed settings for conducted experiments. For evaluation, closed-source models were accessed through their respective vendor APIs, while all open-source checkpoints were downloaded from the HuggingFace Hub. Inference was performed with vLLM on $4 {\times}$ NVIDIA A100 (40GB) GPUs, using a deterministic decoding temperature of $0$ for every experiment.

Supervised fine-tuning (SFT) was conducted on a single compute node equipped with $8 {\times}$ NVIDIA H100 (80GB) GPUs. All tasks and models shared identical training hyper-parameters: learning rate of $1.0 \times 10^{-6}$ with cosine scheduling, global batch size of $16$, $8$ gradient accumulation steps (yielding $128$ training samples per optimization step), and a total of $2$ training epochs, and DeepSpeed ZeRO-3 as the distributed training strategy.

\section{Ethical and Societal Considerations}
\subsection{Ethical Considerations}\label{ethical_consideration}
The PANORAMA dataset consists exclusively of publicly available patent examination documents obtained from the USPTO. As all documents are publicly accessible, individual consent is not required. The dataset does not contain any private or personally identifiable information (PII) or sensitive personal data.

\subsection{Potential Societal Impacts}\label{societal_impact}
We anticipate that PANORAMA will offer a comprehensive view of patent examiners' decision trails and the rationales behind them. Nonetheless, we recognize two potential societal risks. First, because PANORAMA enables LLMs to emulate aspects of patent examination, there is a concern that users might over-rely on AI systems—or even attempt to replace human expertise altogether—in a process that demands nuanced legal judgment and deep technical knowledge. How non-experts should responsibly employ AI in patent practice, and how AI systems can best collaborate with trained examiners, remain open research questions. Second, PANORAMA is confined to U.S. patent law and may not transfer cleanly to other jurisdictions. Models trained on this dataset could also learn and perpetuate biases embedded in historical examination records. Careful evaluation and mitigation strategies are therefore necessary to avoid reinforcing such biases or inadvertently disadvantaging particular groups or industries.

\end{document}